\documentclass[amsmath,prb
showpacs, elsarticle, twocolumn] {revtex4}
\usepackage{bm}
\usepackage{enumerate}
\usepackage{graphicx}
\usepackage[dvips]{epsfig}
\usepackage{epsf}
\usepackage{xcolor}
\makeatletter

\newcommand{\Rmnum}[1]{\expandafter\@slowromancap\romannumeral #1@}

\makeatletter

\begin{document}
\title{De Haas-van Alphen effect and quantum oscillations as a function of temperature in correlated insulators}

\author{Vladimir A. Zyuzin}
\affiliation{L.D. Landau Institute for Theoretical Physics, 142432, Chernogolovka, Russia}
\begin{abstract}
We theoretically study a model of excitonic insulators which show de Haas-van Alphen oscillations as well as periodic dependence of the magnetization on inverse temperature.
The insulating behavior is due to the Coulomb interaction driven hybridization of fermions at the crossing point of their energy bands. We study this hybridization self-consistently and recover known results by A. Allocca and N. Cooper, SciPost Phys. {\bf 12}, 123 (2022). In addition to known results, we show that the hybridization gap decreases with the magnetic field which corresponds to the diamagnetism.
Furthermore, we show that the amplitude of the de Haas-van Alphen is oscillating with inverse temperature with a period defined by a combination of the hybridization gap and magnetic field. We analytically obtain the position and the height of the first and dominant peak of these oscillations.
\end{abstract}
\maketitle

\section{Introduction}
We are motivated by recent experiments where quantum oscillations in Kondo insulator SmB$_{6}$ were observed \cite{LiScience2014,SuchitraScience2015,SuchitraNatPhys2017,SuchitraScience2020}. This material becomes insulating below about $T_{\mathrm{c}}=40 \mathrm{K}$ due to the gapping of the Fermi surface.
Observed quantum oscillations occur only in the magnetization, i.e. de Haas - van Alphen (dHvA) effect, while electric resistivity doesn't show any oscillations (absence of Shubnikov - de Haas effect).
The latter is due to after all insulating behavior, while occurrence of dHvA oscillations in insulators poses a mystery. 
Not to mention that the observed frequency of dHvA oscillations of the insulator is that of the metallic phase of the system, before it turned insulating via a gapping mechanism. 
In addition, the frequency is similar to the one of the LaB$_{6}$ metallic material, which never turns insulating and which has a similar band structure with the metallic phase of the SmB$_{6}$.

There are theories which propose emergence of neutral quasiparticles with Fermi surface that somehow nevertheless couple to the magnetic field to show dHvA effect but not the electric field, or of fermions that fractionalize to either Majoranas or other exotic structures \cite{BaskaranArxiv,SodemannChowdhurySenthil,VarmaPRB}. Whether these theories can explain the experiments \cite{LiScience2014,SuchitraScience2015,SuchitraNatPhys2017,SuchitraScience2020} is currently under debate.
There are theories which obtain dHvA quantum oscillations in insulators based on the non-interacting fermions \cite{KnolleCooperPRL15,TopodHvA2016,AlisultanovJETP2016,PalPRB2016,PalArxiv2022}.
SmB$_{6}$ insulator is known to be strongly correlated (see for example \cite{Hewson}), and it is rather odd to expect that the non-interacting picture will fully explain its physical properties.

%------------------------------------------------------------------------------
\begin{figure}[t] 
\centerline{
\includegraphics[width=0.6 \columnwidth]{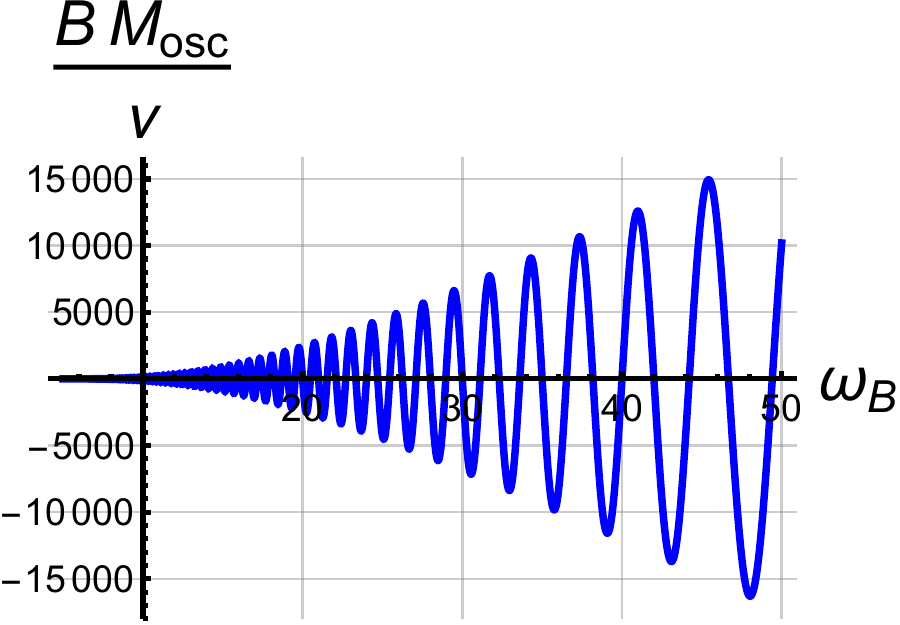}
}
\protect\caption{Plot of the first harmonic of the oscillating part of the torque, i.e. of $M_{osc}B$ where $M_{\mathrm{osc}}$ is given by Eq. (\ref{result1}) for $2\theta_{0} \approx 40K$, effective mass is picked to be $m=0.2 m_{\mathrm{e}}$, at $T=0$, and $\mu m/m_{\mathrm{e}} = 420K$, and $\nu = \frac{m}{2\pi}$ where $m$ is an effective mass of $d-$fermions. }

\label{fig:fig1}   

\end{figure}
%------------------------------------------------------------------------------
In our theory we adopt the Keldysh-Kopaev model of excitonic insulator \cite{KeldyshKopaev}, where two fermions hybridize with each other via the repulsive Coulomb interaction (please see \cite{Miyake,AlloccaCooper}). 
As a result, a gap opens up at their energy band crossing points resulting in an insulator when the Fermi level is at the crossing point.  
Free energy gets minimized as a result of the hybridization.
The magnetic field quantizes the energy spectrum of fermions forming Landau levels. 
Then an intuitive picture of the effect of quantum oscillations in insulators is that when the Landau levels of two fermion branches cross each other, the hybridization is maximised, while when the Landau levels distance apart, the hybridization is minimized. 
To see this effect analytically, the hybridization must be treated self-consistently and obtained from the non-linear equation. 
Finally, since the free energy gain is proportional to the second power of the hybridization, the magnetization and specific heat are expected to be oscillating functions of the magnetic field. 
It turns out that the frequency of oscillations equals to the energy of the energy bands crossing point, and is related to the Fermi energy of the metallic phase before the hybridization (this fact was known since \cite{KnolleCooperPRL15}). 
The response of the system due to the described mechanism is found to be diamagnetic (Landau diamagnetism) and the non-oscillating part of the magnetization is proportional to the magnetic field $B$. 
The oscillating part of the magnetization $M$ at small magnetic fields and at zero temperature can be best described with
\begin{align}\label{result1}
M_{\mathrm{osc}} = - \frac{2\mu\nu e}{\pi mc}  \left(\frac{2\pi \theta_{0}}{\omega_{\mathrm{B}}}\right)^{\frac{1}{2}}
\sin\left(\frac{2\pi \mu}{\omega_{\mathrm{B}}}  \right) e^{-4\pi \frac{\theta_{0}}{\omega_{\mathrm{B}}} },
\end{align}
where $\mu$ is the Fermi energy of the metallic phase and $2\theta_{0}$ is the non-oscillating part of the hybridization gap, $\omega_{\mathrm{B}} = \frac{eB}{mc}$ is the cyclotron frequency, $\nu = \frac{m}{2\pi}$, and $m$ is the effective mass of fermions.
We have set $\hbar \equiv 1$ and $k_{\mathrm{B}} \equiv 1 $ everywhere throughout the text. 
Smallness of the magnetic field is given by the $8\pi \theta_{0} > \omega_{\mathrm{B}}$ condition, which falls in in to the experimental regime.
Despite of the exponential suppression, in reality of the experiments in which $2\theta_{0} \approx 40K$ with the effective mass picked to be $m=0.2 m_{\mathrm{e}}$ (where $m_{\mathrm{e}}$ is bare electron's mass), Eq. (\ref{result1}) gives sizable contribution at small magnetic fields. We plot oscillating part of the torque in Fig. (\ref{fig:fig1}). 
We note that Eq. (\ref{result1}) has been recently derived in \cite{AlloccaCooper}.
Having presented results for zero temperature, in the rest of the paper we outline analytical arguments which lead to them and expand the list of our results to the non-zero temperature.
Each step of the calculations is outlined in details in Appendices to the paper.

%------------------------------------------------------------------------------------------------------------------------------------------------------
%------------------------------------------------------------------------------------------------------------------------------------------------------
\section{Theoretical model}
\begin{figure}[t] 
\centerline{
\begin{tabular}{cc}
\includegraphics[width=0.35 \columnwidth]{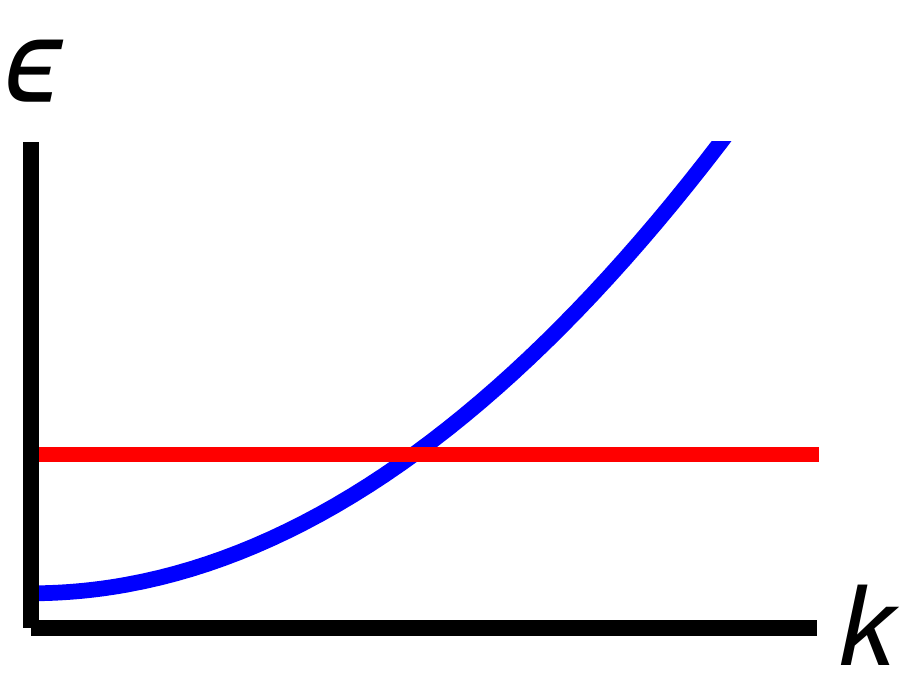}~~~&
\includegraphics[width=0.35 \columnwidth]{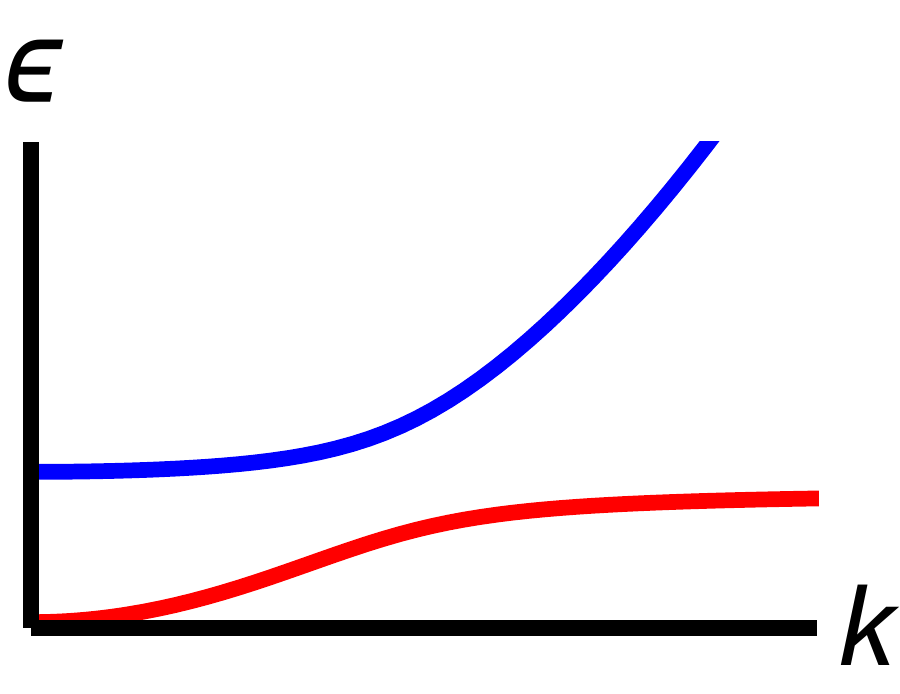}\\
\end{tabular}
}
\protect\caption{Spectrum of fermions before (left) and after (right) the hybridization. The hybridization gap $2\theta$ opens at the crossing point.}

\label{fig:fig2}  

\end{figure}

%------------------------------------------------------------------------------------------------------------------------------------------------------
%------------------------------------------------------------------------------------------------------------------------------------------------------
We consider a model Hamiltonian keeping in mind a SmB$_6$ system. 
Our model system consists of two different fermions, one of which is localized having flat dispersion while the second one is dispersive. 
The Hamiltonian of non-interacting fermions is
\begin{align}\label{Hamiltonian}
H_{0} = \int_{\bf k} \bar{\psi}\left[\begin{array}{cc} \xi_{\bf k} & 0 \\ 0 & \epsilon_{0} \end{array} \right]\psi 
\equiv \int_{\bf k} \bar{\psi} h \psi,
\end{align}
where $\xi_{\bf k} = \frac{{\bf k}^2}{2m}-\mu $, notation $\int_{\bf k} (..) = \int \frac{d{\bf k}}{(2\pi)^2}(..)$ was used, then $\epsilon_{0} =E_{0} - \mu = \mathrm{const}$, which we will set $\epsilon_{0} = 0$ without losing generality, where $\mu$ is the chemical potential,
and
$
\bar{\psi} = \left[ \bar{\phi}_{\mathrm{d}} ,~ \bar{\phi}_{\mathrm{f}} \right], 
$ and the same for 
$\psi$.
In order to present analytic arguments we assume that the system is two-dimensional. 
Three-dimensional case is slightly more analytically involved, but will also be described by the presented below arguments. 
We consider spinless fermions for simplicity, and note that all of the results below will not drastically change if the $f-$ fermion's spectrum had a slight dispersion.
Interaction between the two types of fermions is
\begin{align}\label{interaction}
H_{\mathrm{int}} = U \int_{x} \bar{\phi}_{\mathrm{d}}(x)\phi_{\mathrm{d}}(x) \bar{\phi}_{\mathrm{f}}(x)\phi_{\mathrm{f}}(x),
\end{align}
where $U>0$ corresponds to repulsion.
We decouple interaction Eq. (\ref{interaction}) using the Hubbard-Stratonovich transformation, and keeping in mind the transformation is made within the action, we write the result,
$
-\left(\bar{\theta} + \bar{\phi}_{\mathrm{d}}\phi_{\mathrm{f}} U \right) U^{-1}\left(\theta + U\bar{\phi}_{\mathrm{f}}\phi_{\mathrm{d}} \right)
+
U \bar{\phi}_{\mathrm{d}}\phi_{\mathrm{f}} \bar{\phi}_{\mathrm{f}}\phi_{\mathrm{d}}
=
-\bar{\theta}U^{-1}\theta - \bar{\phi}_{\mathrm{d}}\phi_{\mathrm{f}} \theta - \bar{\theta}\bar{\phi}_{\mathrm{f}}\phi_{\mathrm{d}}.
$
Bilinear in fermion fields term introduces hybridization between $d-$ and $f-$ fermions, described by the Hamiltonian
\begin{align}\label{interaction2}
H_{\mathrm{int};1} = \int_{\bf k} \bar{\psi}\left[\begin{array}{cc} 0 & \theta \\ \bar{\theta} & 0 \end{array} \right]\psi 
\equiv \int_{\bf k} \bar{\psi} h_{\mathrm{int};1} \psi.
\end{align}
Assuming that $\theta$ and $\bar{\theta}$ are constants in space (mean field approximation), the spectrum of fermions described by Hamiltonian $h+h_{\mathrm{int};1}$ becomes
\begin{align}\label{spectrum}
\epsilon_{{\bf k},\pm} = \frac{\xi_{\bf k} + \epsilon_{0}}{2} \pm \sqrt{ \left(\frac{ \xi_{\bf k} - \epsilon_{0}}{2} \right)^2  +\bar{\theta}\theta},
\end{align}
which is schematically plotted in Fig. (\ref{fig:fig2}). There is a gap opening at the cross-section of the energy bands of $d-$ and $f-$ fermions due to the hybridization. 
%-------------------------------------------------------------------------
%-------------------------------------------------------------------------
Next, we need to find the structure of the hybridization. For that, as usual, we integrate our fermions and minimize the action with respect to the hybridization, and get self-consistent equation for the hybridization,
\begin{align}\label{EqHyb}
\theta = U \theta 
\int_{\bf k} 
\frac{{\cal F}_{\epsilon_{{\bf k},+}} 
-
{\cal F}_{\epsilon_{{\bf k},-}} }{\epsilon_{{\bf k},+} - \epsilon_{{\bf k},-}},
\end{align}
where ${\cal F}_{\epsilon}= \tanh\left(\frac{\epsilon}{2T}\right)$ is the Fermi-Dirac distribution function where $T$ is the temperature. Step by step derivation leading to this equation is outlined in Appendix \ref{sectionA}.

A magnetic field perpendicular to the plane of the system is now added. 
It results in the Landau quantization of the energy spectrum, such that $\xi_{\bf k} \rightarrow \xi_{n} = \omega_{\mathrm{B}}(n+\frac{1}{2})-\mu$ must be made in the spectrum, where $n>0$ is an integer denoting corresponding Landau level. 
Other than that none of the steps leading to the self-consistent equation Eq. (\ref{EqHyb}) are changed. 
One may wonder if the interaction Eq. (\ref{interaction}) will acquire a form factor due to the magnetic field.
To see that there is no form factor we keep Eq. (\ref{interaction}) in the coordinate space, perform the Hubbard-Stratonovich transformation, assume mean field and only then change the basis to Landau levels wave-functions.
In this way the Hamiltonian describing hybridization Eq. (\ref{interaction2}) stays the same in the magnetic field. 
We then rewrite Eq. (\ref{EqHyb}) for the system in the applied perpendicular magnetic field,
\begin{align}\label{EqMF}
1 =  U\nu \omega_{\mathrm{B}} \sum_{n} 
\frac{{\cal F}_{\epsilon_{n,+}} 
-
{\cal F}_{\epsilon_{n,-}} }{\sqrt{(\omega_{\mathrm{B}}n+\frac{\omega_{\mathrm{B}}}{2} - \mu)^2 +4\bar{\theta}\theta}},
\end{align}
where now summation is over the Landau levels, and where $\nu = \frac{m}{2\pi}$ is the density of states. 
We note that the summation is over all $n>0$ Landau levels. 
The distribution functions in Eq. (\ref{EqMF}) may cut the sum whenever their difference will be zero.
To sum over the Landau levels, we utilize Poisson summation formula
\begin{align}\label{Poisson}
\sum_{n=0}g(n) = \int_{0}^{\infty}g(x) dx + \sum_{p\neq 0}\int_{0}^{\infty} e^{i2\pi p x} g(x) dx,
\end{align}
where $g(n)$ is some function. It is safe to use the Poisson summation formula in our case as the sum in Eq. (\ref{EqMF}) is over all Landau levels.
In our further derivations we assume $\mu \gg \omega_{\mathrm{B}}$ which allows to set lower limit of certain integrations to $-\infty$ (see \cite{LK,LP} and Appendix \ref{sectionLK}). 
We will first obtain all of the results for $T=0$ and then analyze the $T\neq 0$ case. 
Setting $T=0$ and recalling that the Fermi energy is in the gap, the upper band becomes completely unoccupied meaning ${\cal F}_{\epsilon_{n,+}}=1$, and the lower band becomes completely filled meaning ${\cal F}_{\epsilon_{n,-}} = -1$.
As a result, the non-linear equation in this case can be solved (see Appendix \ref{sectionB} for more details) approximately with
\begin{align}\label{solution1}
\theta =\theta_{0}
 - 2 \theta_{0}  \cos\left(\frac{2\pi \mu}{\omega_{\mathrm{B}}} \right) e^{- \frac{4\pi\theta_{0}}{\omega_{\mathrm{B}}} }   \sqrt{\frac{\omega_{\mathrm{B}}}{2\pi \theta_{0} }}  ,
\end{align}
in the limit $8\pi \frac{\theta_{0}}{\omega_{\mathrm{B}}} > 1$, which allowed us to pick only $p=\pm 1$ in the Poisson summation formula because harmonics with $n>1$ decay with a $e^{-\frac{4\pi\theta_{0}}{\omega_{\mathrm{B}}} n} $ factor. 
Here $\theta_{0} = \sqrt{\mu^2 - \frac{\omega_{\mathrm{B}}^2}{4}} e^{-\frac{1}{4U\nu}}$ is the non-oscillating part of the solution. 
We note that Eq. (\ref{solution1}) was obtained in Ref. \cite{Miyake} for superconductors and in Ref. \cite{AlloccaCooper} for excitonic insulators. 
Having completed standard preliminary steps of the derivation, below we present original results for the magnetization at non-zero temperatures.

%------------------------------------------------------------------------------------------------------------------------------------------------------
%------------------------------------------------------------------------------------------------------------------------------------------------------
%------------------------------------------------------------------------------------------------------------------------------------------------------
\section{Free energy and magnetization}
%------------------------------------------------------------------------------------------------------------------------------------------------------
%------------------------------------------------------------------------------------------------------------------------------------------------------
%------------------------------------------------------------------------------------------------------------------------------------------------------
Here we derive magnetization from the free energy of the system. 
The magnetization is defined as the minus derivative of the free energy with respect to the magnetic field. 
The free energy is given by 
\begin{align}\label{FreeFullMT}
F &= 
-T \nu\omega_{\mathrm{B}} \sum_{n} \ln\left( 1+e^{\frac{-\epsilon_{n;+}}{T}}  \right)
\\
&
-T \nu\omega_{\mathrm{B}} \sum_{n} \ln\left( 1+e^{\frac{-\epsilon_{n;-}}{T}}  \right)
+\frac{\theta^2}{2U}, \nonumber
\end{align}
from where the non-linear self-consistency equation is obtained by setting to zero variation of $F$ with respect to $\theta$. The first two terms are standard while the third term comes from the Hubbard-Stratonovich fields which decoupled the interaction between fermions. Dispersion $\epsilon_{n;\pm}$ is defined in Eq. (\ref{spectrum}) with the substitution $\xi_{\bf k} \rightarrow \xi_{n} = \omega_{\mathrm{B}}(n+\frac{1}{2})-\mu$.

We first calculate the energy at $T=0$. All technical details are outlined in Appendix \ref{sectionC}.
In this case the distribution functions become step functions, and we write  
\begin{align}\label{FreeResult}
F =F_{\mathrm{N}} - \nu\theta_{0}^2 + F_{\Sigma},
\end{align}
the first term is a non-oscillating part of the free energy of a non-interacting Fermi gas. 
$F_{\Sigma}$ is the part of the free energy corresponding to the oscillation in magnetic field of the non-interacting system, i.e. a system with a constant hybridization $\theta_{0}$. 
Recall, that the hybridization is given by Eq. (\ref{solution1}). We have
\begin{align}\label{freeOscBand}
F_{\Sigma} = - \frac{1}{2}\nu\omega_{\mathrm{B}}^2\sum_{p \neq 0}\int_{-\frac{\mu}{\omega_{\mathrm{B}}}-\frac{1}{2}}^{\frac{\Lambda}{\omega_{\mathrm{B}}} - \frac{\mu}{\omega_{\mathrm{B}}}-\frac{1}{2}} &e^{i2\pi p (z+\frac{\mu}{\omega_{\mathrm{B}}}-\frac{1}{2})}
\\
\times
& \sqrt{z^2+\left( \frac{2\theta_{0}}{\omega_{\mathrm{B}}} \right)^2}dz,\nonumber
\end{align}
here $\Lambda$ is the high energy cut-off.
We note that it is not $\frac{\theta^2}{2U}$ that enters the free energy Eq. (\ref{FreeResult}) as it gets cancelled troughout the derivation, but just like in the BCS theory it is rather $-\nu\theta_{0}^2$ (see \cite{Tinkham} for example). Importantly, all possible dominant oscillating terms originating from the Hubbard-Stratonovich term in the free energy vanish. This is the result obtained in Ref. \cite{AlloccaCooper}. 
Overall, the oscillating part of free energy is
\begin{align}\label{freeOSCmt}
\delta F_{\mathrm{osc}} = -
4\nu \theta_{0}^2 \left(  \frac{\omega_{\mathrm{B}}}{2 \pi\theta_{0}} \right)^{\frac{3}{2}}
\cos\left(\frac{2\pi \mu}{\omega_{\mathrm{B}}} \right) e^{- \frac{4\pi\theta_{0}}{\omega_{\mathrm{B}}} },
\end{align}
We observe that Eq. (\ref{freeOSCmt}) can be considered as non-interacting, (when $\theta$ of the Eq. (\ref{interaction2}) type is inserted to Eq. (\ref{Hamiltonian}) by hand), it comes from the oscillating part of $F_{\Sigma}$ which originates from the completely filled $\epsilon_{n;-}$ band, and corresponds to the results of Refs. \cite{KnolleCooperPRL15, PalArxiv2022}.
%It is clear that the latter term is smaller than the former at small magnetic fields defined by the $2\pi \frac{\theta_{0}}{\omega_{\mathrm{B}}} > 1$ condition. 
%When $\frac{\omega_{\mathrm{B}}}{2\pi \theta_{0}} = 1$ the amplitude vanishes, and at larger fields the non-interacting part will become dominant. 
%We note that vanishing \cite{AokiJPSJ2014} and substantially decreasing \cite{Li_PRX2022} amplitude of dHvA oscillations at some particular magnetic field was experimentally observed in strongly correlated systems \cite{AokiJPSJ2014,Li_PRX2022} and was attributed to the field driven metamagnetic transition.
%Proposed here mechanism might be relevant to these and other similar experiments \cite{AokiJPSJ2014,Li_PRX2022}. 
We differentiate the oscillating part of minus the free energy Eq. (\ref{freeOSCmt}) with respect to the magnetic field and obtain oscillating part of the magnetization Eq. (\ref{result1}) by keeping only the term $\propto \mu$.
This is justified as long as $\mu \gg \omega_{\mathrm{B}}$. We repeat the result
\begin{align}\label{result1mt}
M_{\mathrm{osc}} = -  \frac{2\mu\nu e}{\pi mc}  \left(\frac{2\pi \theta_{0}}{\omega_{\mathrm{B}}}\right)^{\frac{1}{2}}
\sin\left(\frac{2\pi \mu}{\omega_{\mathrm{B}}}  \right) e^{-4\pi \frac{\theta_{0}}{\omega_{\mathrm{B}}} },
\end{align}
which is the non-interacting result of Refs. \cite{KnolleCooperPRL15, PalArxiv2022}.

From Eq. (\ref{FreeResult}) we observe that the part of the non-oscillating magnetic susceptibility of the system originating from the hybridization mechanism is diamagnetic. 
This is precisely due to the magnetic field dependence of the non-oscillating part of the hybridization $\theta_{0}$ defined after Eq. (\ref{solution1}).
Indeed, the non-oscillating part of the magnetization is $M = -\partial_{B}(-\nu\theta_{0}^2) = -\nu\frac{\omega_{\mathrm{B}}}{2}e^{-\frac{1}{2U\nu}}$.
Physics of the obtained diamagnetism is that the system wants to expel the magnetic field in order to minimize the energy associated with the Coulomb repulsion.
Another point is that calculations suggest that we could have had any quadratic in magnetic field structure under the square root in the definition of $\theta_{0}$, for example $\sqrt{\mu^2 - \frac{\omega_{\mathrm{B}}^2}{4} + a\omega_{\mathrm{B}}}$, where $a$ is some energy, then there would be spontaneous magnetic order in the system characterized by a finite magnetization. 
This is experimentally not the case \cite{LiScience2014}, hence the choice of $a=0$.  
Moreover, we wish to point out that $-\frac{\omega_{\mathrm{B}}^2}{4}$ in $\theta_{0} = \sqrt{\mu^2 - \frac{\omega_{\mathrm{B}}^2}{4}} e^{-\frac{1}{4U\nu}}$ is due to the lowest Landau level.

%-----------------------------------------------------------------------------------------------------------------
%-----------------------------------------------------------------------------------------------------------------
\section{Temperature dependence}
\begin{figure}[h] 
\centerline{
\begin{tabular}{cc}
\includegraphics[width=0.45 \columnwidth]{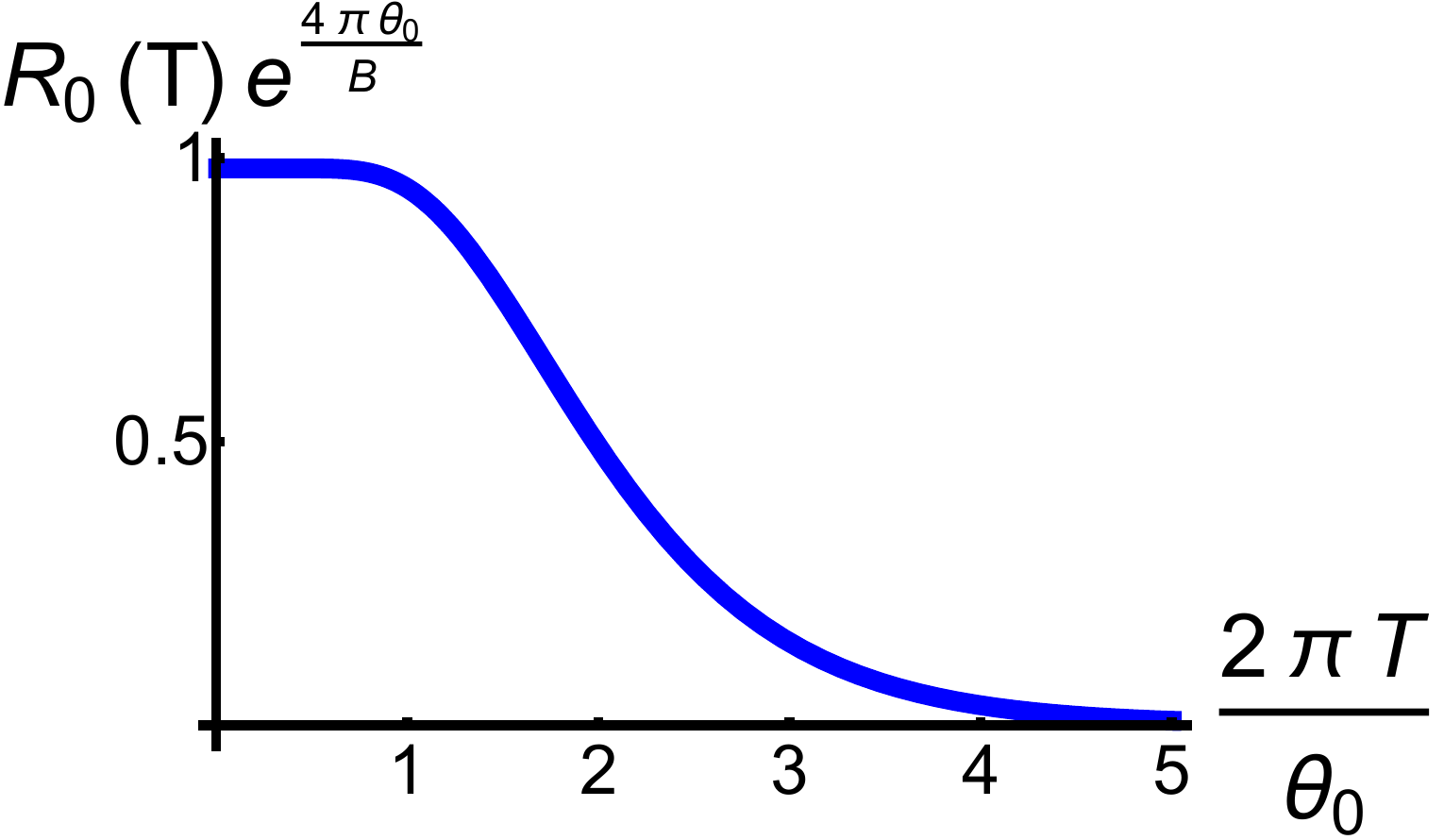}~~&
\includegraphics[width=0.45 \columnwidth]{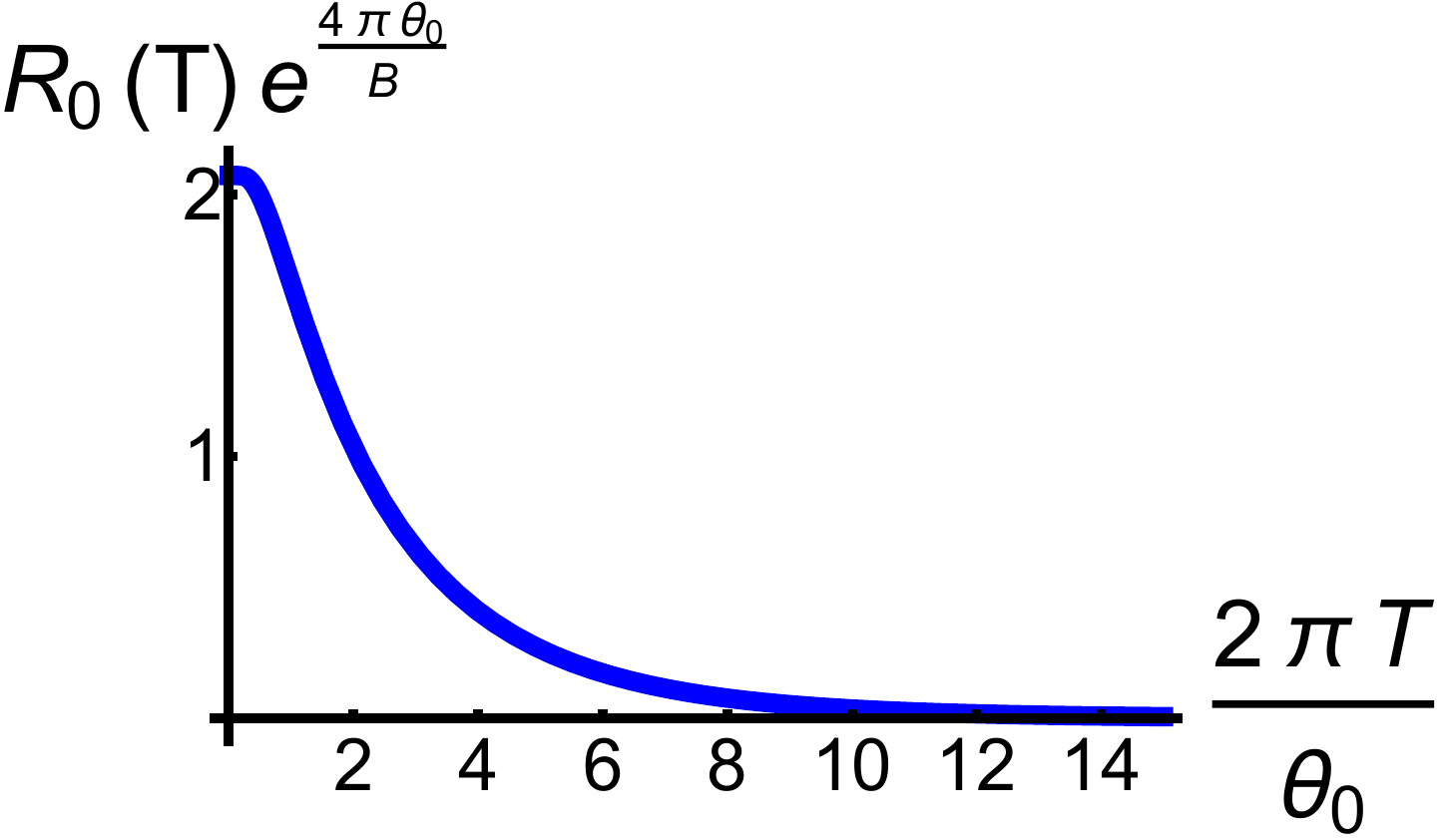}
\end{tabular}
}
\protect\caption{Plot of the numerically estimated expression Eq. (\ref{Rn}) for $n=0$ times the $e^{\frac{4\pi\theta_{0}}{\omega_{\mathrm{B}}}}$ factor for left $\frac{2\theta_{0}}{\omega_{\mathrm{B}}} = 1$ and right for $\frac{2\theta_{0}}{\omega_{\mathrm{B}}} = 0.2$. 
It is approximately consistent with standard Lifshits-Kosevich structure.}

\label{fig:fig3}  

\end{figure}

%-----------------------------------------------------------------------------------------------------------------
%-----------------------------------------------------------------------------------------------------------------
Let us now discuss temperature dependence of the amplitude of the dHvA oscillations.
The non-linear equation at finite temperature is given in Eq. (\ref{EqHyb}), where the temperature $T$ is in the Fermi-Dirac distribution functions ${\cal F}_{x} = \tanh\left( \frac{x}{2T} \right)$.
With all the technical steps presented in the Appendix \ref{sectionD} to the paper, we here briefly outline the steps.
Summation over the Landau levels in Eq. (\ref{EqHyb}) is performed with the help
of the Poisson summation formula Eq. (\ref{Poisson}), and the equation is rewritten as
\begin{align}
1=\nu U \omega_{\mathrm{B}}\sum_{n} {\cal R}_{n},
\end{align}
where $n=0,\pm1,\pm2,..$ and where we defined
\begin{align} \label{R+1_mt}
&{\cal R}^{(n)}(T) =
\frac{e^{i2\pi n\left( \frac{\mu}{\omega_{\mathrm{B}}}-\frac{1}{2}\right)} }{\omega_{\mathrm{B}}}\int_{-\infty}^{+\infty} \frac{e^{i2\pi n z}}{\sqrt{z^2+\left(\frac{2\theta}{\omega_{\mathrm{B}}}\right)^2}} 
\\
\times
&\left[1- \frac{1}{e^{\frac{-z+\sqrt{z^2+\left(\frac{2\theta}{\omega_{\mathrm{B}}}\right)^2}}{2T/\omega_{\mathrm{B}}}  }+1} - \frac{1}{e^{\frac{z+\sqrt{z^2+\left(\frac{2\theta}{\omega_{\mathrm{B}}}\right)^2}}{2T/\omega_{\mathrm{B}}}  }+1}\right]dz.
\nonumber
\end{align}
The distribution functions in the expression above have residues, which are essentially Matsubara frequencies, and the integral over the Poisson summation variable $z$ in Eq. (\ref{R+1_mt}) is calculated with the help of these residues. 
Next, the resulting sum over the Matsubara frequencies is performed by another Poisson summation formula. 
Then the obtained solution for the hybridization Eq. (\ref{solution1}) at zero temperature gets updated in accord with the mentioned steps as
\begin{align}\label{thetaT}
\theta = \theta_{0} - 2 \theta_{0}   \cos\left(\frac{2\pi \mu}{\omega_{\mathrm{B}}} \right)  R(T),
\end{align}
where we have approximated $\theta_{0}$ as temperature independent, which is the case way below the transition, function
$R(T) =\sum_{n=-\infty}^{+\infty} R_{n}(T)$ is a result of application of the Poisson summation formula to the sum over the Matsubara frequencies, 
\begin{align}\label{Rn}
R_{n}(T)=\frac{1}{2}\int_{0}^{+\infty}\frac{e^{i2\pi n x}}{x+\frac{1}{2}} e^{-\frac{2\pi \theta_{0}}{\omega_{\mathrm{B}}} \left[ \frac{1}{2\pi \frac{T}{\theta_{0}}\left(x+\frac{1}{2} \right)} + 2\pi \frac{T}{\theta_{0}}\left(x+\frac{1}{2} \right)\right]}dx,
\end{align}
where $x+\frac{1}{2}$ is reminiscent of the Matsubara frequency. We plot $R_{0}(T)$ in Fig. \ref{fig:fig3} to suggest that except for the shoulder it is consistent with standard Lifshits-Kosevich formula \cite{LK,LP}.
%--------------------------------------------------------------------------------------------------------------------------------------------------------------------
%-----------------------------------------------------------------------------------------------------------------------------------------------------
It is now instructive to perform the steepest descent approximation in evaluating the $R_{0}(T)$, we get the following limiting expressions
$
R^{\mathrm{sd}}_{0}(T=0) = e^{-\frac{4\pi \theta_{0}}{\omega_{\mathrm{B}}}}\frac{1}{2}\sqrt{\frac{\omega_{\mathrm{B}}}{2\theta_{0}}}, 
$
and
$
R^{\mathrm{sd}}_{0}(T\rightarrow\infty) \approx \frac{\omega_{\mathrm{B}}}{8\pi^2 T} e^{- \frac{4\pi \theta_{0}}{\omega_{\mathrm{B}}}} e^{-\pi \frac{\theta_{0}}{2\omega_{\mathrm{B}}}\left( \frac{2\pi T}{\theta_{0}} \right)^2}.
$
Indeed, the $T=0$ limit, except for the numerical coefficient difference, $\frac{1}{\sqrt{\pi}}$ in Eq. (\ref{solution1}) and $\frac{1}{2}$ above, because of the saddle point approximation of the latter (see Appendix \ref{sectionD} for more details), is consistent with the Eq. (\ref{solution1}), and the $T\rightarrow \infty$ decay is consistent with the Lishits-Kosevich expression \cite{LK,LP}.

\begin{figure}[t] 
\centerline{
\begin{tabular}{cc}
\includegraphics[width=0.45 \columnwidth]{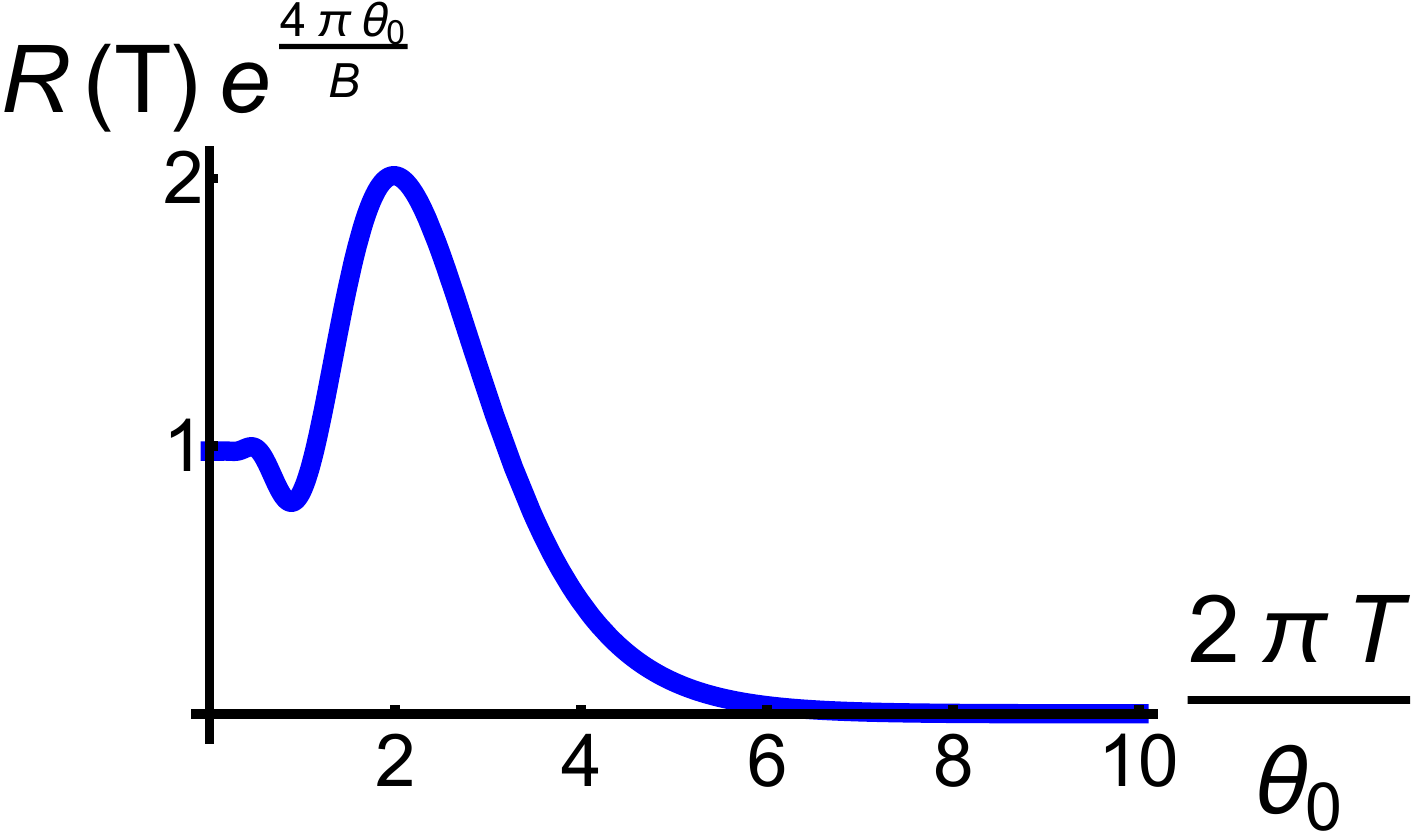} &
\includegraphics[width=0.45 \columnwidth]{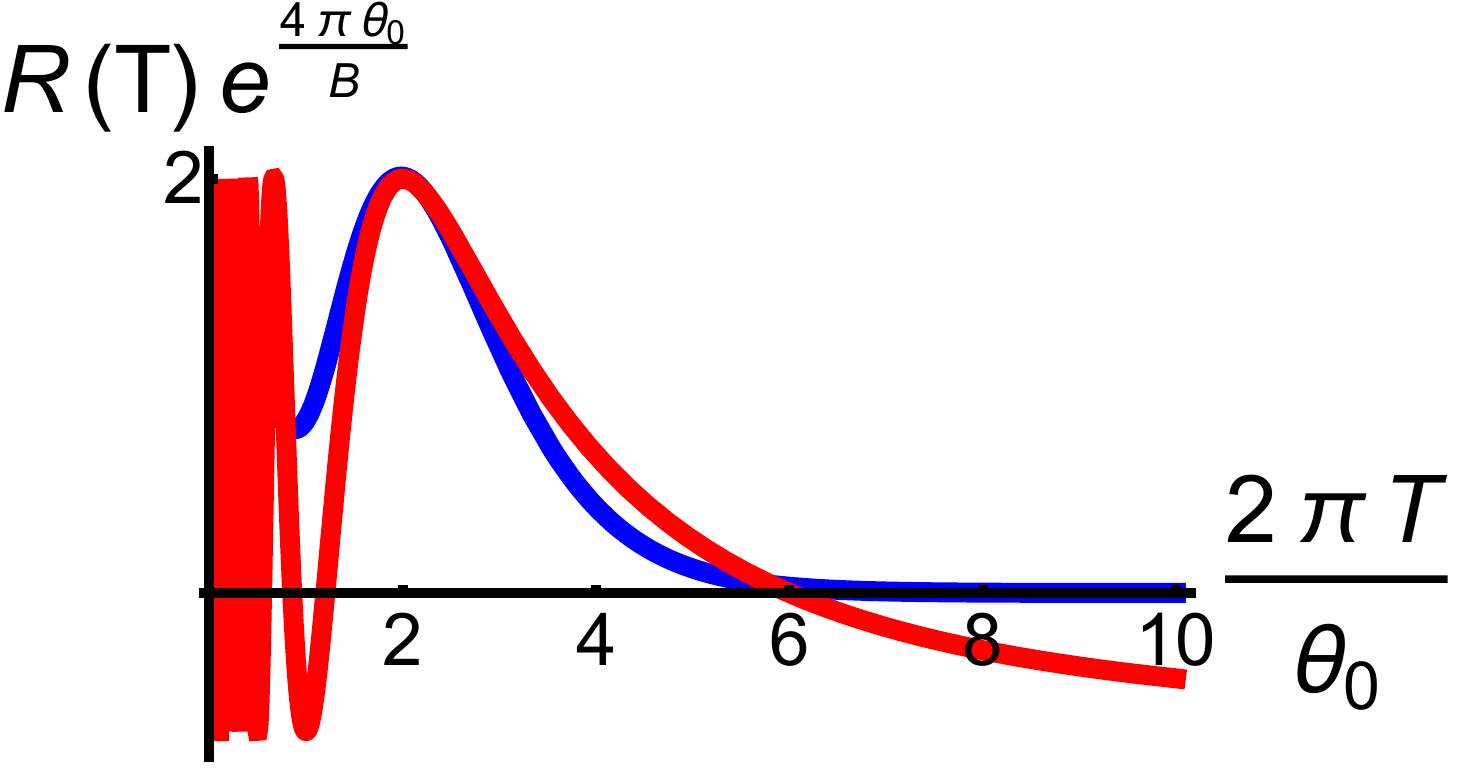} 
\end{tabular}
}
\protect\caption{Left: plot of the expression $R(T)$ for $\frac{2\theta_{0}}{\omega_{\mathrm{B}}} = 1$. 
Right: the same but with $\left[ 1 - 2\cos\left( \frac{\theta_{0}}{T}\right) \right]\frac{1}{3}$ in red plotted together.
}

\label{fig:fig4}  

\end{figure}

\begin{figure}[t] 
\centerline{
\begin{tabular}{cc}
\includegraphics[width=0.45 \columnwidth]{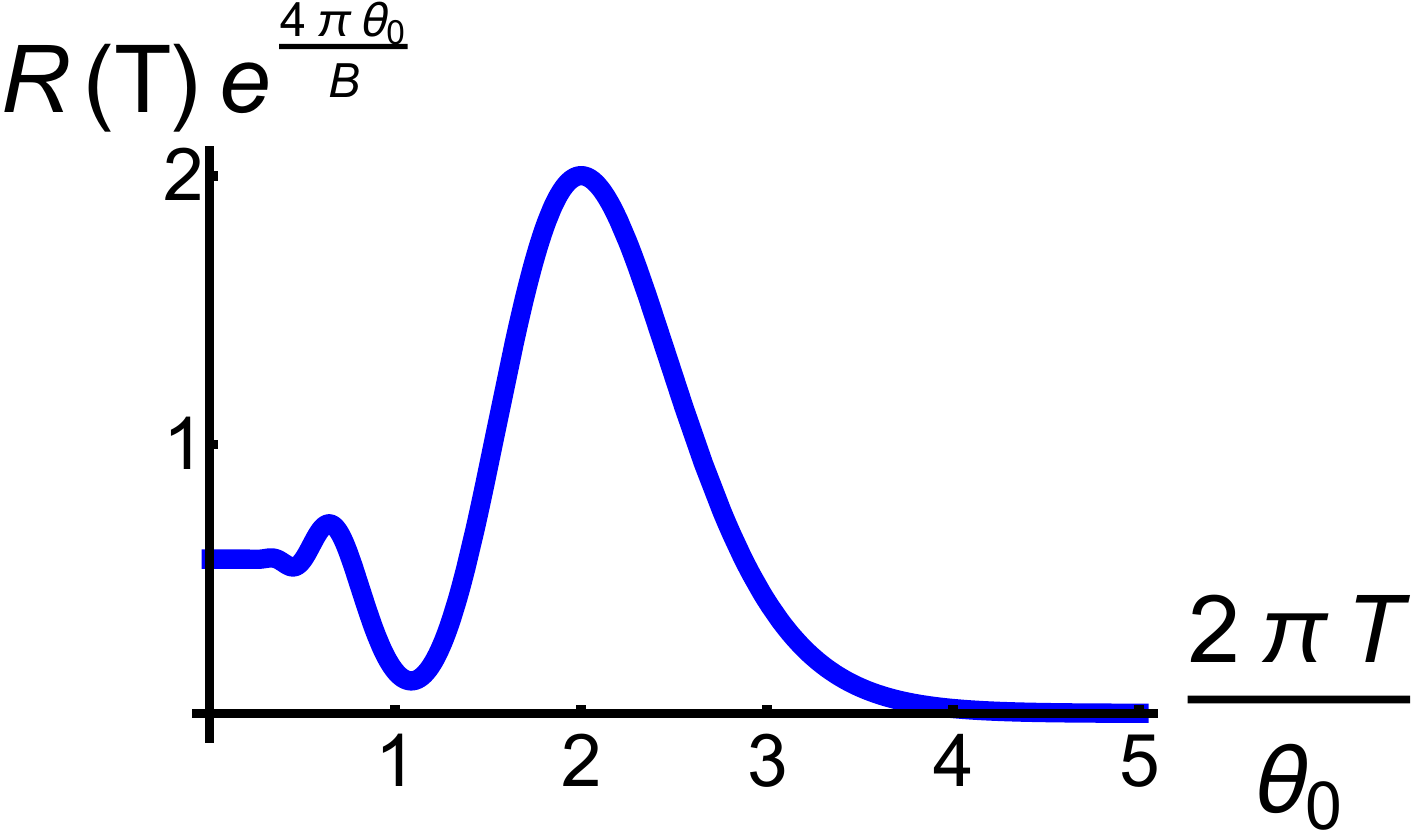} ~~&
\includegraphics[width=0.45 \columnwidth]{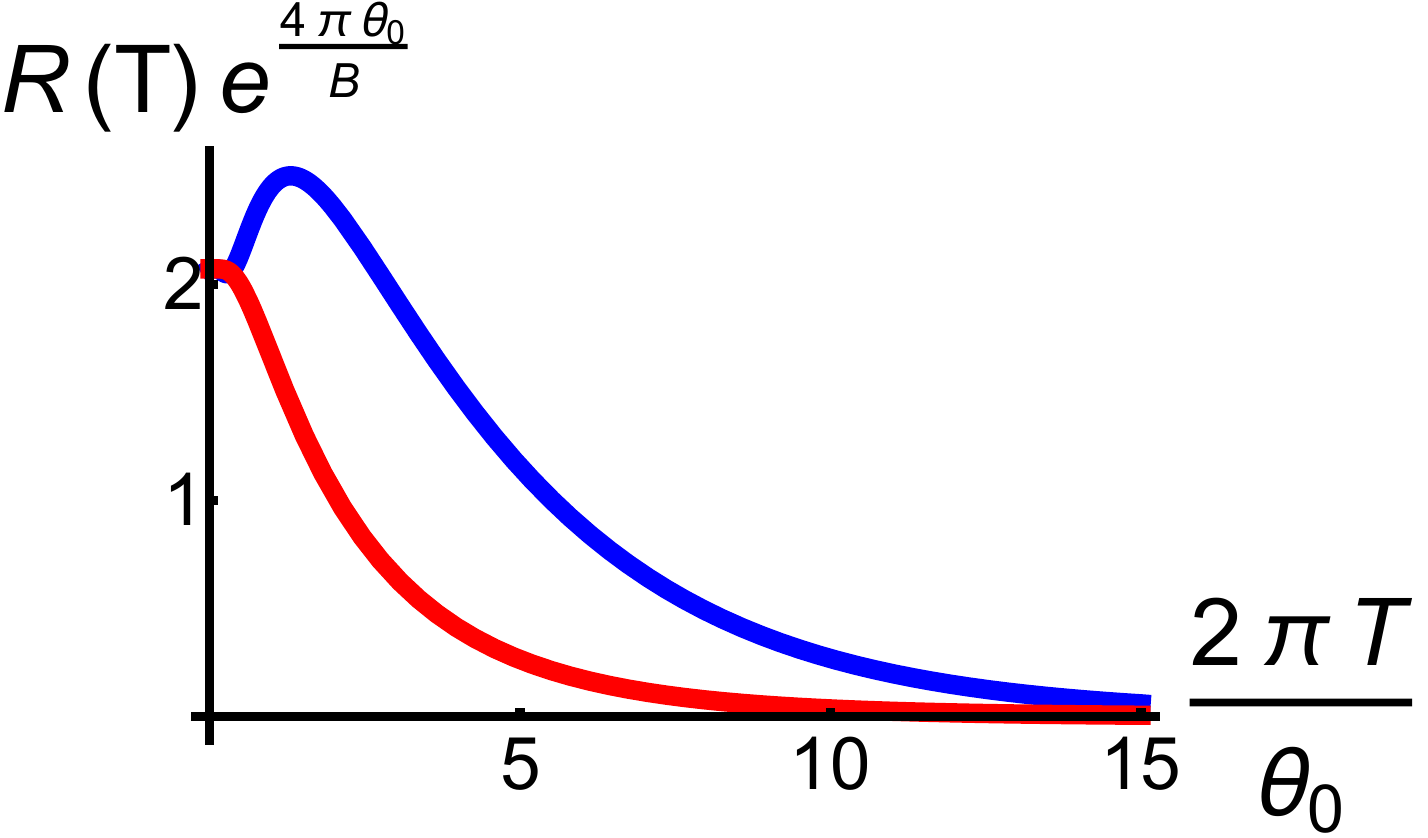}
\end{tabular}
}
\protect\caption{
Left: plot of the expression for $R(T) e^{\frac{4\pi\theta_{0}}{\omega_{\mathrm{B}}}}$ for $\frac{2\theta_{0}}{\omega_{\mathrm{B}}} = 3$, which makes it $R(T) e^{6\pi}$, as a proof of principle. 
The magnitude is exponentially suppressed but with that more oscillation cycles, as compared with the first one if going from large temperatures, become visible. 
Position of the first peak is consistent with analytical predictions Eq. (\ref{peak2}).
Right: in blue is the plot of $R(T)$ which has a peak structure, and in red we repeat right plot of Fig. (\ref{fig:fig3}), both for $\frac{2\theta_{0}}{\omega_{\mathrm{B}}} = 0.2$. 
}
\label{fig:fig5}  
\end{figure}

Having understood $R_{0}(T)$ let us now add higher harmonics. 
We work in the regime when the solution for the gap Eq. (\ref{thetaT}) is valid.
We plot $R(T)$ as a function of temperature in Figs. (\ref{fig:fig4}) and (\ref{fig:fig5}). 
In order to describe depicted dependence we can approximate the integral with the steepest descent method and obtain
in the vicinity of the peak,
\begin{align}\label{Rosc}
R^{\mathrm{sd}}(T) \approx R^{\mathrm{sd}}_{0}(T)\left[ 1+2\sum_{n=1}\cos\left(\frac{\theta_{0}}{T} \bar{z} n  -\pi n \right) e^{-\frac{2\pi^2 n^2}{\vert g(\bar{z}) \vert}} \right] ,
\end{align}
where the saddle point is at
\begin{align}
\bar{z} = \frac{4\pi \theta_{0}}{ \omega_{\mathrm{B}}+\sqrt{\omega_{\mathrm{B}}^2+(4\pi \theta_{0})^2}},
\end{align} 
and $g(\bar{z}) = \left( \frac{2\pi T}{\theta_{0}}\right)^2\frac{1}{\bar{z}^3}\left( \bar{z} - \frac{4\pi \theta_{0}}{\omega_{\mathrm{B}}}\right)$, and a first maximum in the temperature dependence of Eq. (\ref{Rosc}) occurs at $\frac{\theta_{0}}{T_{\mathrm{peak}}} \bar{z} = \pi$ which reads
\begin{align}\label{peak1}
T_{\mathrm{peak}} =\frac{4\theta_{0}^2}{{\omega_{\mathrm{B}}+\sqrt{\omega_{\mathrm{B}}^2+\left(4\pi \theta_{0}\right)^2}}}.
\end{align}
We note that as a result of a summation over the Matsubara frequency an oscillation of the dHvA oscillations amplitude with the inverse temperature of the hybridization $\theta$ appears, which can be observed in Eq. (\ref{Rosc}). 
Some sums are oscillating some not. For example, a sum over the Landau levels is oscillating.
To the best of our knowledge most sums over the Matsubara frequencies are not oscillating.

Let us first discuss the case of relatively small magnetic fields, $4\pi\theta_{0}>\omega_{\mathrm{B}}$. 
In particular, in Fig. (\ref{fig:fig4}) we plot $R(T)$ for $2\theta_{0} = \omega_{\mathrm{B}}$. 
The position of the peak defined in Eq. (\ref{peak1}) can be approximated to be
\begin{align}\label{peak2}
T_{\mathrm{peak}} = \frac{\theta_{0}}{\pi}.
\end{align}
To justify our approximation, we plot $\propto \cos\left( \frac{\theta_{0}}{T} \right)$ together with $R(T)$ in Fig. (\ref{fig:fig4}). 
The approximation Eq. (\ref{Rosc}) works well in the vicinity of the peak, away from it one has to include more terms to make sure the function is positive-definite. 
As a proof of principle, when the parameter $\frac{2\theta_{0}}{\omega_{\mathrm{B}}}$ increases, the amplitude of Eq. (\ref{Rosc}) exponentially decreases due to the decaying $e^{-4\pi \frac{\theta_{0}}{\omega_{\mathrm{B}}}}$ term, however, with that more oscillation cycles appear. 
For example, see Fig. (\ref{fig:fig5}) for the plot for $\frac{2\theta_{0}}{\omega_{\mathrm{B}}} = 3$, where the position of the peak is consistent with analytical predictions Eq. (\ref{peak2}). 

%------------------------------------------------------------------------------------------------------------------------

Let us estimate the height of the peak at $T_{\mathrm{peak}} = \frac{\theta_{0}}{\pi}$.
Numerics suggest, see Fig. (\ref{fig:fig4}), that the peak has a value of almost $e^{-\frac{4\pi\theta_{0}}{\omega_{\mathrm{B}}}}$, let us show it is indeed the case. 
We found it convenient to not perform the steepest descent for this task, but rather do the following transformations after application of the Poisson summation formula,
\begin{align}
R(T_{\mathrm{peak}}) 
=
\frac{1}{2}
\sum_{n} e^{-i\pi n} \int_{0}^{+\infty}e^{i n\pi e^{z}} e^{- \frac{4\pi\theta_{0}}{\omega_{\mathrm{B}}} \cosh(z)}dz,
\end{align}
now goes our approximation, because $\cosh(z)$ has a minimum at $z=0$ (reminiscent of steepest descent approximation made above), $e^{z}\sim 1+z$ in the vicinity of $z=0$. 
We then write
\begin{align}
&
\sum_{n} e^{-i\pi n} \int_{0}^{+\infty}e^{i n\pi e^{z}} e^{-\frac{4\pi\theta_{0}}{\omega_{\mathrm{B}}}  \cosh(z)}dz 
\\
&
\approx 2 \sum_{m}\int_{0}^{+\infty}\delta(z-m) e^{-\frac{4\pi\theta_{0}}{\omega_{\mathrm{B}}} \cosh(z)}dz .
\nonumber
\end{align}
For $\frac{2\theta_{0}}{\omega_{\mathrm{B}}}\geq1$ only $m=0$ term is significant in the sum. 
Therefore, we conclude that 
\begin{align}\label{height}
R(T_{\mathrm{peak}}) \approx  e^{-\frac{4\pi\theta_{0}}{\omega_{\mathrm{B}}} },
\end{align}
which is indeed plotted in Fig. (\ref{fig:fig4}) and in the left plot in Fig. (\ref{fig:fig5}).

%-------------------------------------------------------------------------------------------------------------------------
To conclude the derivations, the non-zero temperature expression for the magnetization reads as
\begin{align}\label{result2}
M_{\mathrm{osc}} =&  -\frac{2\mu \nu e}{\pi mc}  \left(\frac{2\pi \theta_{0}}{\omega_{\mathrm{B}}}\right)^{\frac{3}{2}}R(T) 
\sin\left( \frac{2\pi\mu}{\omega_{\mathrm{B}}} \right),
\end{align}
where $R(T)$ is given by primarily Eq. (\ref{Rn}), and with Eqs. (\ref{Rosc}) approximating the behavior in the vicinity of the peak. We note that we have calculated $R(T)$ from the gap self-consistency equation. The same structure of $R(T)$ will be obtained in the extension of Eq. (\ref{freeOscBand}) to non-zero temperatures.

Let us now discuss relatively large magnetic fields $\omega_{\mathrm{B}} \sim 4\pi\theta_{0}$ and $\omega_{\mathrm{B}} > 4\pi\theta_{0}$. We plot $R(T)$ in Fig. (\ref{fig:fig5}) right for $\frac{2\theta_{0}}{\omega_{\mathrm{B}}} = 0.2$. 
Unfortunately, the steepest descent method fails to work well in this regime. We, nevertheless, still can predict the existence of the peak in the amplitude due to the oscillations.
From the right plot of Fig. (\ref{fig:fig5}) we observe that for $\frac{2\theta_{0}}{\omega_{\mathrm{B}}}=0.2$ the position of the peak is at $\frac{2\pi T^{(\mathrm{plot})}_{\mathrm{peak}}}{\theta_{0}} \approx 1.32$ which should be compared with the corresponding quantity from the Eq. (\ref{peak1}), which is $\frac{2\pi T^{(\mathrm{analytics})}_{\mathrm{peak}}}{\theta_{0}} \approx 0.96$. 
It turns out that for whatever reason a better approximation is $T^{*}_{\mathrm{peak}} = \frac{2\theta_{0}^2}{\omega_{\mathrm{B}}}$ which is a high field limit of Fig. (\ref{fig:fig5}) and which gives  $\frac{2\pi T^{*}_{\mathrm{peak}}}{\theta_{0}} \approx 1.25$.

%-----------------------------------------------------------------------------------------------------------------------------------------------------
%-----------------------------------------------------------------------------------------------------------------------------------------------------

%-----------------------------------------------------------------------------------------------------------------
%-----------------------------------------------------------------------------------------------------------------

%-----------------------------------------------------------------------------------------------------------------------------------------------------

In addition to the discussed above, we find a new regime of oscillations occurring in rather strong magnetic fields, $\omega_{\mathrm{B}}>8\pi\theta_{0}$. 
There in the course of the oscillations the hybridization gap vanishes at the minimum of the cycle and then emerges back at the peak. 
We speculate that such an oscillation can undergo with a phase separation to metallic and insulating domains.
Unfortunately, it can't be experimentally found in the SmB$_{6}$ material due to the physical parameters of the system, however, InAs/GaSb quantum wells where value of the Coulomb interaction can be controlled and where quantum oscillations in the insulating regime were observed \cite{ExpInAs/GaSb1,ExpInAs/GaSb2}, are very promising candidates. 
%-----------------------------------------------------------------------------------------------------------------

We note that in previous works \cite{KnolleCooperPRL15,PalArxiv2022} it has been numerically shown that non-interacting insulating fermion system shows a deviation from the Lifshits-Kosevich temperature dependence of the dHvA oscillation amplitude at small temperatures, and a peak in the amplitude has been numerically plotted in these papers. 
However, it is only in the present paper it was recognized that the deviation from Lifshits-Kosevich formula is actually an oscillation of the amplitude with inverse temperature, and the peak can be associated with the first cycle of the oscillation. 
Note that there are not that many instances of oscillations of physical observables with temperature. 
We are aware of one theoretical prediction of occurrence of mesoscopic oscillations of conductance as a function of temperature in the disordered conductors due to the mesoscopic fluctuations \cite{SpivakZyuzinCobden}. 
These oscillations are different from the ones predicted in this paper.

In regards of topology, here we were working with ordinary insulator. There are theoretical claims that SmB$_6$ may be a topological insulator \cite{DzeroSunGalitskiColeman}.
Then, if instead we were to think of a topological insulator, we can fit our model to Volkov-Pankratov model of topological insulators \cite{SovietTI1,SovietTI2} by promoting the hybridization gap to momentum and spin dependences as $\theta \rightarrow \theta ({\bf k}\cdot{\bm \sigma})$, as well as adding $k-$ dependence to the mass in \cite{SovietTI1,SovietTI2}.
Although, the hybridization gap will now be proportional to the Landau level index, at the Fermi level it can be approximated as a constant, i.e. $\propto\theta \omega_{\mathrm{B}} n_{\mathrm{F}}$,  where integer $n_{\mathrm{F}}$ is defined from $\omega_{\mathrm{B}}(n_{\mathrm{F}}+\frac{1}{2}) = \mu$, and all the predictions of this paper hold. 
The only complication is to find such an interaction that will result in momentum and spin dependences of the hybridization, and solve the self-consistency equation with this interaction. If necessary, we leave the details to future research.

\section{Conclusion}
To conclude we have studied de Haas-van Alphen oscillations in correlated insulators. 
We studied a mechanism when the osillations originate from the magnetic field dependence of the hybridization. The hybridization is due to the Coulomb repulsion and is treated self-consistently. We have recovered the results by  \cite{AlloccaCooper}, namely that the dominant contribution to the oscillations is due to the non-interacting picture \cite{KnolleCooperPRL15,PalArxiv2022}. In addition we have shown that the non-oscillating contribution to the free energy originating from the mean-field treatment of the hybridization gap is diamagnetic. This corresponds to the decay of the hybridization gap in magnetic field.

Moreover, we showed that the amplitude of dHvA oscillations is itself a periodic function of inverse temperature. Analytically it shows up to three visible cycles of oscillations.
The position of the peak in the amplitude dependence on the temperature is analytically estimated in Eqs. (\ref{peak1}) and (\ref{peak2}) from where one can extract the value of the hybridization gap. 
The height of the peak is analytically estimated in Eq. (\ref{height})
In experiments \cite{SuchitraScience2015, SuchitraNatPhys2017,SuchitraScience2020} deviation from the Lifshits-Kosevich expression was observed at small temperatures.
Our work might be relevant to these experiments. 

Finally, we checked that the peak shown in Fig. (\ref{fig:fig4}) does not survive if the spectrum of $f-$fermions in Eq. (\ref{Hamiltonian}) was of the hole-like type with the same mass as the $d-$fermions, i.e. when $\epsilon_{0} \rightarrow - \xi_{\bf k}$. 
In this case the temperature dependence of the dHvA amplitude is of the Lifshits-Kosevich structure shown in Fig (\ref{fig:fig3}). 
On the other hand, slight particle-like or hole-like dispersion of $f-$fermions does not alter the peak structure Fig. (\ref{fig:fig4}) at all. 
Analysis of the situation when $f-$fermions have a slight particle-like spectrum will be a subject of the subsequent work by the author.

\section{Acknowledgements.}
The author thanks I.S. Burmistrov, A.M. Finkel'stein, M.M. Glazov, A. Kamenev, D.G. Yakovlev, and A.Yu. Zyuzin for helpful discussions. The author thanks A. Allocca and N. Cooper for pointing out a mistake in the first version of this paper.
VAZ is grateful to Pirinem School of Theoretical Physics and to Weizmann Institute of Science for hospitality during the Summer of 2022, and in particular to Y. Oreg and A. Stern for support. 
VAZ is supported by the Foundation for the Advancement of Theoretical Physics and Mathematics BASIS.

%--------------------------------------------------------------------------------------------

\appendix

\begin{widetext}

\section{Theoretical model}\label{sectionA}
Throughout the Supplemental Material we use $\hbar \equiv 1$ and $k_{\mathrm{B}} \equiv 1$ convention.
\begin{align}
H_{0} = \int_{\bf k} \bar{\psi}\left[\begin{array}{cc} \xi_{\bf k} & 0 \\ 0 & \epsilon_{0} \end{array} \right]\psi 
\equiv \int_{\bf k} \bar{\psi} h \psi,
\end{align}
where $\xi_{\bf k} = \frac{{\bf k}^2}{2m}-\mu $ and $\epsilon_{0} =E_{0} - \mu = \mathrm{const}$ (we will later assume $\epsilon_{0} = 0$), where $\mu$ is the chemical potential.
Arguments presented below will also be valid for the inverted band structure.
Spinors are defined as
\begin{align}
\bar{\psi} = \left[ \bar{\phi}_{\mathrm{d}} ,~ \bar{\phi}_{\mathrm{f}} \right], 
~~~
\psi = \left[\begin{array}{c} \phi_{\mathrm{d}} \\ \phi_{\mathrm{f}} \end{array} \right].
\end{align}
Interaction between the two types of fermions is
\begin{align}
H_{\mathrm{int}} 
= U \int_{x} \bar{\phi}_{\mathrm{d}}(x)\phi_{\mathrm{d}}(x) \bar{\phi}_{\mathrm{f}}(x)\phi_{\mathrm{f}}(x) 
= - U\int_{x} \bar{\phi}_{\mathrm{d}}(x)\phi_{\mathrm{f}}(x) \bar{\phi}_{\mathrm{f}}(x)\phi_{\mathrm{d}}(x),
\end{align}
where $U>0$ corresponds to repulsion. Hubbard-Stratonovich transformation of the action reads
\begin{align}
-\left(\bar{\theta} + \bar{\phi}_{\mathrm{d}}\phi_{\mathrm{f}} U \right) U^{-1}\left(\theta + U\bar{\phi}_{\mathrm{f}}\phi_{\mathrm{d}} \right)
+
U \bar{\phi}_{\mathrm{d}}\phi_{\mathrm{f}} \bar{\phi}_{\mathrm{f}}\phi_{\mathrm{d}}
=
-\bar{\theta}U^{-1}\theta - \bar{\phi}_{\mathrm{d}}\phi_{\mathrm{f}} \theta - \bar{\theta}\bar{\phi}_{\mathrm{f}}\phi_{\mathrm{d}}.
\end{align}
In order to obtain self-consistent equation for the hybridization we employ the Keldysh technique (for example see Ref. \cite{Kamenev}),
although, the task at hand can be formulated in Matsubara space. It is our personal choice to work in Keldysh technique.
In the Main Text in order not to burden the reader we don't even mention that we formulated our theory in the Keldysh space.
\begin{align}
iS_{0} =i \int_{C}dt~\bar{\psi}(x) (\epsilon-h )\psi(x)
&
= 
i \int_{\bf r} \left[\int_{-\infty}^{+\infty} dt \bar{\psi}_{+}(x)(\epsilon - h)\psi_{+}(x)
-
\int_{-\infty}^{+\infty} dt \bar{\psi}_{-}(x)(\epsilon - h)\psi_{-}(x) \right],
\\
&
\equiv
i \int_{\bf r} \int_{-\infty}^{+\infty} dt \hat{\bar{\Psi}}(x)(\epsilon - \hat{H}_{0})\hat{\Psi}(x),
\end{align}
where we have performed Larkin-Ovchinnikov rotation defining
\begin{align}
\hat{\bar{\Psi}} = \hat{\bar{\psi}}\hat{L}^{-1}, ~~\hat{\Psi} = \hat{L}\hat{\sigma}_{3}\hat{\psi},
\end{align}
where $\hat{\bar{\psi}} = (\bar{\psi}_{+},~\bar{\psi}_{-})$, and 
\begin{align}
\hat{L} = \frac{1}{\sqrt{2}}\left[\begin{array}{cc} 1 & -1 \\ 1 & 1 \end{array} \right],
~~~
\hat{\sigma}_{3} = \left[\begin{array}{cc} 1 & 0 \\ 0 & -1 \end{array} \right]
\end{align}
are matrices acting in Keldysh space.
\begin{align}
iS_{\mathrm{int};1} 
=-i \int_{C}dt~ 
H_{\mathrm{int};1}
&
= -i \int_{\bf r} \int_{-\infty}^{+\infty} dt \hat{\bar{\Psi}}(x)
\left[
\left(\theta_{\mathrm{cl}}\hat{\sigma}_{0} + \theta_{\mathrm{q}}\hat{\sigma}_{1}\right)\tau^{+}
+ 
\left(\bar{\theta}_{\mathrm{cl}}\hat{\sigma}_{0} + \bar{\theta}_{\mathrm{q}}\hat{\sigma}_{1}\right)\tau^{-}
\right]\hat{\Psi}(x)
\\
&
= - i
\int_{\bf r}
\int_{-\infty}^{+\infty} dt \hat{\bar{\Psi}}(x)
\left[
\hat{\theta} \tau^{+}
+ 
\hat{\bar{\theta}}\tau^{-}
\right]\hat{\Psi}(x)
\end{align}
where 
\begin{align}
\tau^{+} = \left[\begin{array}{cc} 0 & 1 \\ 0 & 0 \end{array} \right], ~~~
\tau^{-} = \left[\begin{array}{cc} 0 & 0 \\ 1 & 0 \end{array} \right]
\end{align}
are matrices acting in d-f space, and
\begin{align}
\hat{\sigma}_{0} = \left[\begin{array}{cc} 1 & 0 \\ 0 & 1 \end{array} \right], ~~~
\hat{\sigma}_{1} = \left[\begin{array}{cc} 0 & 1 \\ 1 & 0 \end{array} \right]
\end{align}
are Pauli matrices acting in Keldysh space. 
We have defined classical and quantum combinations of the interaction fields, namely $\theta_{\mathrm{cl}/\mathrm{q}} = \frac{1}{2}(\theta_{+} \pm \theta_{-})$ and $\bar{\theta}_{\mathrm{cl}/\mathrm{q}} = \frac{1}{2}(\bar{\theta}_{+} \pm \bar{\theta}_{-})$. For convenience we will be calling $\theta_{\mathrm{cl}} \equiv \theta$ and $\bar{\theta}_{\mathrm{cl}} \equiv \bar{\theta}$.
\begin{align}
iS_{\mathrm{int};2} = - \frac{i}{2}\int_{x} \mathrm{Tr}_{\mathrm{K}}\left[\hat{\bar{\theta}}(x)U^{-1}\hat{\sigma}_{1}\hat{\theta}(x)\right],
\end{align}
where $\mathrm{Tr}_{\mathrm{K}}$ is the trace over the Keldysh components of the matrices.

Action describing the fermions is
\begin{align}
iS_{0} + iS_{\mathrm{int};1} =  i
\int_{\bf r}\int_{-\infty}^{+\infty} dt \left\{
\hat{\bar{\Psi}}(x)\hat{\sigma}_{0}
\left[
\begin{array}{cc} 
\epsilon - \hat{\xi}_{\bf k} & -\theta \\
-\bar{\theta} & \epsilon -\epsilon_{0}
\end{array}
\right]
\hat{\Psi}(x)
+
\hat{\bar{\Psi}}(x)\hat{\sigma}_{1}
\left[
\begin{array}{cc} 
0 & -\theta_{\mathrm{q}} \\
-\bar{\theta}_{\mathrm{q}} & 0
\end{array}
\right]
\hat{\Psi}(x)
\right\},
\end{align}
where $\hat{\xi}_{\bf k}$ stands for operator acting on the coordinate. Whenever, Fourier transformation is performed, $\hat{\xi}_{\bf k} \rightarrow \xi_{\bf k} = \frac{{\bf k}^2}{2m} - \mu$.
Integrating fermions out, we get
\begin{align}
iS_{0}+iS_{\mathrm{int};1} = \int_{\bf x} \mathrm{Tr}\ln\left[ 1- \hat{G}\hat{h}_{\mathrm{int};1;q} \right],
\end{align}
where $\hat{H}_{\mathrm{int};1;q}$ contains quantum componet of the interaction with the Hubbard-Stratonovich field,
\begin{align}
\hat{h}_{\mathrm{int};1;q} = \hat{\sigma}_{1}
\left[
\begin{array}{cc} 
0 & \theta_{\mathrm{q}} \\
\bar{\theta}_{\mathrm{q}} & 0
\end{array}
\right]
\end{align}
We recall, 
\begin{align}
\hat{G}(\epsilon ,{\bf k} ) = \hat{U}_{\epsilon}\hat{g}_{\epsilon}(\epsilon ,{\bf k} )\hat{U}_{\epsilon} = \left[\begin{array}{cc} G^{\mathrm{R}}(\epsilon ,{\bf k} ) & G^{\mathrm{K}}(\epsilon ,{\bf k} ) \\ 0 & G^{\mathrm{A}}(\epsilon ,{\bf k} )  \end{array} \right],
\end{align}
where 
\begin{align}
\hat{g}(\epsilon ,{\bf k} ) =  \left[\begin{array}{cc} G^{\mathrm{R}}(\epsilon ,{\bf k} ) & 0 \\ 0 & G^{\mathrm{A}}(\epsilon ,{\bf k} )  \end{array} \right],
\end{align}
and
\begin{align}
\hat{U}_{\epsilon} = \left[\begin{array}{cc} 1 & {\cal F}_{\epsilon} \\ 0 & -1  \end{array} \right],
\end{align}
where ${\cal F}_{\epsilon} = \tanh\left( \frac{\epsilon}{2T} \right)$ is the Fermi-Dirac distribution function with $T$ being the temperature. Keldysh part of the Green function is
\begin{align}
G^{\mathrm{K}}(\epsilon ,{\bf k} ) = {\cal F}_{\epsilon}G^{\mathrm{R}}(\epsilon ,{\bf k} ) - G^{\mathrm{A}}(\epsilon ,{\bf k} ) {\cal F}_{\epsilon}.
\end{align}
Details of the Green function are 
\begin{align}
G^{\mathrm{R}/\mathrm{A}}(\epsilon, {\bf k}) = \frac{1}{(\epsilon\pm i0 -\epsilon_{{\bf k},+})(\epsilon\pm i0 -\epsilon_{{\bf k},-})}
\left[\begin{array}{cc} \epsilon - \epsilon_{0}  & \theta \\ \bar{\theta} & \epsilon - \xi_{\bf k}\end{array}\right],
\end{align}
where 
\begin{align}
\epsilon_{{\bf k},\pm} = \frac{\xi_{\bf k} + \epsilon_{0}}{2} \pm \sqrt{ \left(\frac{ \xi_{\bf k} - \epsilon_{0}}{2} \right)^2  +\bar{\theta}\theta}
\end{align}
is the spectrum. Then, for example,
\begin{align}
\mathrm{Tr}\left[ \tau^{-}\hat{\sigma}_{1}\hat{G}(\epsilon, {\bf k}) \right] 
= 
\mathrm{Tr}\left[ \tau^{-}G^{\mathrm{K}}\right]
&
= 
\theta
{\cal F}_{\epsilon} 
\left[ 
\frac{1}{(\epsilon + i0 -\epsilon_{{\bf k},+})(\epsilon + i0 -\epsilon_{{\bf k},-})}
-
\frac{1}{(\epsilon - i0 -\epsilon_{{\bf k},+})(\epsilon - i0 -\epsilon_{{\bf k},-})}
\right]
\\
&
=
-
\frac{\theta}{\epsilon_{{\bf k},+} - \epsilon_{{\bf k},-} }
{\cal F}_{\epsilon} 
2\pi i
\left[ 
\delta(\epsilon - \epsilon_{{\bf k},+})
-
\delta(\epsilon - \epsilon_{{\bf k},-})
\right],
\end{align}
such that 
\begin{align}
\int_{\epsilon}\mathrm{Tr}\left[ \tau^{-}\hat{\sigma}_{1}\hat{G}(\epsilon, {\bf k}) \right] 
=
-
\frac{\theta}{\epsilon_{{\bf k},+} - \epsilon_{{\bf k},-} }
 i
\left( 
{\cal F}_{\epsilon_{{\bf k},+}} 
-
{\cal F}_{\epsilon_{{\bf k},-}} 
\right).
\end{align}
From where we derive self-consistent equation for the hybridization. For that we vary the action over the $\bar{\theta}_{\mathrm{q}}$ and set it to zero,
\begin{align}
\frac{\delta }{\delta \bar{\theta}_{\mathrm{q}}}\left( iS_{0}+iS_{\mathrm{int};1}+iS_{\mathrm{int};2}\right) 
&=\frac{\delta }{\delta \bar{\theta}_{\mathrm{q}}}\left\{ \int_{\bf x} \mathrm{Tr}\ln\left[ 1- \hat{G}\hat{h}_{\mathrm{int};1;q} \right] - \frac{i}{2}\int_{x} \mathrm{Tr}_{\mathrm{K}}\left[\hat{\bar{\theta}}(x)U^{-1}\hat{\sigma}_{1}\hat{\theta}(x)\right]\right\} 
\\
&= - \int_{{\bf k},\epsilon} \mathrm{Tr}\left[ \tau^{-}\hat{\sigma}_{1}\hat{G}(\epsilon, {\bf k}) \right] -i U^{-1}\theta = 0,
\end{align}
and we get
\begin{align}
\theta = U \theta 
\int_{\bf k} 
\frac{{\cal F}_{\epsilon_{{\bf k},+}} 
-
{\cal F}_{\epsilon_{{\bf k},-}} }{\epsilon_{{\bf k},+} - \epsilon_{{\bf k},-}},
\end{align}
where we have
\begin{align}
\epsilon_{{\bf k},+} - \epsilon_{{\bf k},-}
= 2 \sqrt{ \left(\frac{ \xi_{\bf k} - \epsilon_{0}}{2} \right)^2  +\bar{\theta}\theta}.
\end{align}
Simple check shows ${\cal F}_{\epsilon_{{\bf k},+}} 
-
{\cal F}_{\epsilon_{{\bf k},-}} >0 $ and $\epsilon_{{\bf k},+} - \epsilon_{{\bf k},-} > 0$ such that the equation can be satisfied with a non-trivial $\theta$.

%--------------------------------------------------------------
%--------------------------------------------------------------
%--------------------------------------------------------------
%--------------------------------------------------------------
\section{Zero temperature}\label{sectionB}
%--------------------------------------------------------------
%--------------------------------------------------------------
%--------------------------------------------------------------
%--------------------------------------------------------------
In magnetic field and at $T=0$ (also recall we are working in two-dimensions) the equation reads
\begin{align}
1 = 2 U\nu \omega_{\mathrm{B}} \sum_{n} 
\frac{1 }{\sqrt{(\omega_{\mathrm{B}}n + \frac{\omega_{\mathrm{B}}}{2} - \mu)^2 +4\bar{\theta}\theta}},
\end{align}
where factor of $2$ is due to the sum of the distribution functions, $\nu = \frac{m}{2\pi}$ and $\omega_{\mathrm{B}}= \frac{eB}{mc}$ is the cyclotron frequency.
In deriving the equation we have performed Hubbard-Stratonovich decoupling of the four fermion interaction (in the way done in the previous subsection) before going to the Landau levels basis. 
In this way possible form-factors don't appear. 

We use Poisson summation formula
\begin{align}\label{PoissonSM}
\sum_{n \geq 0}g(n) = \int_{0}^{\infty} g(x) dx + \sum_{p\neq 0}\int_{0}^{\infty} e^{i2\pi p x}g(x) dx. 
\end{align}
in order to estimate the sum.
Non-oscillating term is
\begin{align}\label{nonoscillatingSM}
\int_{0}^{\Lambda} \frac{dx}{\sqrt{(x-f)^2+b^2}} \approx \ln\left[ \frac{4(\Lambda-f)f}{b^2} \right],
\end{align}
where $\Lambda$ is the upper cut-off, $f = \frac{\mu}{\omega_{\mathrm{B}}} - \frac{1}{2}$, and we assumed $b<\Lambda-f$ and $b<f$.
The other integral requires some care.
We wish to study the lowest harmonic of oscillation and pick $p=\pm 1$ from the sum.
Here is the step-by-step calculation,
\begin{align}
\int_{0}^{+\infty}\frac{\cos(2\pi x)}{\sqrt{(x-f)^2+b^2}} \approx 
\int_{-\infty}^{+\infty}\frac{\cos(2\pi (x+f))}{\sqrt{x^2+b^2}}.
\end{align}
Consider now 
\begin{align}\label{integral2D}
I_{+}  = \int_{C_{+}} \frac{e^{i2\pi(z+f)}}{\sqrt{z^2+b^2}}dz
\end{align}
over the contour $C_{+}$ shown in the Fig. (\ref{fig:figSM1}). 

%--------------------------------------------------------------------------------------------------------------------------------------------------------------------
\begin{figure}[t] 
\centerline{
\includegraphics[width=0.3 \columnwidth]{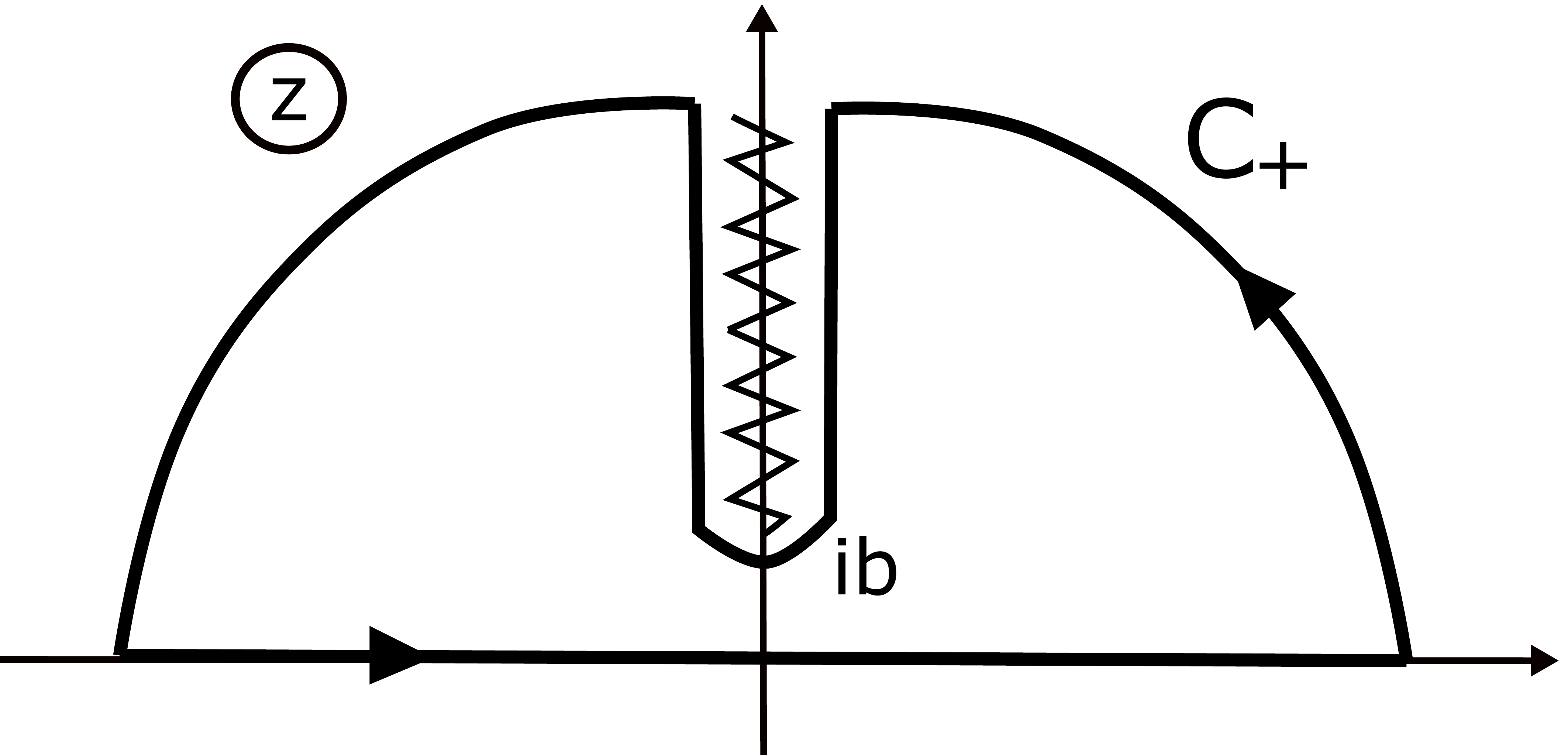}
}
\protect\caption{Contour of integration $C_{+}$ used in integral Eq. (\ref{integral2D}). There is a branch cut from $ib$ to $i\infty$.}

\label{fig:figSM1}  

\end{figure}
%--------------------------------------------------------------------------------------------------------------------------------------------------------------------

Then
\begin{align}
I_{+} = \int_{-\infty}^{+\infty} \frac{e^{i2\pi(z+f)}}{\sqrt{z^2+b^2}}dz + \int_{i\infty}^{ib}\frac{e^{i2\pi(z+f)}}{\sqrt{(z+0)^2+b^2}}dz + \int_{ib}^{i\infty}\frac{e^{i2\pi(z+f)}}{\sqrt{(z-0)^2+b^2}}dz = 0,
\end{align}
and the integrals over the arcs vanished due to their exponential decay as the radius goes to infinity.
In the expression above notation $\pm 0$ should be understood from which side of the real axis, positive or negative, the variable $z$ approaches zero. This is needed to pick the correct list of the multivalued square root.
Therefore, 
\begin{align}\label{IntegralOsc}
\int_{-\infty}^{+\infty} \frac{e^{i2\pi(z+f)}}{\sqrt{z^2+b^2}}dz 
&= - \int_{i\infty}^{ib}\frac{e^{i2\pi(z+f)}}{\sqrt{(z+0)^2+b^2}}dz - \int_{ib}^{i\infty}\frac{e^{i2\pi(z+f)}}{\sqrt{(z-0)^2+b^2}}dz \\
&=i\int_{b}^{\infty}\frac{e^{i2\pi f - 2\pi z}}{\sqrt{-(z-i0)^2+b^2}}dz - i\int_{b}^{\infty}\frac{e^{i2\pi f - 2\pi z}}{\sqrt{-(z+i0)^2+b^2}}dz \\
&=2e^{i2\pi f }\int_{b}^{\infty}\frac{e^{- 2\pi z}}{\sqrt{z^2-b^2}}dz \\
&= 2e^{i2\pi f }\int_{1}^{\infty}\frac{e^{- 2\pi b z}}{\sqrt{z^2-1}}dz \\
& =  2e^{i2\pi f }e^{- 2\pi b }\int_{0}^{\infty}\frac{e^{- 2\pi b z}}{\sqrt{z(z+2)}}dz \\
&\approx 4e^{i2\pi f }e^{-2\pi b } \mathrm{arcsinh}\left( \sqrt{\frac{1}{4\pi b}} \right)\\
&=4e^{i2\pi f }e^{-2\pi b } \ln\left( \sqrt{\frac{1}{4\pi b}} + \sqrt{\frac{1}{4\pi b} + 1} \right).
\end{align}
Therefore, 
\begin{align}
2\int_{-\infty}^{+\infty}\frac{\cos(2\pi (x+f))}{\sqrt{x^2+b^2}} dx = 8\cos(2\pi f)e^{-2\pi b }   \ln\left( \sqrt{\frac{1}{4\pi b}} + \sqrt{\frac{1}{4\pi b} + 1} \right)
\end{align}
\begin{align}\label{equation_theta}
1= 2U\nu\left\{  2\ln\left[ \frac{2\sqrt{(\Lambda-f)f}}{b}\right] + 8 \cos(2\pi f) e^{-2\pi b } \ln\left( \sqrt{\frac{1}{4\pi b}} + \sqrt{\frac{1}{4\pi b} + 1} \right) \right\} , 
\end{align}
where we remind that $b=\frac{2\theta}{\omega_{\mathrm{B}}}$.

%-------------------------------------------------
\subsubsection{$8\pi \frac{\theta}{\omega_{\mathrm{B}}} >1$ case}
%-------------------------------------------------
In this case we expand the logarithm as 
\begin{align}
2\int_{-\infty}^{+\infty}\frac{\cos(2\pi (x+f))}{\sqrt{x^2+b^2}} dx = 8\cos(2\pi f)e^{-2\pi b }  \sqrt{\frac{1}{4\pi b}} .
\end{align}
This is now a small correction to the non-linear equation Eq. (\ref{equation_theta}), 
\begin{align}
1= 4U\nu\left\{  \ln\left[ \frac{\sqrt{(\Lambda-f)f\omega_{\mathrm{B}}^2}}{\theta} \right] + 4 \cos(2\pi f)e^{-4\pi \frac{\theta}{\omega_{\mathrm{B}}} }   \sqrt{\frac{\omega_{\mathrm{B}}}{8 \pi \theta}}  \right\},
\end{align}
 and can be treated perturbatively.
Assume we found
\begin{align}
1= 4U\nu\ln\left[ \frac{\sqrt{(\Lambda-f)f\omega_{\mathrm{B}}^2}}{\theta} \right],
\end{align}
which is 
\begin{align}
\theta_{0} = \sqrt{\mu^2 - \frac{\omega_{\mathrm{B}}^2}{4}} e^{-\frac{1}{4U\nu}},
\end{align}
where $fB = \mu - \frac{\omega_{\mathrm{B}}}{2}$, and
where we set $\Lambda-\mu = \mu$ in order to make sure that the magnetization is absent when the magnetic field is zero, i.e. there is no spontaneous magnetic order in the system.
We comment on that in more details in the subsection C below.

In order to find oscillating part of the hybridization we perturb $\theta = \theta_{0}+\delta\theta$ and get
\begin{align}
\delta\theta =  2\theta_{0}  \cos\left(2\pi \frac{\mu}{\omega_{\mathrm{B}}} - \pi \right)e^{-4\pi \frac{\theta_{0}}{\omega_{\mathrm{B}}} }   \sqrt{\frac{\omega_{\mathrm{B}}}{2\pi \theta_{0}}}  ,
\end{align}
and the solution reads
\begin{align}\label{solution1SM}
\theta =\theta_{0} +  2 \theta_{0}  \cos\left(2\pi \frac{\mu}{\omega_{\mathrm{B}}} - \pi \right) e^{-4\pi \frac{\theta_{0}}{\omega_{\mathrm{B}}} }   \sqrt{\frac{\omega_{\mathrm{B}}}{2\pi \theta_{0} }} .
\end{align}

%-------------------------------------------------
\subsubsection{$8\pi \frac{\theta}{\omega_{\mathrm{B}}} \sim 1$ case}
%-------------------------------------------------
For example, we keep in mind a situation when $b=0.1$, then $4\pi b = 8\pi \frac{\theta}{\omega_{\mathrm{B}}} \approx 1.2$.
Then, we estimate last term in the brackets in the right-hand side of Eq. (\ref{equation_theta}) to be 
\begin{align}
4 e^{-2\pi b } \ln\left( \sqrt{\frac{1}{4\pi b}} + \sqrt{\frac{1}{4\pi b} + 1} \right) \approx 1.6,
\end{align}
which is still a small correction as compared to the first term in the brackets of the right-hand side of Eq. (\ref{equation_theta}).
We just can not expand the logarithm anymore. We get the solution
\begin{align}\label{solution2SM}
\theta =\theta_{0} +  4 \theta_{0}  \cos\left(2\pi \frac{\mu}{\omega_{\mathrm{B}}} - \pi \right) e^{-4\pi \frac{\theta_{0}}{\omega_{\mathrm{B}}} }  \ln\left( \sqrt{\frac{\omega_{\mathrm{B}}}{8\pi \theta_{0}}} + \sqrt{\frac{\omega_{\mathrm{B}}}{8\pi \theta_{0}} + 1} \right) .
\end{align}

%-------------------------------------------------
\subsubsection{$8\pi \frac{\theta}{\omega_{\mathrm{B}}} <1$ case}
%-------------------------------------------------
The non-linear equation reads
\begin{align}
1= 2U\nu \left\{  2\ln\left[ \frac{2\sqrt{(\Lambda-f)f}}{b}\right] + 4 \cos(2\pi f) e^{-2\pi b } \ln\left(\frac{1}{\pi b}\right)  \right\} , 
\end{align}
where, recall $b=\frac{2\theta}{\omega_{\mathrm{B}}}$ and $f \omega_{\mathrm{B}}= \mu - \frac{\omega_{\mathrm{B}}}{2}$, and $\omega_{\mathrm{B}} \Lambda = 2\mu$ to make sure there is no spontaneous magnetization in the system.
We then derive a solution for the hybridization,
\begin{align}\label{solution3SM}
\theta = \sqrt{\mu^2 - \frac{\omega_{\mathrm{B}}^2}{4}   }  e^{-\frac{1 + 8U\nu \cos\left(2\pi \frac{\mu}{\omega_{\mathrm{B}}} - \pi \right) e^{-4\pi \frac{\theta}{\omega_{\mathrm{B}}} }\ln\left[\pi\frac{\sqrt{4\mu^2 - \omega_{\mathrm{B}}^2 }}{\omega_{\mathrm{B}}} \right]}{4U\nu\left[1+2\cos\left(2\pi \frac{\mu}{\omega_{\mathrm{B}}} - \pi \right)
e^{-4\pi \frac{\theta}{\omega_{\mathrm{B}}}  } \right]} }.
\end{align}
When $\cos\left(2\pi \frac{\mu}{\omega_{\mathrm{B}}} - \pi \right) = 0$, the gap is 
\begin{align}
\theta = \sqrt{\mu^2 - \frac{\omega_{\mathrm{B}}^2}{4}   } e^{-\frac{1}{4U\nu}}.
\end{align}
Moreover, it is clear that the solution is on the boarderline of its existance. 
Namely, when the sign of the argument of the natural exponent changes sign to positive, the solution no longer obeys $8\pi \frac{\theta}{\omega_{\mathrm{B}}} <1$ and $\theta < \mu$ smallness conditions which were assumed in the course of the derivation. 
Even next harmonics of the oscillation which were omitted in the calculation can't help the argument to not change the sign to positive.
Hence, large magnetic fields causes a phase transition by either closing the gap or changing the solution to another limit, i.e. to $8\pi \frac{\theta}{\omega_{\mathrm{B}}} >1$ case, when the oscillations are rather suppressed and the argument changes sign back to negative. 
It is quite likely this process is accompanied with a phase separation of the two possibilities. 
We anticipate that the oscillations are going to be antisymmetric with respect to highest and lowest phases of the oscillation, besides the existence of the higher harmonics which become less exponentially suppressed.

%--------------------------------------------------------------
%--------------------------------------------------------------
%--------------------------------------------------------------
%--------------------------------------------------------------
%--------------------------------------------------------------
%--------------------------------------------------------------

%--------------------------------------------------------------
\section{Free energy and magnetization}\label{sectionC}
%--------------------------------------------------------------
Here we wish to derive the free energy $F$ within the Keldysh technique. 
We will be following Ref. \cite{SchwieteFinkelstein}.
\begin{align}
F= {\cal U} - TS,
\end{align}
where $S = \int dT \frac{1}{T} \frac{d{\cal U}}{dT}$ is entropy, and the internal energy is denoted by ${\cal U}$ (not to be confused with the fermion interaction $U$ in our model) and is typically ${\cal U} =\int_{x}{\cal F}_{x} A_{x}$, where $A_{x}$ is a function of $x$ and ${\cal F}_{x} = \tanh\left[\frac{f(x)}{2T}\right]$ is the distribution function, where $f(x)$ is fermion dispersion, 
then the entropy
\begin{align}
S = \frac{1}{T}\int_{x}A_{x} {\cal F}_{x}  - \frac{1}{T}\int_{x}A_{x} \left\{ 1+\frac{2T}{f(x)}\ln\left[ 1+e^{-\frac{f(x)}{T}}\right] \right\},
\end{align}
and therefore,
\begin{align}\label{freeAepsilon}
F = \int_{x}A_{x}\left\{ 1+\frac{2T}{f(x)}\ln\left[ 1+e^{-\frac{f(x)}{T}}\right] \right\}
\rightarrow
\int_{x}A_{x} \frac{2T}{f(x)}\ln\left[ 1+e^{-\frac{f(x)}{T}}\right] .
\end{align}
Here $1$ in the square brackets should be dropped in order to define a proper free energy, hence the right arrow. 
Then we can define the free energy by figuring out the internal energy, which can be done within the Keldysh technique. 
For that, following the lines of Ref. \cite{SchwieteFinkelstein}, we introduce heat density sources in to the action (not to be confused with the entropy) as
\begin{align}
&
iS_{\eta} + iS_{0} +
iS_{\mathrm{int},1} 
+iS_{\mathrm{int};2} 
= 
\\
&
\frac{i}{2}\int_{t,{\bf r}}
\left\{ i  {\bar \Psi}(x) \frac{1}{1+\hat{\eta}} \partial_{t}\Psi(x)
- i \left[\partial_{t} {\bar \Psi}(x)\right] \frac{1}{1+\hat{\eta}} \Psi(x) \right\}
\\
&
+
i \int_{t,{\bf r}}
\left\{
\hat{\bar{\Psi}}(x)\hat{\sigma}_{0}
\left[
\begin{array}{cc} 
- \hat{\xi}_{\bf k} & 0\\
0 & - \epsilon_{0}
\end{array}
\right]
\hat{\Psi}(x)
+
\hat{\bar{\Psi}}(x)
\left[
\begin{array}{cc} 
0 & -\theta \hat{\sigma}_{0}  -\theta_{\mathrm{q}}\hat{\sigma}_{1} \\
-\theta \hat{\sigma}_{0}  -\bar{\theta}_{\mathrm{q}}\hat{\sigma}_{1} & 0
\end{array}
\right]
\frac{1}{1+\hat{\eta}} 
\hat{\Psi}(x)
\right\}
\\
&
- 
\frac{i}{2} \int_{t,{\bf r}} \mathrm{Tr}_{\mathrm{K}}\left[\hat{\bar{\theta}}(x)U^{-1}\hat{\sigma}_{1}\frac{1}{1+\hat{\eta}}\hat{\theta}(x)\right],
\end{align}
where $\hat{\eta} = \eta_{1}\hat{\sigma}_{0} + \eta_{2}\hat{\sigma}_{1}$ are the heat density sources, and $\hat{\xi}_{\bf k}$ stands for operator acting on the coordinate. 
Heat density is given by varying the partition function with respect to the source, 
\begin{align}
{\cal U} \equiv \langle h_{\mathrm{cl}} \rangle = \frac{i}{2} \left. \frac{\delta Z}{\delta \eta_{2}} \right|_{\eta_{1}=\eta_{2}=0}.
\end{align}
Then let us find out what $A_{x}$ defined in Eq. (\ref{freeAepsilon}) equals to in our model.
After fermions are integrated out the action becomes
\begin{align}
iS = \int_{\epsilon}\int_{{\bf k}} \mathrm{Tr}\ln\left[ 1- \hat{G}\hat{h}_{\mathrm{int};q}- \hat{G}\left[
\begin{array}{cc} 
 \epsilon & -\theta \\
-\theta &  \epsilon
\end{array}
\right]\hat{\eta}\right],
\end{align}
where $\hat{G}$ and $\hat{h}_{\mathrm{int};q}$ were defined in the Appendix \ref{sectionA}. 
Then
\begin{align}
\frac{i}{2}  &
 \left. \frac{\delta }{\delta \eta_{2}} 
 \left( iS_{\eta} + iS_{0} + iS_{\mathrm{int},1} \right)\right|_{\eta_{1}=\eta_{2}=0; \theta_{\mathrm{q}} = 0}
 \\
 &
=\frac{i}{2}(2\pi i)
\int_{\epsilon}\int_{\bf k}
\left[ 2\epsilon^2 -\epsilon (\epsilon_{0}+\xi_{\bf k}) - 2\theta^2  \right]\left[ \delta(\epsilon - \epsilon_{{\bf k};+}) - \delta(\epsilon - \epsilon_{{\bf k};-}) \right] 
\frac{{\cal F}_{\epsilon}}{\epsilon_{{\bf k};+} - \epsilon_{{\bf k};-} }
\\
&
= 
-\frac{1}{2}\int_{\bf k}\epsilon_{{\bf k};+}{\cal F}_{\epsilon_{{\bf k};+}}
-\frac{1}{2}\int_{\bf k}\epsilon_{{\bf k};-}{\cal F}_{\epsilon_{{\bf k};-}}
+\theta^2
\int_{\bf k}\frac{{\cal F}_{\epsilon_{{\bf k};+}}-{\cal F}_{\epsilon_{{\bf k};-}}}{\epsilon_{{\bf k};+} - \epsilon_{{\bf k};-} }
\\
&
=
-\frac{1}{2}\int_{\bf k}\epsilon_{{\bf k};+}{\cal F}_{\epsilon_{{\bf k};+}}
-\frac{1}{2}\int_{\bf k}\epsilon_{{\bf k};-}{\cal F}_{\epsilon_{{\bf k};-}}
+\frac{\theta^2}{U},
\end{align}
where in the last equality sign we utilized self-consistent gap equation. Another term reads
\begin{align}
\frac{i}{2}  
 \left. \frac{\delta }{\delta \eta_{2}} 
 iS_{\mathrm{int},2} \right|_{\eta_{1}=\eta_{2}=0 ;\theta_{\mathrm{q}} = 0} = - \frac{\theta^2}{2U}.
\end{align}
Summing the two we get for the internal energy
\begin{align}
{\cal U} = 
\frac{i}{2}  &
 \left. \frac{\delta }{\delta \eta_{2}} 
 \left( iS_{\eta} + iS_{0} + iS_{\mathrm{int},1} +  iS_{\mathrm{int},2} \right)\right|_{\eta_{1}=\eta_{2}=0; \theta_{\mathrm{q}} = 0}
=
-\frac{1}{2}\int_{\bf k}\epsilon_{{\bf k};+}{\cal F}_{\epsilon_{{\bf k};+}}
-\frac{1}{2}\int_{\bf k}\epsilon_{{\bf k};-}{\cal F}_{\epsilon_{{\bf k};-}}
+\frac{\theta^2}{2U},
\end{align}
which gives the correct expression for the free energy upon calculating entropy, namely
\begin{align}\label{FreeFull}
F = -T\int_{\bf k}\ln\left[ 1+e^{\frac{-\epsilon_{{\bf k};+}}{T}}  \right]
-T\int_{\bf k}\ln\left[ 1+e^{\frac{-\epsilon_{{\bf k};-}}{T}}  \right]
+\frac{\theta^2}{2U}
\end{align}

We study the free energy in the presence of the magnetic field and at $T=0$, in which case $e^{\frac{-\epsilon_{n;+}}{T}} = 0$ and $e^{\frac{-\epsilon_{n;-}}{T}} \gg 1$ since the $-\epsilon_{n;-} > 0$ for any $n$, and from Eq. (\ref{FreeFull}) we get,
\begin{align}\label{free}
F = \frac{\omega_{\mathrm{B}}}{2}\nu\sum_{n} \left(\omega_{\mathrm{B}}n+\frac{\omega_{\mathrm{B}}}{2} - \mu \right)   - \frac{\omega_{\mathrm{B}}}{2}\nu\sum_{n} \sqrt{\left(\omega_{\mathrm{B}}n+\frac{\omega_{\mathrm{B}}}{2} - \mu \right)^2 +4\theta^2} + \frac{\theta^2}{2U}.
\end{align}
The sum over the Landau levels in the second term will be made again with the help of Poisson summation formula Eq. (\ref{PoissonSM}) .
\begin{align}\label{freeALL}
&
F=F_{0}+F_{\Sigma}, \\
\label{free0A}
&
F_{0} =\frac{1}{2}\nu \omega_{\mathrm{B}}^2 \int_{-f}^{\Lambda - f} zdz - \frac{1}{2}\nu \omega_{\mathrm{B}}^2\int_{-f}^{\Lambda - f}  \sqrt{z^2+b^2}dz + \frac{\theta^2}{2U},\\
& \label{freeOSC}
F_{\Sigma} = - \frac{1}{2}\nu\omega_{\mathrm{B}}^2\sum_{p \neq 0}\int_{-f}^{\Lambda - f} e^{i2\pi p (z+f)} \sqrt{z^2+b^2}dz.
\end{align}
In the limit $8\pi\frac{\theta_{0}}{\omega_{\mathrm{B}}}>1$ we calculate
\begin{align}
F_{0}
&
=\frac{1}{4}\nu\left[ (\Lambda - f)^2 \omega_{\mathrm{B}}^2 - f^2 \omega_{\mathrm{B}}^2\right]
 -\frac{1}{4}\nu\left\{ (\Lambda - f)^2 \omega_{\mathrm{B}}^2 + f^2 \omega_{\mathrm{B}}^2  + 4\theta^2  +4 \theta^2 \ln\left[\frac{\omega_{\mathrm{B}}^2(\Lambda - f)f}{\theta^2}\right] \right\} + \frac{\theta^2}{2U}
\\
&
=\frac{1}{4}\nu\left[ (\Lambda - f)^2 \omega_{\mathrm{B}}^2 - f^2 \omega_{\mathrm{B}}^2\right]
 -\frac{1}{4}\nu\left[ (\Lambda - f)^2 \omega_{\mathrm{B}}^2 + f^2 \omega_{\mathrm{B}}^2  + 4\theta^2  - 16 \theta^2 \cos(2\pi f) e^{-4\pi \frac{\theta}{\omega_{\mathrm{B}}}} \sqrt{\frac{\omega_{\mathrm{B}}}{2\pi \theta }} \right] -\frac{\theta^2}{2U} + \frac{\theta^2}{2U}
 \label{free01}
\\
&
= - \frac{\nu}{2}\left(\mu - \frac{\omega_{\mathrm{B}}}{2}\right)^2 - \nu\theta^2 + 4\nu \theta^2 \cos(2\pi f) e^{-4\pi \frac{\theta}{\omega_{\mathrm{B}}}} \sqrt{\frac{\omega_{\mathrm{B}}}{2\pi \theta }}
\label{free02}
\\
&
\approx - \frac{\nu}{2}\left(\mu - \frac{\omega_{\mathrm{B}}}{2}\right)^2 -  \nu\theta_{0}^2. 
\label{free0}
\end{align}
where, recall, $\Lambda \omega_{\mathrm{B}} = 2\mu$ and $f \omega_{\mathrm{B}} = \mu - \frac{\omega_{\mathrm{B}}}{2}$, and where we have used self-consistent equation in obtaining $- \frac{\theta^2}{2U}$ term in the first equality sign.
We have applied results of $8\pi\frac{\theta_{0}}{\omega_{\mathrm{B}}}>1$ limit, but other limits will also lead to the result in the last line Eq. (\ref{free0}).
This calculation of the free energy is in line with the BCS theory, i.e the energy gain is equal to $-\nu\theta_{0}^2$.
We note that the response from the non-interacting term (when hybridization is treated as a constant rather than self-consistently) is diamagnetic, although it is not immediately clear from Eq. (\ref{free0}).

Now the sum over the harmonics. We again pick only $p=\pm 1$ from the sum.
Then we are left with the integral, which in the small magnetic field regime, i.e. $\frac{1}{2\pi b} <1$, is estimated to be
\begin{align}
\int_{-\infty}^{+\infty} \sqrt{z^2+b^2} e^{i2\pi(z+f)}dz \rightarrow - e^{i2\pi f} e^{-2\pi b} \frac{4b \omega_{\mathrm{B}}^2}{\pi} \sqrt{\frac{1}{4\pi b}\left( 1+\frac{1}{4\pi b} \right)},
\end{align}
where by $\rightarrow$ we extracted only oscillating part, and we remind $b = \frac{2\theta_{0}}{\omega_{\mathrm{B}}}$.
We have used integration by parts once in order to get rid of divergence at infinity. After that we utilized the same contour integration as was used in Eq. (\ref{IntegralOsc}).
In the integration by parts we have omitted the boundary terms in it. 
Let us examine whether omitted boundary terms can contribute anything oscillating from oscillatory dependence of $\theta$. 
They are
\begin{align}
&
\left( \left.\frac{1}{2\pi i} \sqrt{z^2 + b^2} e^{i2\pi z} \right|_{0}^{\infty} - \left.\frac{1}{2\pi i} \sqrt{z^2 + b^2} e^{i2\pi z} \right|_{0}^{\infty} \right)\frac{e^{i2\pi f}}{2} + \mathrm{c.c.}
\\
=
&
\left(
\frac{1}{2\pi i} \sqrt{\Lambda^2 + b^2} e^{i2\pi \Lambda} 
 - 
\frac{1}{2\pi i} \sqrt{\Lambda^2 + b^2} e^{-i 2\pi \Lambda}
 \right)
 \frac{e^{i2\pi f}}{2} + \mathrm{c.c.},
\end{align}
where $\Lambda \rightarrow \infty$.
We see there is no extra oscillating term $\propto\cos\left( 2\pi \frac{\mu}{\omega_{\mathrm{B}}}\right)$ in the free energy from these boundary terms. Therefore it was safe to ignore them. 

Therefore, for $4\pi b > 1$ we get for the oscillating part of free energy,
\begin{align}
F_{\Sigma} &= -\nu\omega_{\mathrm{B}}^2\int_{-\infty}^{+\infty}\sqrt{z^2+b^2}\cos(2\pi (z+f)) dz 
\rightarrow  \nu\frac{\omega_{\mathrm{B}}^2}{\pi^2}\sqrt{\frac{2\pi \theta_{0}}{\omega_{\mathrm{B}}}} \cos\left(2\pi \frac{\mu}{\omega_{\mathrm{B}}} - \pi \right)e^{-4\pi \frac{\theta_{0}}{\omega_{\mathrm{B}}} } .
\label{freeSigma}
\end{align}
%Oscillating term from $-\nu\theta^2$ is obtained by plugging in Eq. (\ref{solution1SM}) there,
%\begin{align}
%F_{0} = -\nu\theta^2 
%\approx 
%-\nu\theta_{0}^2 - 4\nu \theta_{0}^2   \cos\left(2\pi \frac{\mu}{\omega_{\mathrm{B}}} - \pi \right) e^{-4\pi 
%\frac{\theta_{0}}{\omega_{\mathrm{B}}} } \sqrt{\frac{\omega_{\mathrm{B}}}{2\pi \theta_{0} }}. 
%\label{free0b}
\%end{align}
Here is why we have been choosing $\Lambda \omega_{\mathrm{B}} = 2\mu$ in all of our calculations. 
The non-oscillating part of the free energy obtained from Eq. (\ref{free0}) reads $-\nu\theta_{0}^2 =- \nu\left(\mu^2 - \frac{\omega_{\mathrm{B}}^2}{4}\right) e^{-\frac{1}{2U\nu}}$, and hence there is no term here propotional to the magnetic field in free energy. 
If there was a term propotional to the magnetic field, the system would have had a spontaneous magnetization, which is experimentally not the case.
We also conclude from this argument that the response of the system is diamagnetic, because non-oscillating part of the magnetization is $M_{0}=- \partial_{B}(-\nu\theta_{0}^2) = - \frac{\omega_{\mathrm{B}}}{2}\frac{e}{mc} \nu e^{-\frac{1}{2U\nu}}$. 
The system wants to expel the magnetic field such that the Coulomb energy of the system is minimized. This is first key results of the present work.

We compare now the two oscillating terms in the free energy, the one came from Eq. (\ref{free0}), which is due to the Coulomb interaction, and the other from the non-interacting picture given by Eq. (\ref{freeSigma}),
\begin{align}\label{freeResultSM}
\delta F_{\mathrm{osc}} 
=-
4\nu \theta_{0}^2 \sqrt{\frac{\omega_{\mathrm{B}}}{2\pi \theta_{0} }}\left(   -  \frac{\omega_{\mathrm{B}}}{2 \pi\theta_{0}} \right)\cos\left(2\pi \frac{\mu}{\omega_{\mathrm{B}}} - \pi \right) e^{-4\pi \frac{\theta_{0}}{\omega_{\mathrm{B}}} }.
\end{align}

Oscillating part of the magnetization reads
\begin{align}
M_{\mathrm{osc}} 
= -\frac{\partial \delta F_{\mathrm{osc}} }{\partial B} 
&\approx   
- \frac{2\mu\nu}{\pi} \frac{e}{mc} \left(\frac{2\pi \theta_{0}}{\omega_{\mathrm{B}}}\right)^{\frac{3}{2}}
\left(   - \frac{\omega_{\mathrm{B}}}{2 \pi\theta_{0}} \right) 
\sin\left(2\pi \frac{\mu}{\omega_{\mathrm{B}}} - \pi \right) e^{-4\pi \frac{\theta_{0}}{\omega_{\mathrm{B}}} }
\\
&
= -\frac{2\mu\nu}{\pi} \frac{e}{mc} \left(\frac{2\pi \theta_{0}}{\omega_{\mathrm{B}}}\right)^{\frac{1}{2}}
\sin\left(2\pi \frac{\mu}{\omega_{\mathrm{B}}}  \right) e^{-4\pi \frac{\theta_{0}}{\omega_{\mathrm{B}}} },
\label{magnetizationSM}
\end{align}
where only the $\propto \mu$ term was kept because it is the largest parameter here.
Clearly, in Eq. (\ref{magnetizationSM}) we have repeated the results obtained in Ref. \onlinecite{AlloccaCooper}. Namely, that the oscillating part of the magnetization doesn't depend on the oscillations of the hybridization gap in significant way. Main part of the sscillations is due to the magnetization of the non-interacting (when $\theta_{0}$ is treated as a constant) system. There might be contributions to the free energy and magnetization originating from the oscillations of the gap which are proportional to a square of the oscillating part of the gap, see Ref. \onlinecite{AlloccaCooper} for details. Such terms are subleading to the result Eq. (\ref{magnetizationSM}). We stress that in our published paper we have accidentally and unfortunatelly missed the term in red in Eq. (\ref{free01}) and (\ref{free02}). The rest of our paper is not affected by this mistake.

%------------------------------------------------------------------------------
\begin{figure}[h] 
\centerline{
\begin{tabular}{cc}
\includegraphics[width=0.3 \columnwidth]{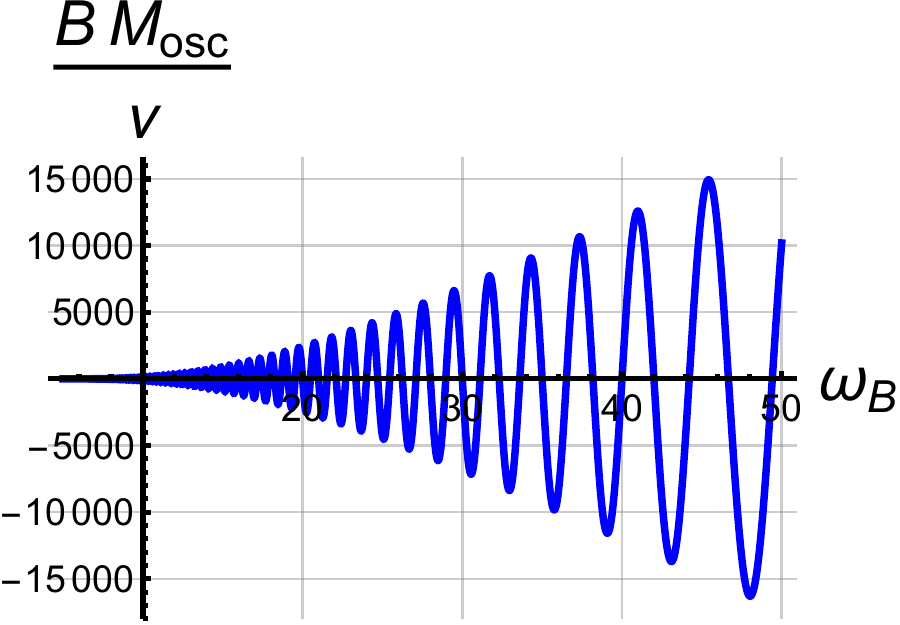}~~~&
\includegraphics[width=0.3 \columnwidth]{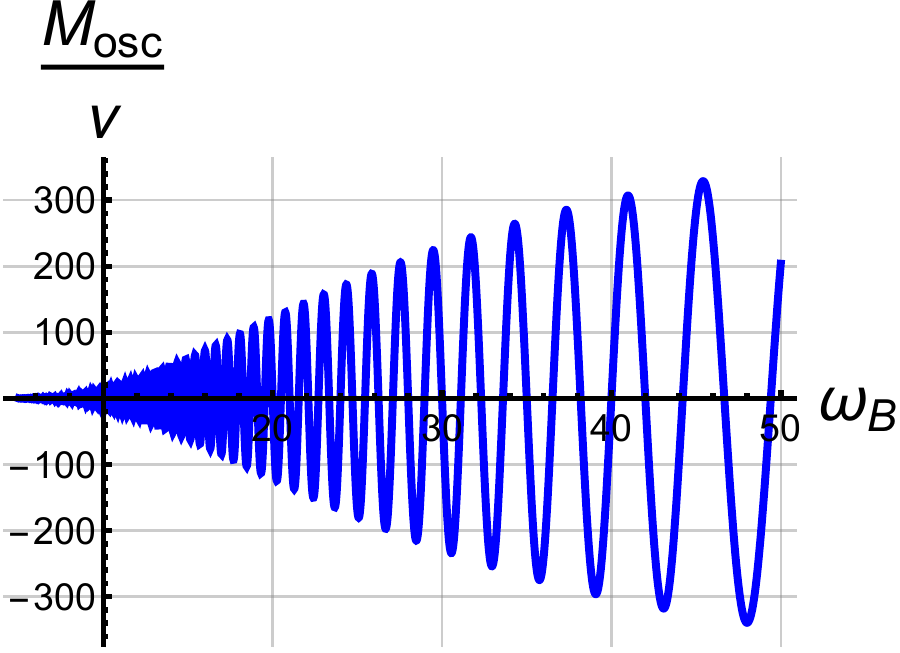}~~~\\
\includegraphics[width=0.3 \columnwidth]{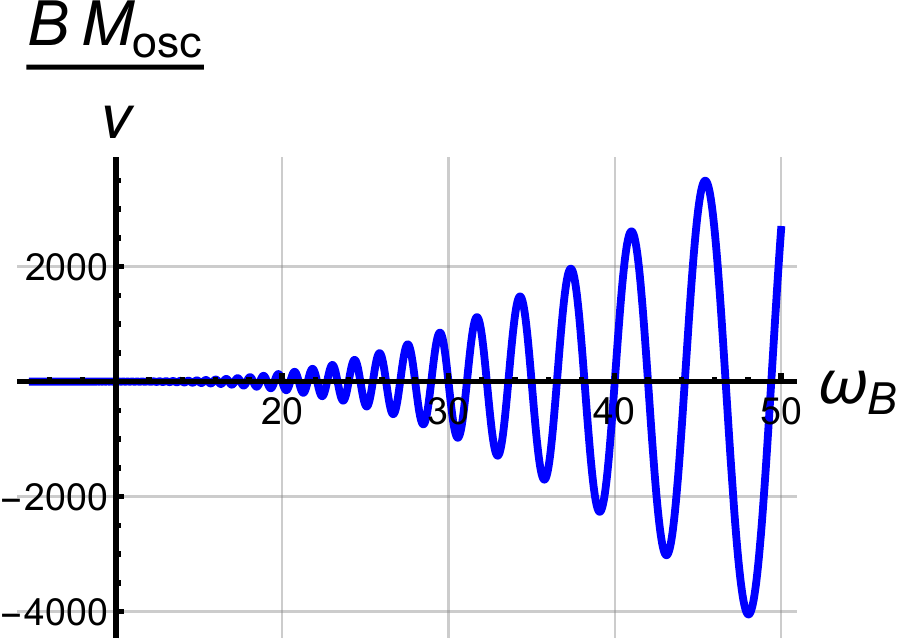}~~~&
\includegraphics[width=0.3 \columnwidth]{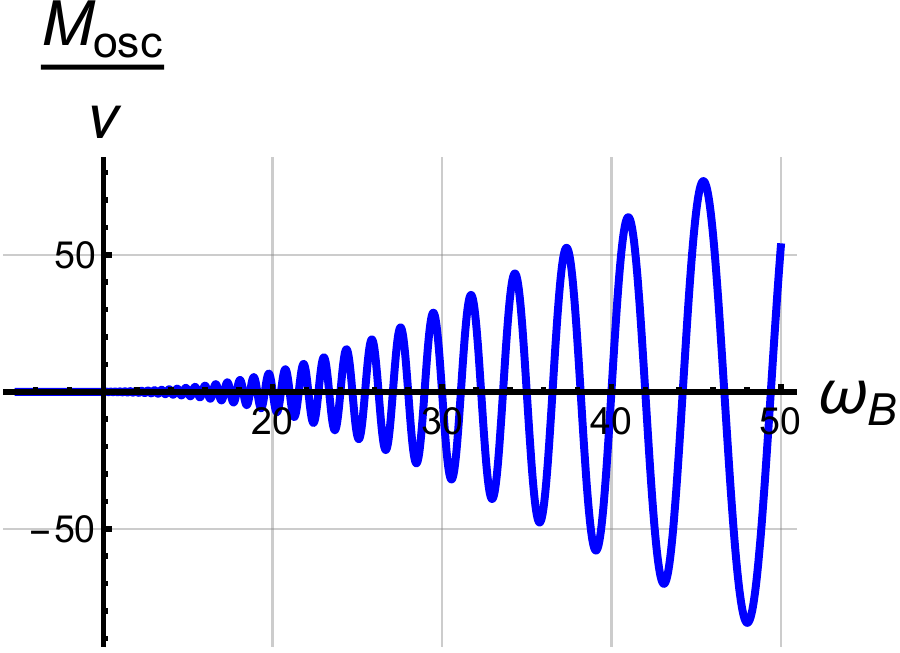}~~~
\end{tabular}
}
\protect\caption{Left: plot of the first harmonic of the oscillating part of the torque, i.e. of $M_{osc}B$ where $M_{\mathrm{osc}}$ is given by Eq. (\ref{magnetizationSM}) for $2\theta_{0} \approx 40K$. Top: effective mass is $m=0.2 m_{\mathrm{e}}$; Bottom: $m=0.4 m_{\mathrm{e}}$. Right: magnetization for the same parameters. All plots are for $T=0$ and $\mu m/m_{\mathrm{e}} = 420K$.}

\label{fig:figSM2}   

\end{figure}
%------------------------------------------------------------------------------

%--------------------------------------------------------------
%--------------------------------------------------------------
%--------------------------------------------------------------
\section{Temperature dependence}\label{sectionD}
%--------------------------------------------------------------
%--------------------------------------------------------------
%--------------------------------------------------------------
Here we study temperature dependence of oscillations obtained in previous subsections.
For that we update the non-linear equation to account for the temperature,
\begin{align}
1 =  U \nu \omega_{\mathrm{B}} \sum_{n} 
\frac{{\cal F}_{\epsilon_{n,+}} 
-
{\cal F}_{\epsilon_{n,-}} }{\sqrt{(\omega_{\mathrm{B}}n+\frac{\omega_{\mathrm{B}}}{2} - \mu)^2 +4\bar{\theta}\theta}}
\end{align}
We follow the same steps as were done in the $T=0$ case. 
We first introduce notations by rewriting the equation
\begin{align}
1=U \nu \omega_{\mathrm{B}}\left[{\cal R}^{(0)}(T) +  \sum_{n \neq 0} {\cal R}^{(n)}_{\mathrm{osc}}(T) \right].
\end{align}
Below we fill focus on the stufy of ${\cal R}^{(n=\pm 1)}_{\mathrm{osc}}(T)$. Let us first discuss ${\cal R}^{(0)}(T)$.
In case of finite temperature the integral Eq. (\ref{nonoscillatingSM}) gets modified as
\begin{align}\label{temp1}
\int_{0}^{\Lambda} \frac{dx}{\sqrt{(x-f)^2+b^2}} \rightarrow \int_{-f}^{\Lambda-f} \frac{dz}{\sqrt{z^2+b^2}}  \left[ 1- \frac{1}{e^{\frac{-z+\sqrt{z^2+b^2}}{2T/\omega_{\mathrm{B}}}  }+1} - \frac{1}{e^{\frac{z+\sqrt{z^2+b^2}}{2T/\omega_{\mathrm{B}}}  }+1}\right]dz
\end{align}
where we used ${\cal F}_{-x} = -{\cal F}_{x}$ identity after shifting the $z$ in $\mu - \frac{\omega_{\mathrm{B}}}{2}$ to write down the first fraction in the way it is written in the square brackets, and also of use was ${\cal F}_{x} = 1 -\frac{2}{e^{\frac{x}{T}}+1}$. 
Here, again $b=\frac{2\theta_{0}^2}{\omega_{\mathrm{B}}}$ and $f=\frac{\mu}{\omega_{\mathrm{B}}} - \frac{1}{2}$.
\begin{align}
&
- \int_{-f}^{\Lambda-f} \frac{dz}{\sqrt{z^2+b^2}}  \left[ \frac{1}{e^{\frac{-z+\sqrt{z^2+b^2}}{2T/\omega_{\mathrm{B}}}  }+1} + \frac{1}{e^{\frac{z+\sqrt{z^2+b^2}}{2T/\omega_{\mathrm{B}}}  }+1}\right]dz
\\
&
=
-\left.\frac{1}{2}\ln\left[ \frac{z+\sqrt{z^2+b^2}}{\sqrt{z^2+b^2}-z}\right]\left[ \frac{1}{e^{\frac{-z+\sqrt{z^2+b^2}}{2T/\omega_{\mathrm{B}}}  }+1} + \frac{1}{e^{\frac{z+\sqrt{z^2+b^2}}{2T/\omega_{\mathrm{B}}}  }+1}\right] \right|_{-f}^{\Lambda-f}
\label{byparts}
\\
&
+  \frac{1}{2}\int_{-f}^{\Lambda-f}dz\ln\left[ \frac{z+\sqrt{z^2+b^2}}{\sqrt{z^2+b^2}-z}\right]
\partial_{z}\left[ \frac{1}{e^{\frac{-z+\sqrt{z^2+b^2}}{2T/\omega_{\mathrm{B}}}  }+1} + \frac{1}{e^{\frac{z+\sqrt{z^2+b^2}}{2T/\omega_{\mathrm{B}}}  }+1}\right]
\\
&
\approx
-2\ln\left[ \frac{2\sqrt{(\Lambda-f)f}}{b} \right]\frac{1}{e^{\frac{\omega_{\mathrm{B}} b^2}{4Tf}}+1}
\\
&
= - 2\ln\left[ \frac{\sqrt{\mu^2-\frac{\omega_{\mathrm{B}}^2}{4}}}{\theta_{0}}\right]\frac{1}{e^{\frac{\theta_{0}^2}{T\mu}}+1},
\end{align}
where we kept only Eq. (\ref{byparts}) after integrating by parts, as the other is less singular at $z=0$, and assumed $\Lambda -f \approx f$ in the distribution function when taking the upper and lower limits.
This assumption is valid at small temperatures, when the derivative over the distribution function is almost a delta function. The upper and lower limits are fixed and are never equal to infinity.

Now, the most interesting part. The integral corresponding to oscillations at zero temperature Eq. (\ref{IntegralOsc}) gets modified as
\begin{align}
\int_{-\infty}^{+\infty} \frac{e^{i2\pi(z+f)}}{\sqrt{z^2+b^2}}dz  
\rightarrow
\int_{-\infty}^{+\infty} \frac{e^{i2\pi(z+f)}}{\sqrt{z^2+b^2}} \left[ 1- \frac{1}{e^{\frac{-z+\sqrt{z^2+b^2}}{2T/\omega_{\mathrm{B}}}  }+1} - \frac{1}{e^{\frac{z+\sqrt{z^2+b^2}}{2T/\omega_{\mathrm{B}}}  }+1}\right]dz.
\end{align}
Therefore, 
\begin{align}
\theta &=\theta_{0} +  2 \theta_{0}  \cos\left(2\pi \frac{\mu}{\omega_{\mathrm{B}}} - \pi \right) e^{-4\pi \frac{\theta_{0}}{\omega_{\mathrm{B}}} }   \sqrt{\frac{\omega_{\mathrm{B}}}{2\pi \theta_{0} }} 
\\
&\rightarrow
\theta_{0} + 2 \theta_{0}  \cos\left(2\pi \frac{\mu}{\omega_{\mathrm{B}}} - \pi \right) R(T),
\label{R(T)sm}
\end{align}
where $R(T)$ is obtained below and $R(T=0) = e^{-4\pi \frac{\theta_{0}}{\omega_{\mathrm{B}}} }   \sqrt{\frac{\omega_{\mathrm{B}}}{2\pi \theta_{0} }}$ is the exact result at $T=0$ obtained above.
We approximated $\theta_{0}$ to be an independent on temperature which is a good approximation at temperatures way below the transition.

We define the $+1$ harmonic to be
\begin{align} \label{R+1}
{\cal R}^{(+1)}_{\mathrm{osc}}(T) =
&
\frac{1}{\omega_{\mathrm{B}}}\int_{-\infty}^{+\infty} \frac{e^{i2\pi(z+f)}}{\sqrt{z^2+b^2}} \left[1- \frac{1}{e^{\frac{-z+\sqrt{z^2+b^2}}{2T/\omega_{\mathrm{B}}}  }+1} - \frac{1}{e^{\frac{z+\sqrt{z^2+b^2}}{2T/\omega_{\mathrm{B}}}  }+1}\right]dz \\
=
&
\frac{1}{\omega_{\mathrm{B}}} e^{i2\pi f}\int_{-\infty}^{+\infty} e^{i2\pi b \sinh(y)} \left[1- \frac{1}{e^{\frac{\theta_{0}}{T} e^{-y} }+1} - \frac{1}{e^{\frac{\theta_{0}}{T}e^{y}  }+1}\right]dy 
\label{W} 
\\
\equiv
&
\frac{1}{\omega_{\mathrm{B}}}e^{i2\pi f}\left( W_{-} + W_{+}\right),
\end{align}
where we made $z=b\sinh(y)$ change of variable, which resulted in $dz = b\cosh(y)dy$, $\sqrt{z^2+b^2} = b\cosh(y)$ and $-z+\sqrt{z^2+b^2} = be^{-y}$ while $z+\sqrt{z^2+b^2} = be^{y}$, and $\frac{\omega_{\mathrm{B}}b}{2T} = \frac{\theta_{0}}{T}$.
Harmonic with $-1$ is obtained from Eq. (\ref{R+1}) by complex conjugation, i.e. ${\cal R}^{(-1)}_{\mathrm{osc}}(T)  = \left[{\cal R}^{(+1)}_{\mathrm{osc}}(T)\right]^{*} $. 
%--------------------------------------------------------------------------------------------------------------------------------------------------------------------
\begin{figure}[h] 
\centerline{
\includegraphics[width=0.3 \columnwidth]{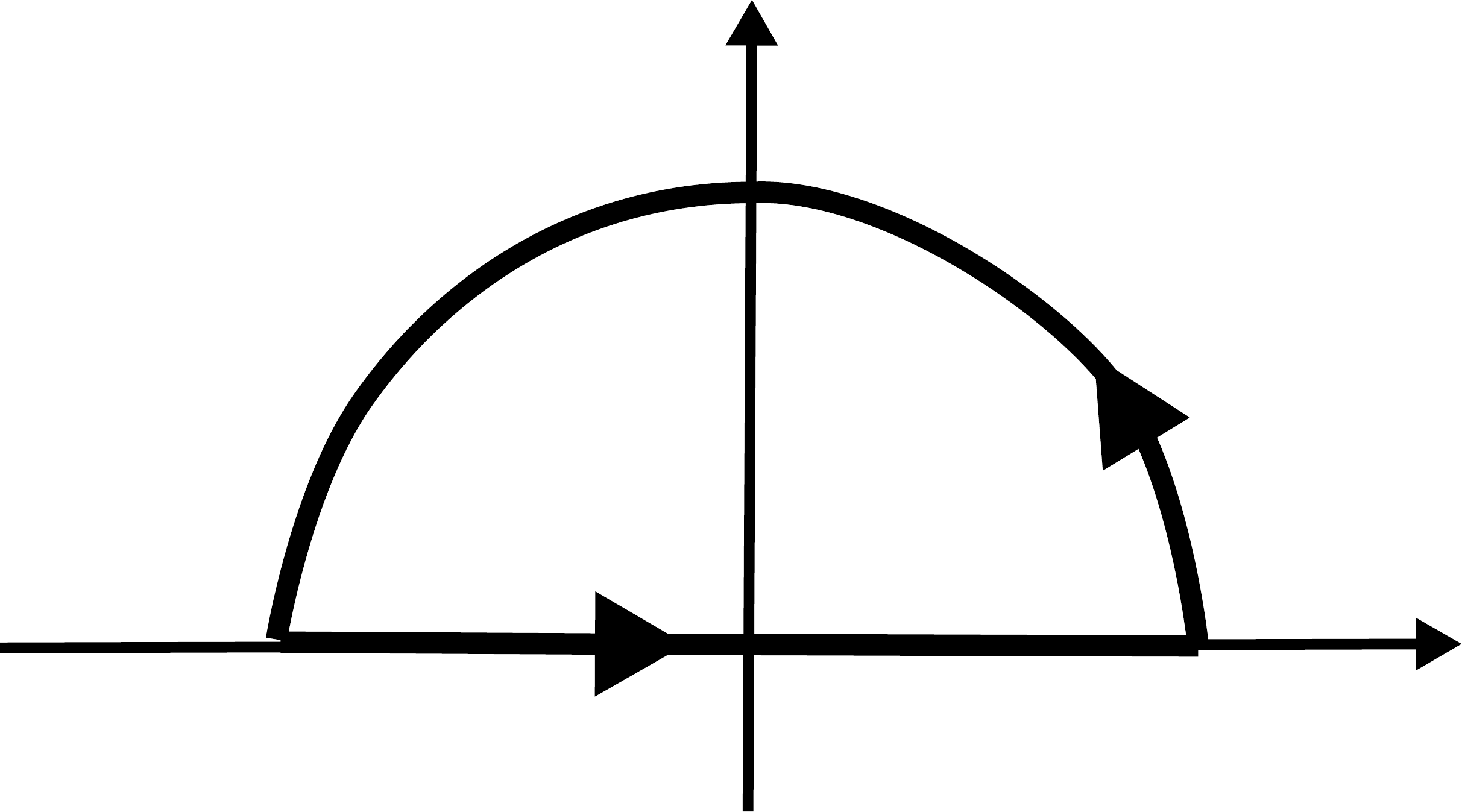}
}
\protect\caption{Contour of integration used to calculate $W_{-}$ defined in Eq. (\ref{W}) and Eq. (\ref{W-}). The contour is closed on the top part of the complex plane in order to ensure convergence. The residues are picked from inside of the contour.}

\label{fig:figSM4}  

\end{figure}
%--------------------------------------------------------------------------------------------------------------------------------------------------------------------
We introduce contour integration shown in Fig. (\ref{fig:figSM4}) and obtain the expression for $W_{-}$ defined in Eq. (\ref{W}),
\begin{align}\label{W-}
W_{-}
&
= \int_{-\infty}^{+\infty}  e^{i2\pi b \sinh(y)} \left[ \frac{1}{2} - \frac{1}{e^{\frac{\theta_{0}}{T} e^{-y} }+1}\right] dy 
\\
&
=
\frac{1}{2}\int_{-\infty}^{+\infty}  e^{i2\pi b \sinh(y)} \frac{e^{\frac{\theta_{0}}{T} e^{-y} }-1}{e^{\frac{\theta_{0}}{T} e^{-y} }+1} dy
\\
& = 2\pi i \sum_{ \mathrm{res}} (...) .
\end{align}
The expression for $W_{+}$ is obtained from $W_{-}$ by complex conjugation, since  
\begin{align}
W_{+}
 &= \int_{-\infty}^{+\infty}  e^{i2\pi b \sinh(y)} \left[ \frac{1}{2} - \frac{1}{e^{\frac{\theta_{0}}{T} e^{y} }+1}\right] dy  
 \\
 &
 = \int_{-\infty}^{+\infty}  e^{-i2\pi b \sinh(y)} \left[ \frac{1}{2} - \frac{1}{e^{\frac{\theta_{0}}{T} e^{-y} }+1}\right] dy  
 \\
 &
 = W_{-}^{*},
\end{align}
where in the second equality sign we changed $y \rightarrow -y$.

We found it better to come back to $\tanh$ in order to take care of the divergence at infinity.
The chosen contour of integration shown in Fig. (\ref{fig:figSM4}) is such that the integral over the arc converges to zero because of the exponent in the numerator of the integrand in Eq. (\ref{W-}).
The residues of the integrand of $W_{-}$ are at $y=y_{n}$ and defined as
\begin{align}
e^{-y_{n}} = 2\pi i \frac{T}{\theta_{0}}\left(n+\frac{1}{2} \right),
\end{align}
and according to the contour Fig. (\ref{fig:figSM4}) they are with $n<0$, such that $\mathrm{Im}[y_{n}] > 0$ and real part can be of any sign.
Then
\begin{align}
&
i\sinh(y_{n}) = \frac{1}{2}\left[ \frac{1}{2\pi \frac{T}{\theta_{0}}\left(n+\frac{1}{2} \right)} + 2\pi \frac{T}{\theta_{0}}\left(n+\frac{1}{2} \right)\right], \\
&
\left. \partial_{y}e^{\frac{\theta_{0}}{T} e^{-y} } \right|_{y=y_{n}}= 2\pi i \left(n+\frac{1}{2}\right). \\
&
\frac{1}{2}\left(e^{\frac{\theta_{0}}{T} e^{-y_{n}} }-1 \right) = -1.
\end{align}
Then
\begin{align}
W_{-} = 2\pi i \sum_{ \mathrm{res}} (...) 
& = - \frac{1}{2} \sum_{n<0}\frac{1}{n+\frac{1}{2}} e^{\pi b\left[ \frac{1}{2\pi \frac{T}{\theta_{0}}\left(n+\frac{1}{2} \right)} + 2\pi \frac{T}{\theta_{0}}\left(n+\frac{1}{2} \right)\right]} \\
&=  \frac{1}{2} \sum_{n\geq 0}\frac{1}{n+\frac{1}{2}} e^{-\pi b\left[ \frac{1}{2\pi \frac{T}{\theta_{0}}\left(n+\frac{1}{2} \right)} + 2\pi \frac{T}{\theta_{0}}\left(n+\frac{1}{2} \right)\right]}. 
\label{Wresult}
\end{align}
It is not clear how to analytically sum up the series above, however, in order to gain some insight, we can approximate the series by applying the Poisson summation formula and performing the steepest descent method.
We can then numerically plot the result of summation and compare it with the approximate analytics.

%-----------------------------------------------------------------------------------------------------------
\subsubsection{Poisson summation and steepest descent method}
%-----------------------------------------------------------------------------------------------------------
To sum the series Eq. (\ref{Wresult}) we again employ Poisson summation formula Eq. (\ref{PoissonSM}). 
\begin{align}
W_{-} 
& = 
\frac{1}{2}
\sum_{n\geq 0}\frac{1}{n+\frac{1}{2}} e^{-\pi b\left[ \frac{1}{2\pi \frac{T}{\theta_{0}}\left(n+\frac{1}{2} \right)} + 2\pi \frac{T}{\theta_{0}}\left(n+\frac{1}{2} \right)\right]} 
\\
&
=
\frac{1}{2}
\int_{0}^{\infty}\frac{1}{x+\frac{1}{2}}e^{-\pi b\left[ \frac{1}{2\pi \frac{T}{\theta_{0}}\left(x+\frac{1}{2} \right)} + 2\pi \frac{T}{\theta_{0}}\left(x+\frac{1}{2} \right)\right]} dx
+ 
\frac{1}{2}
\sum_{n \neq 0}\int_{0}^{\infty}\frac{e^{i2\pi n x}}{x+\frac{1}{2}}e^{-\pi b\left[ \frac{1}{2\pi \frac{T}{\theta_{0}}\left(x+\frac{1}{2} \right)} + 2\pi \frac{T}{\theta_{0}}\left(x+\frac{1}{2} \right)\right]} dx
\\
& 
\equiv  R_{0}(T) + R_{1}(T) + R_{-1}(T) + ... .
\end{align}
The non-oscillating part of the formula reads
\begin{align}\label{R0}
R_{0}(T) = \frac{1}{2}\int_{0}^{+\infty} \frac{dx}{x+\frac{1}{2}}  e^{-\pi b\left[ \frac{1}{2\pi \frac{T}{\theta_{0}}\left(x+\frac{1}{2} \right)} + 2\pi \frac{T}{\theta_{0}}\left(x+\frac{1}{2} \right)\right]} ,
\end{align}
we numerically estimate it and plot it as a function of temperature in Fig. (\ref{fig:figSM5}). 
%--------------------------------------------------------------------------------------------------------------------------------------------------------------------
\begin{figure}[h] 
\centerline{
\begin{tabular}{ccc}
\includegraphics[width=0.3 \columnwidth]{fig3.pdf}~~&
\includegraphics[width=0.3 \columnwidth]{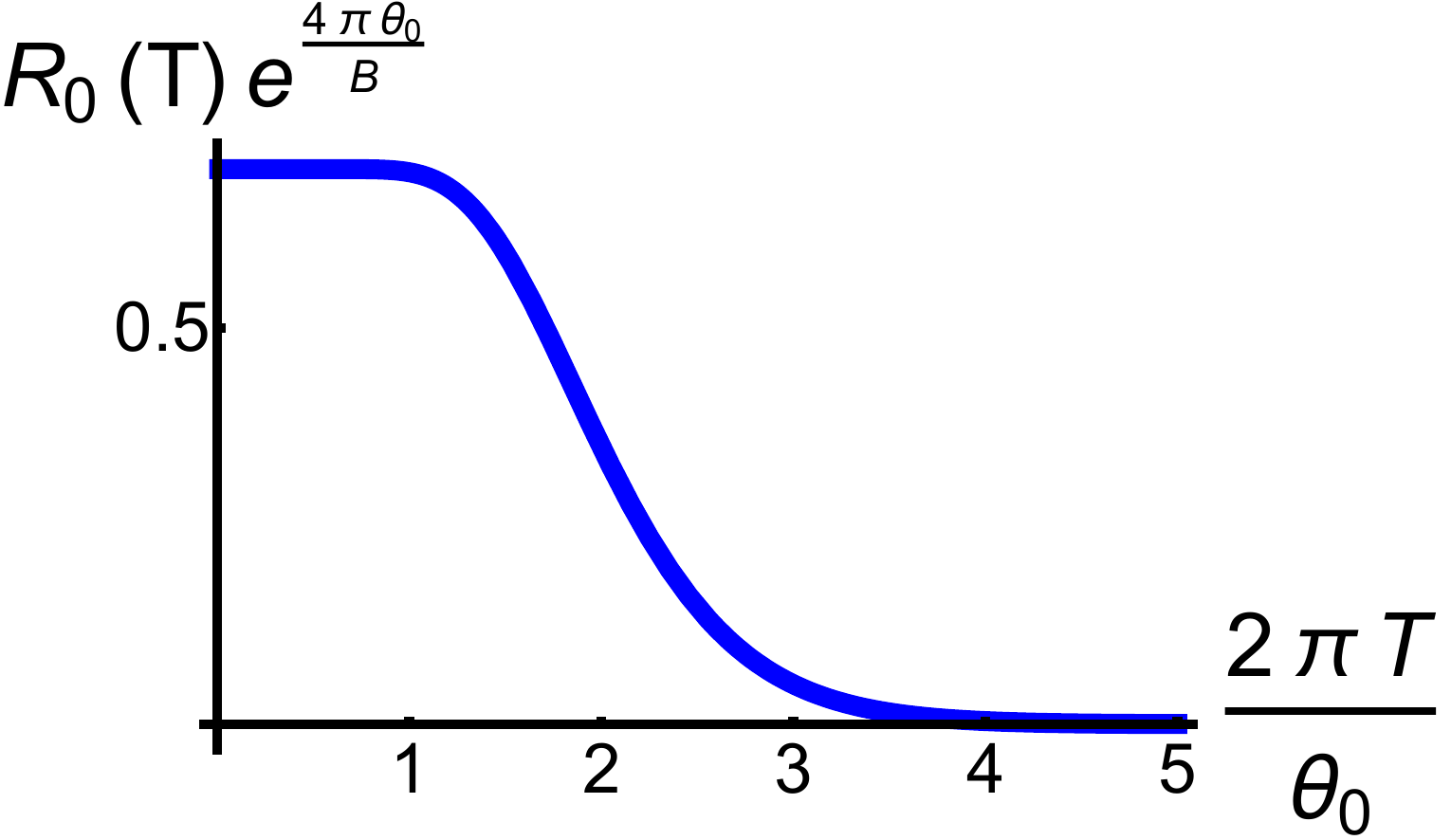}~~&
\includegraphics[width=0.3 \columnwidth]{fig3b.pdf}
\end{tabular}
}

\protect\caption{Plot of the numerically estimated expression Eq. (\ref{R0}), left for $b= \frac{2\theta_{0}}{\omega_{\mathrm{B}}} = 1$, center for $b= \frac{2\theta_{0}}{\omega_{\mathrm{B}}} = 2$, and right for $b= \frac{2\theta_{0}}{\omega_{\mathrm{B}}} = 0.2$.
It is consistent with standard Lifshits-Kosevich structure \cite{LK,LP}.}

\label{fig:figSM5}  

\end{figure}
%--------------------------------------------------------------------------------------------------------------------------------------------------------------------
We can estimate the integral using steepest descent method,
\begin{align}
R_{0}^{\mathrm{sd}}(T) = \frac{1}{2}
\frac{1}{ \frac{\theta_{0}}{2\pi T} \bar{z}}
e^{-\pi b \left( \frac{1}{\bar{z}} + \bar{z} \right)}
\int_{0}^{+\infty} e^{-\frac{1}{2}\vert g(\bar{z}) \vert \left( x + \frac{1}{2} -\frac{\theta_{0}}{2\pi T}\bar{z} \right)^2}dx,
\end{align}
where we defined
\begin{align}
&
\bar{z}= \sqrt{\left( \frac{1}{2\pi b}\right)^2 +1} - \frac{1}{2\pi b} = \frac{2\pi b}{1 + \sqrt{1 + (2\pi b)^2}  },\\
&
g(\bar{z})= \left( \frac{2\pi T}{\theta_{0}}\right)^2 \left.\partial^2_{z}\left[ -\ln(z)-\pi b\left( \frac{1}{z}+z \right) \right] \right|_{z=\bar{z}} 
= \left( \frac{2\pi T}{\theta_{0}}\right)^2  \left(\frac{1}{\bar{z}^2} - 2\pi b\frac{1}{\bar{z}^3}\right).
\end{align}
In Fig. (\ref{fig:figSM6}) we plot justification for the steepest descent method. The original curve is plotted in blue, while the curve corresponding to the steepest descent approximation is shown in red.  
This figure suggests that the position of the extremum is captured correctly, but for large magnetic fields (right plot in Fig. (\ref{fig:figSM6})) the approximation only qualitatively describes the integral.
We will show in the following that the position of the extremum defines period of quantum oscillations with inverse temperature.

%------------------------------------------------------------------------------------------------------------------------------------------------
\begin{figure}[h] 
\centerline{
\begin{tabular}{cc}
\includegraphics[width=0.4 \columnwidth]{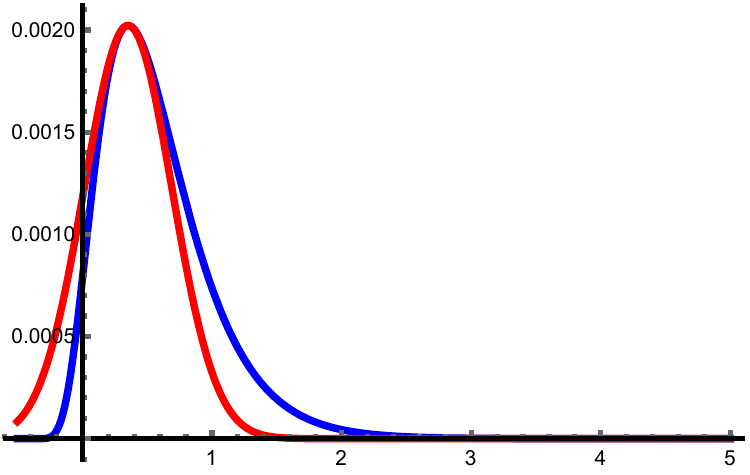} ~~&
\includegraphics[width=0.4 \columnwidth]{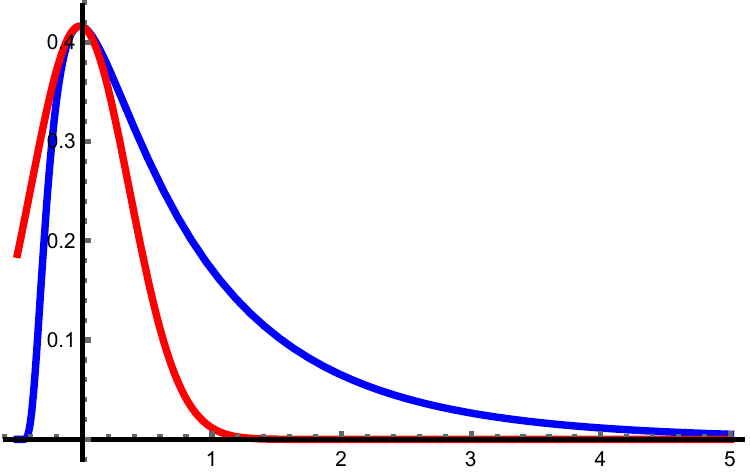}
\end{tabular}
}
\protect\caption{Approximating the integral within the steepest descent method. 
Left: for $\frac{2\theta_{0}}{\omega_{\mathrm{B}}} = 1$ and $2\pi \frac{T}{\theta_{0}} = 1$. Right:  for $\frac{2\theta_{0}}{\omega_{\mathrm{B}}} = 0.2$ and $2\pi \frac{T}{\theta_{0}} = 1$.
Red curve is the Gaussian obtained within the steepest descent method, while blue corresponds to the original curve.
In both cases the position of the saddle, which defines oscillations with inverse temperature, is perfectly approximated. For large magnetic fields corresponding to the right plot, steepest descent method does not do a good job approximating the magnitude of the integral. It works well for small magnetic fields corresponding to the left plot. 
}

\label{fig:figSM6}  

\end{figure}
%------------------------------------------------------------------------------------------------------------------------------------------------

{\bf (i)} We first consider $2\pi b > 1$ case. The steepest descent approximation is shown in the left plot in Fig. (\ref{fig:figSM6}). 
Then $\bar{z}\approx 1$ and $\vert g(1) \vert = 2\pi b \left( \frac{2\pi T}{\theta_{0}}\right)^2$, and we get
\begin{align}
R_{0}^{\mathrm{sd}}(T) = \frac{1}{2} \frac{2\pi T}{\theta_{0}} e^{-2\pi b}
\int_{0}^{+\infty} e^{- \pi b \left( \frac{2\pi T}{\theta_{0}}\right)^2  \left( x + \frac{1}{2} -\frac{\theta_{0}}{2\pi T} \right)^2 }dx,
\end{align}
which we plot in Fig. (\ref{fig:figSM5}). 
At $T=0$ the integral over $x$ can be approximated as Gaussian, i.e. from $-\infty$ to $+\infty$ in our case by shifting $x  \rightarrow x + \frac{1}{2} -\frac{\theta_{0}}{2\pi T}$ and setting $-\infty$ in the lower limit,
then
\begin{align}
R_{0}^{\mathrm{sd}}(T=0) \approx \frac{e^{-2\pi b}}{2\sqrt{ b}}.
\end{align}
At $T\rightarrow \infty$
\begin{align}
R_{0}^{\mathrm{sd}}(T\rightarrow \infty) \approx  
\frac{e^{-2\pi b}}{2\sqrt{ b}}\mathrm{Erfc}\left(\sqrt{\pi b} \frac{2\pi T}{\theta_{0}}\frac{1}{2} \right)
\approx
\frac{\theta_{0}}{2\pi T}
\frac{e^{-2\pi b}}{\pi b}
e^{-\frac{\pi b}{4}\left( \frac{2\pi T}{\theta_{0}} \right)^2},
\end{align}
where is the complimentary error function, whose expansion at large values of argument we used. 

Let us check our approximation with the exact results at $T=0$. 
We defined $R(T)$ in Eq. (\ref{R(T)sm}), then within the steepest descent approximation we have 
\begin{align}
R(T= 0 ) \approx R_{0}^{\mathrm{sd}}(T=0) = \frac{e^{-2\pi b}}{2\sqrt{ b}} = e^{-2\pi \frac{2\theta_{0}}{\omega_{\mathrm{B}}}} \frac{1}{2}\sqrt{\frac{\omega_{\mathrm{B}}}{2\theta_{0}}},
\end{align}
while the exact expression reads as 
\begin{align}
R(T=0) = e^{-2\pi \frac{2\theta_{0}}{\omega_{\mathrm{B}}}} \frac{1}{\sqrt{\pi}} \sqrt{\frac{\omega_{\mathrm{B}}}{2\theta_{0}}},
\end{align}
the two expressions are in good agreement since $\sqrt{\pi} \approx 1.77$. 
This slight discrepancy is in the fact that saddle point approximation is not exact.

Now we pick the $\pm n$ harmonics in the formula Eq. (\ref{PoissonSM}),
\begin{align}
R_{+n}(T) + R_{-n}(T)
& 
=
\frac{1}{2}
\int_{0}^{+\infty}e^{i2\pi n x}\frac{1}{x+\frac{1}{2}} e^{-\pi b\left[ \frac{1}{2\pi \frac{T}{\theta_{0}}\left(x+\frac{1}{2} \right)} + 2\pi \frac{T}{\theta_{0}}\left(x+\frac{1}{2} \right)\right]}dx + \mathrm{c.c.} 
\\
&
\approx
\frac{1}{2}
\frac{1}{ \frac{\theta_{0}}{2\pi T} \bar{z}}
e^{-\pi b \left( \frac{1}{\bar{z}} + \bar{z} \right)}
\int_{0}^{+\infty} e^{i2\pi n x} e^{-\frac{1}{2}\vert g(\bar{z}) \vert \left( x + \frac{1}{2} -\frac{\theta_{0}}{2\pi T}\bar{z} \right)^2}dx + \mathrm{c.c.} 
\\
&
=
\frac{1}{2}
\frac{1}{ \frac{\theta_{0}}{2\pi T} \bar{z}}
e^{-\pi b \left( \frac{1}{\bar{z}} + \bar{z} \right)}
e^{-\frac{2\pi^2 n^2}{\vert g(\bar{z}) \vert}}
e^{ i n\left(\frac{\theta_{0}}{T}\bar{z} - \pi \right)}
\int_{0}^{+\infty}  e^{-\frac{1}{2}\vert g(\bar{z}) \vert \left( x + \frac{1}{2} -\frac{\theta_{0}}{2\pi T}\bar{z} - i \frac{2\pi n}{\vert g(\bar{z}) \vert}\right)^2}dx + \mathrm{c.c.} 
\\
&
=
2\cos\left[n\left(\frac{\theta_{0}}{T}\bar{z} - \pi \right)\right]
 e^{-\frac{2\pi^2 n^2}{\vert g(\bar{z}) \vert}}
R_{0}^{\mathrm{sd}}(T).
\end{align}
Since steepest descent method gives $\bar{z} = 1$, we predict that there is a peak in $R(T)$ which occurs at 
\begin{align}\label{SMpeak}
T_{\mathrm{peak}} = \frac{\theta_{0}}{\pi}.
\end{align}
In Fig. (\ref{fig:figSM7}) we plot $R(T)$ for $b=1$, $b=3$ and $b=4$. There the peak indeed occurs at a value determined in Eq. (\ref{SMpeak}).
%---------------------------------------------------------------------------------------------------------------------------------------------
\begin{figure}[h] 
\centerline{
\begin{tabular}{cc}
\includegraphics[width=0.4 \columnwidth]{fig4a.pdf} ~~&
\includegraphics[width=0.4 \columnwidth]{fig4b.pdf} ~~\\
\includegraphics[width=0.4 \columnwidth]{fig5.pdf} ~~&
\includegraphics[width=0.4 \columnwidth]{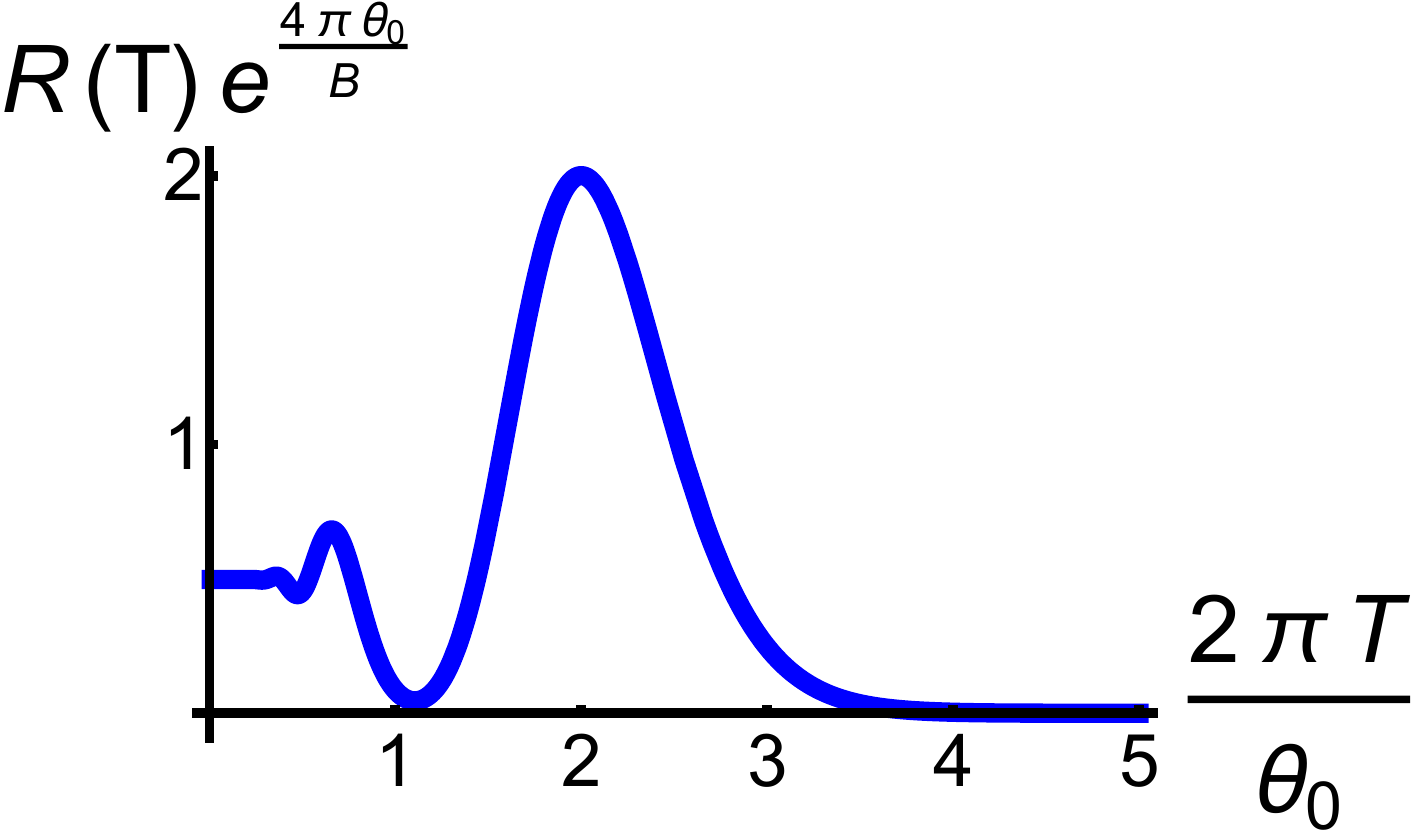}
\end{tabular}
}
\protect\caption{Top row, left: plot of the expression Eq. (\ref{Wresult}) times $e^{2\pi b}$ factor for $b= \frac{2\theta_{0}}{\omega_{\mathrm{B}}} = 1$. Top row, right: the same but with $\left[ 1 - 2\cos\left( \frac{\theta_{0}}{T}\right) \right]\frac{2}{3}$ (red) plotted together.
Now goes a proof of principle.
Bottom row, left is the plot of the expression Eq. (\ref{Wresult}) for $b= \frac{2\theta_{0}}{\omega_{\mathrm{B}}} = 3$. 
Bottom row, right is the plot of the expression Eq. (\ref{Wresult}) for $b= \frac{2\theta_{0}}{\omega_{\mathrm{B}}} = 4$. 
The magnitude more and more exponentially suppressed but with that more oscillation cycles, as compared with the first one if going from large temperatures, become visible.
In all of the figures the height of the peak equals to approximately $1$ (not universal). 
}

\label{fig:figSM7}  

\end{figure}
%--------------------------------------------------------------------------------------------------------------------------------------------------------------------

Let us estimate the height of the peak at $T_{\mathrm{peak}} = \frac{\theta_{0}}{\pi}$.
Numerics suggest, see Fig. (\ref{fig:figSM7}), that the peak has a value of almost $e^{-2\pi b}$, let us show it is indeed the case. 
We found it convenient to not perform the steepest descent for this task, but rather do the following transformations after application of the Poisson summation formula,
\begin{align}
R(T_{\mathrm{peak}}) = 
&
\frac{1}{2}\sum_{n}\int_{0}^{+\infty}e^{i2\pi n x}\frac{1}{x+\frac{1}{2}} e^{-\pi b\left[ \frac{1}{2\pi \frac{T_{\mathrm{peak}}}{\theta_{0}}\left(x+\frac{1}{2} \right)} + 2\pi \frac{T_{\mathrm{peak}}}{\theta_{0}}\left(x+\frac{1}{2} \right)\right]}dx 
\\
&
=
\frac{1}{2}
\sum_{n} e^{-i\pi n} \int_{\ln\left( 2\pi\frac{T_{\mathrm{peak}}}{\theta_{0}}\frac{1}{2}\right)}^{+\infty}e^{i n \frac{\theta_{0}}{T_{\mathrm{peak}}} e^{z}} e^{-2\pi b \cosh(z)}dz 
\\
&
=
\frac{1}{2}
\sum_{n} e^{-i\pi n} \int_{0}^{+\infty}e^{i n\pi e^{z}} e^{-2\pi b \cosh(z)}dz,
\end{align}
now goes our approximation, because $\cosh(z)$ has a minimum at $z=0$ (reminiscent of steepest descent approximation made above), $e^{z}\sim 1+z$ in the vicinity of $z=0$. 
Therefore, 
\begin{align}
\sum_{n} e^{-i\pi n} \int_{0}^{+\infty}e^{i n\pi e^{z}} e^{-2\pi b \cosh(z)}dz 
&
\approx
\sum_{n} \int_{0}^{+\infty}e^{i n\pi z} e^{-2\pi b \cosh(z)}dz 
\\
&
= 2 \sum_{m}\int_{0}^{+\infty}\delta(z-m) e^{-2\pi b \cosh(z)}dz 
\\
&
= 2 e^{-2\pi b } + 4 e^{-2\pi b \cosh(1) } + 4 e^{-2\pi b \cosh(2) } +..
\\
&
\approx
 2 e^{-2\pi b } +  4 e^{-2\pi b (1+0.5)  } + 4 e^{-2\pi b (1+2.75)  }.
\end{align}
For $b=1$ we have $4e^{-2\pi b*0.5   } \approx 0.17$ and the other terms in the series are practically zero.
For $b=3$ we have $4e^{-2\pi b*0.5   } \approx 3*10^{-4}$ which is also practically zero.
Therefore, we conclude that 
\begin{align}
R(T_{\mathrm{peak}}) \approx  e^{-2\pi b },
\end{align}
which is indeed numerically plotted in Fig. (\ref{fig:figSM7}), recalling that $b=\frac{2\theta_{0}}{\omega_{\mathrm{B}}}$. 
This result, we remind, is valid for $2\pi b=\frac{4\pi\theta_{0}}{\omega_{\mathrm{B}}}>1$.

%------------------------------------------------------------------------------------------------------------------------------------------------
{\bf (ii)} Let us now consider $2\pi b \sim 1$ and $2\pi b <1 $ cases. 

%------------------------------------------------------------------------------------------------------------------------------------------------
\begin{figure}[h] 
\centerline{
\begin{tabular}{cc}
\includegraphics[width=0.4 \columnwidth]{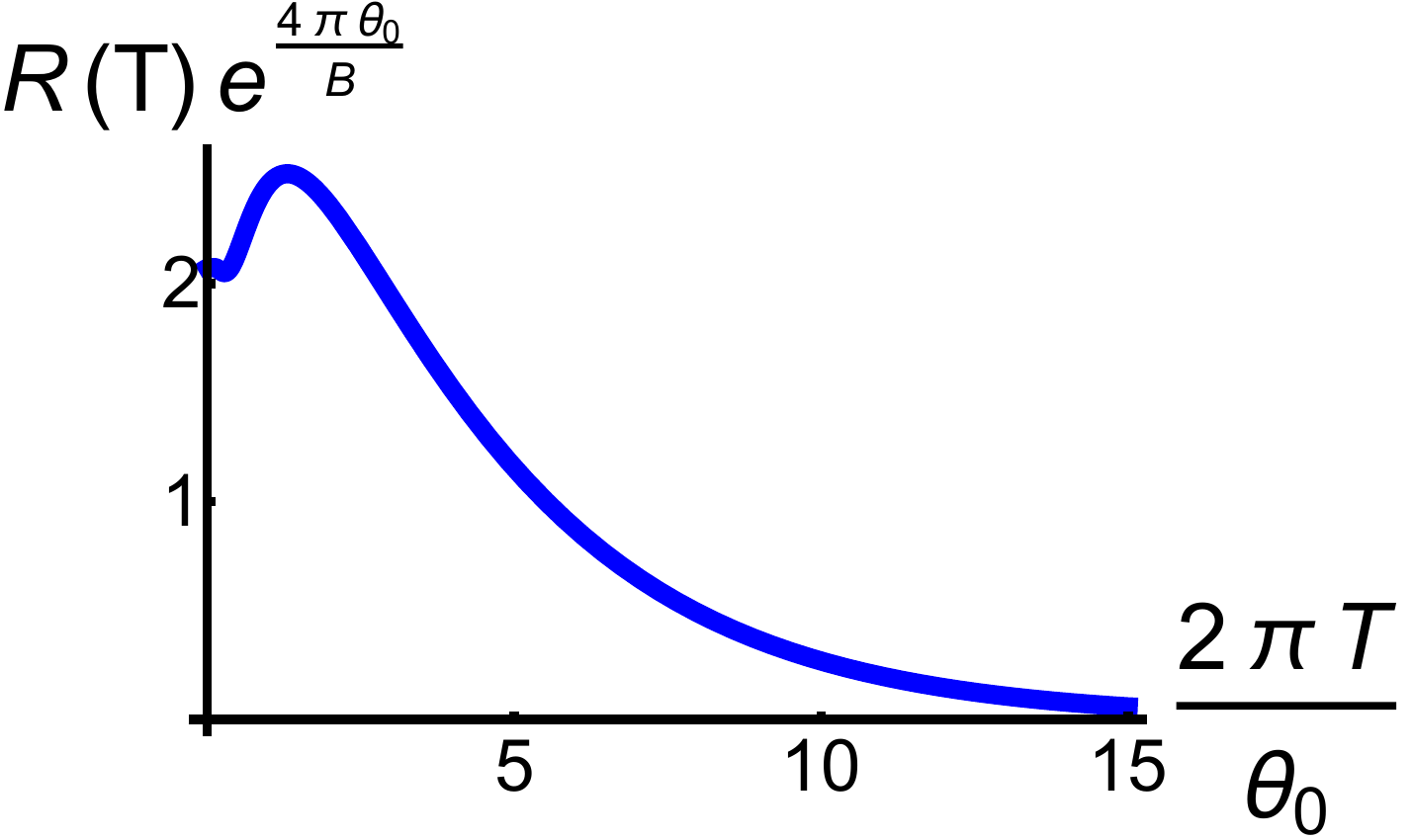} ~~&
\includegraphics[width=0.4 \columnwidth]{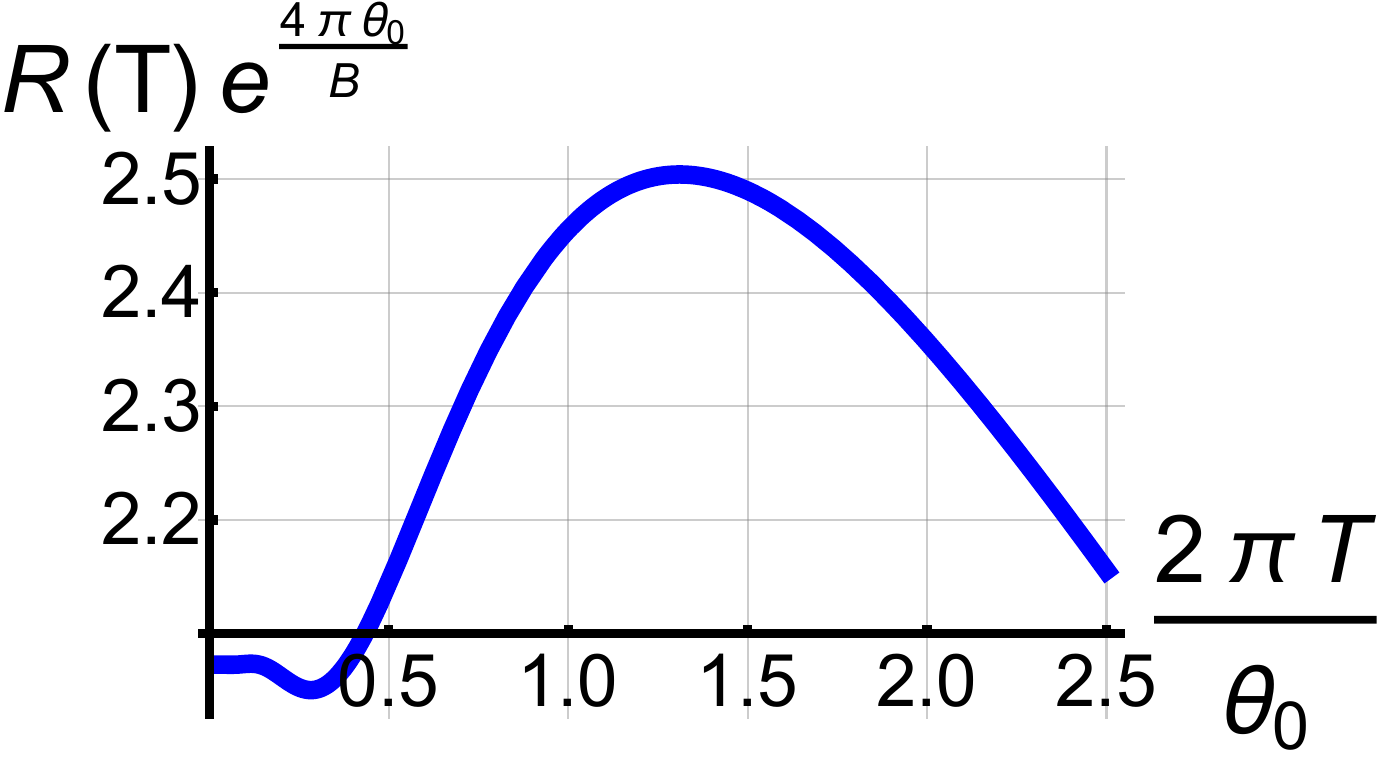} ~~\\
\includegraphics[width=0.4 \columnwidth]{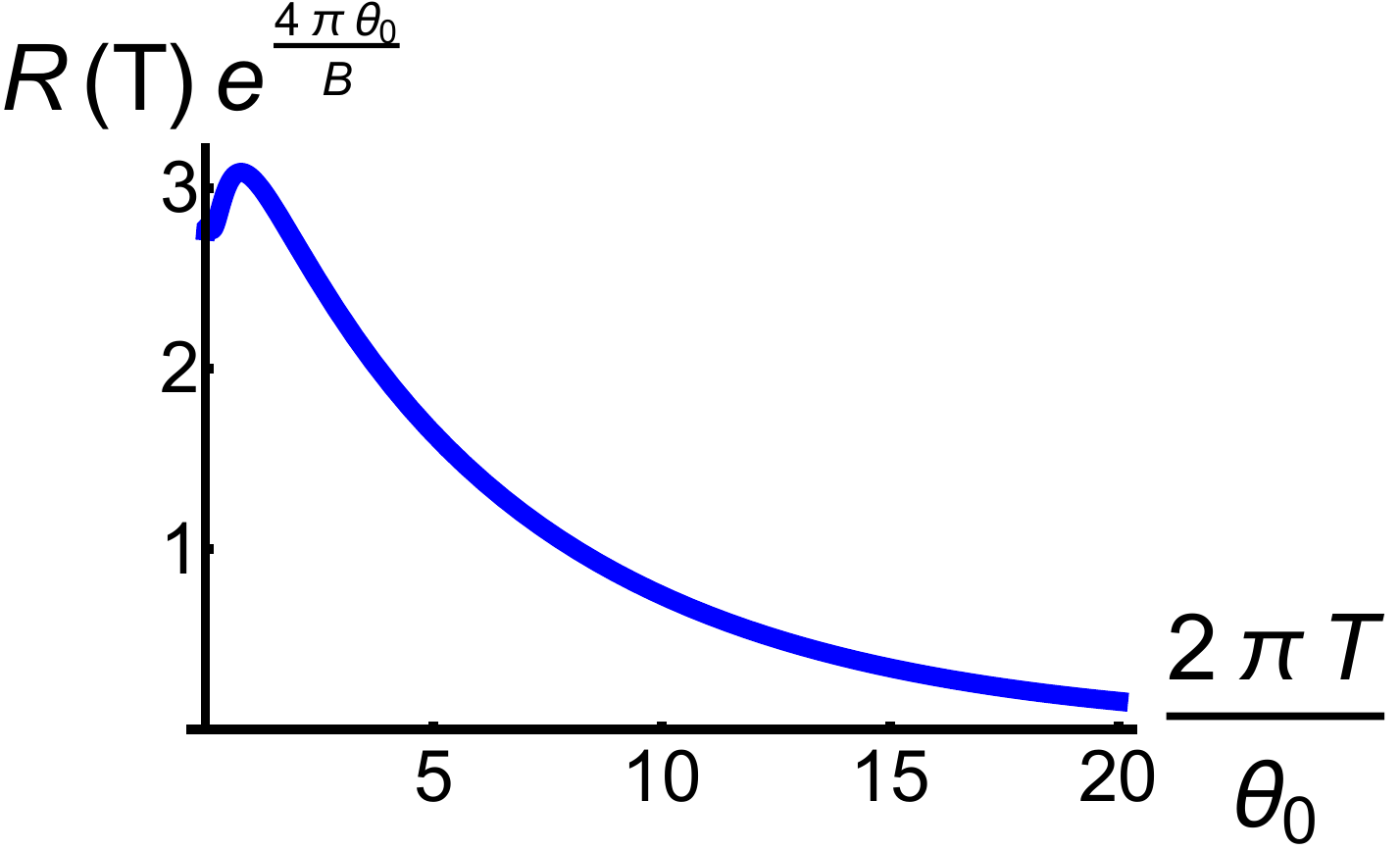} ~~&
\includegraphics[width=0.4 \columnwidth]{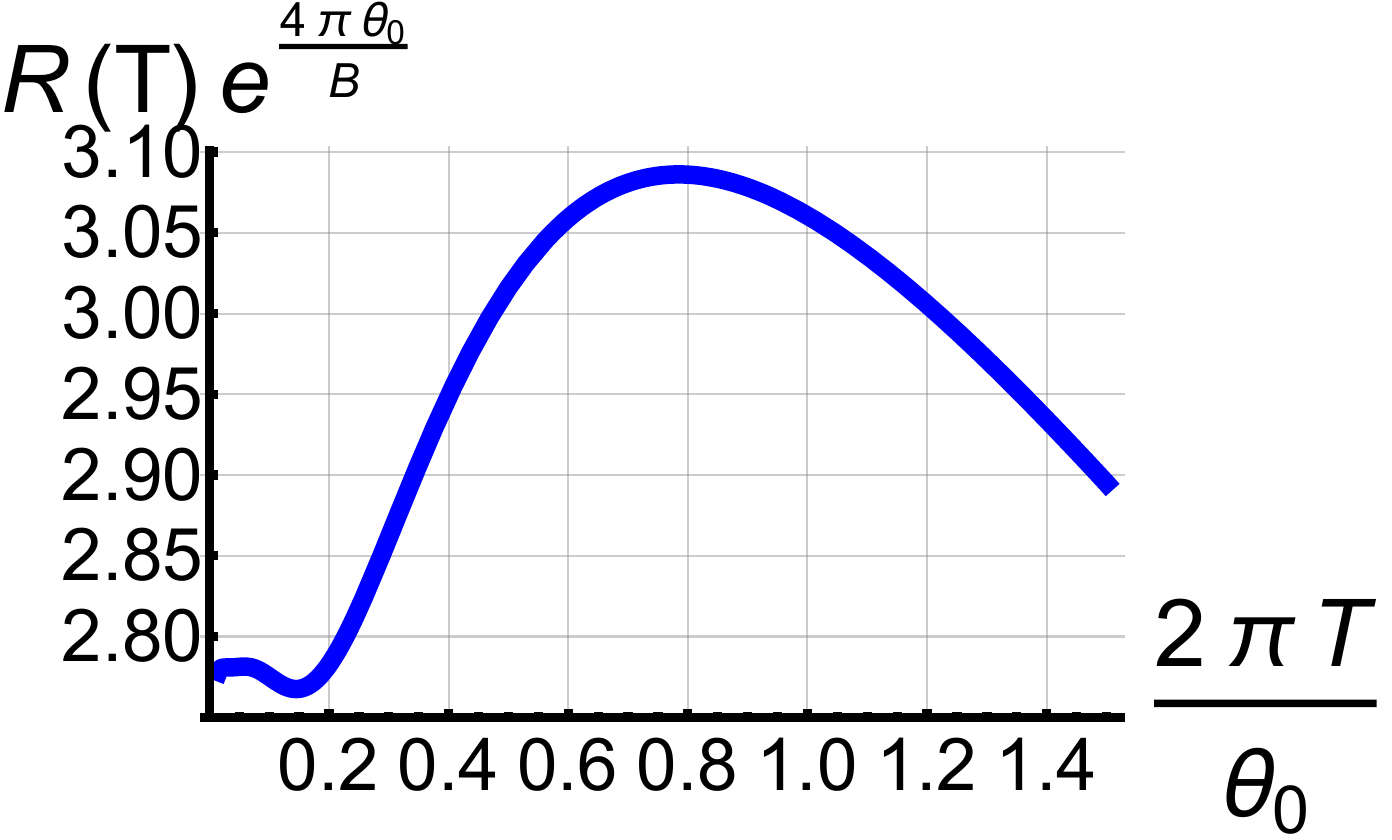} 
\end{tabular}
}
\protect\caption{Left: plot of $R(T)e^{\frac{4\pi \theta_{0}}{\omega_{\mathrm{B}}}}$ and right is the zoom in in to the peak for an eye determination of the position of the peak. 
Top is for $\frac{2\theta_{0}}{\omega_{\mathrm{B}}}=0.2$ while bottom is for $\frac{2\theta_{0}}{\omega_{\mathrm{B}}}=0.1$.
}

\label{fig:figSM8}  

\end{figure}
%------------------------------------------------------------------------------------------------------------------------------------------------
In particular we want to first focus on $b=0.2$ case as we think it might be relevant to the experiment.
For this value the steepest descent saddle point is defined by the most general expression,
\begin{align}
&
\bar{z} = \frac{2\pi b}{1+\sqrt{1+(2\pi b)^2}},
\\
&
g(\bar{z}) = \left( \frac{2\pi T}{\theta_{0}}\right)^2  \left(\frac{1}{\bar{z}^2} - 2\pi b\frac{1}{\bar{z}^3}\right).
\end{align}
Calculations and Fig. (\ref{fig:figSM6}b) suggest we have to be careful with the steepest descent and take in to account higher orders of approximation.
In particular, in Fig. (\ref{fig:figSM6}b) we see that the position of the saddle point is captured well with the approximation, however, the tail isn't.
The tail is going to contribute to the amplitude of the integral. We don't know how to estimate the tail. Let us see if the position of the peak in $R(T)$ is captured by the steepest descent.
\begin{align}
R_{+n}(T) + R_{-n}(T)  \approx 2\cos\left[n\left(2\pi \frac{\theta_{0}^2}{\omega_{\mathrm{B}}T} - \pi \right)\right]
 e^{-\frac{2\pi^2 n^2}{\vert g(\bar{z}) \vert}}R_{0}^{\mathrm{sd}}(T),
\end{align}
the peak occurs at 
\begin{align}\label{peakLarge}
T_{\mathrm{peak}} = \frac{4\theta_{0}^2}{\omega_{\mathrm{B}}}\frac{1}{{1+\sqrt{1+\left(\frac{4\pi \theta_{0}}{\omega_{\mathrm{B}}} \right)^2}}}  = \frac{4\theta_{0}^2}{{\omega_{\mathrm{B}}+\sqrt{\omega_{\mathrm{B}}^2+\left(4\pi \theta_{0}\right)^2}}},
\end{align}
which for small values of $\frac{4\pi \theta_{0}}{\omega_{\mathrm{B}}}$ becomes $T_{\mathrm{peak}} = \frac{2\theta_{0}^2}{\omega_{\mathrm{B}}}$.
Let us check that our approximation works. 
From right plots of Fig. (\ref{fig:figSM8}) we observe that for $\frac{2\theta_{0}}{\omega_{\mathrm{B}}}=0.2$ the position of the peak is at $\frac{2\pi T^{(\mathrm{plot})}_{\mathrm{peak}}}{\theta_{0}} \approx 1.32$ which should be compared with the corresponding quantity from the Eq. (\ref{peakLarge}), which is $\frac{2\pi T^{(\mathrm{analytics})}_{\mathrm{peak}}}{\theta_{0}} \approx 0.96$. It turns out that for whatever reason better approximation is $T^{*}_{\mathrm{peak}} = \frac{2\theta_{0}^2}{\omega_{\mathrm{B}}}$, which gives  $\frac{2\pi T^{*}_{\mathrm{peak}}}{\theta_{0}} =2\pi b \approx 1.25$. For $\frac{2\theta_{0}}{\omega_{\mathrm{B}}}=0.1$ we have $\frac{2\pi T^{(\mathrm{plot})}_{\mathrm{peak}}}{\theta_{0}} \approx 0.78$ which is compared with $\frac{2\pi T^{(\mathrm{analytics})}_{\mathrm{peak}}}{\theta_{0}} \approx 0.57$ and $\frac{2\pi T^{*}_{\mathrm{peak}}}{\theta_{0}} =2\pi b \approx 0.62$. We conclude that the steepest descent approximation is a poor approximation for $\frac{4\pi \theta_{0}}{\omega_{\mathrm{B}}} <1 $ case.

%-----------------------------------------------------------------------------------------------------------
\subsubsection{Alternative calculation of Eq. (\ref{Wresult})  }
%-----------------------------------------------------------------------------------------------------------
It is instructive to perform a different calculation of the Eq. (\ref{Wresult}) suggested to us by I.S. Burmistrov.
\begin{align}
W_{-} =  \sum_{n\geq 0}\frac{1}{n+\frac{1}{2}} e^{-\pi b\left[ \frac{1}{2\pi \frac{T}{\theta_{0}}\left(n+\frac{1}{2} \right)} + 2\pi \frac{T}{\theta_{0}}\left(n+\frac{1}{2} \right)\right]},
\end{align}
where $b=\frac{2\theta_{0}}{\omega_{\mathrm{B}}}$.
We use 
\begin{align}
\int_{-\infty}^{+\infty} \cos(x) e^{-ax^2}dx = \sqrt{\frac{\pi}{a}}e^{-\frac{1}{4a}}
\end{align}
to transform the $ e^{-\pi b\left[ \frac{1}{2\pi \frac{T}{\theta_{0}}\left(n+\frac{1}{2} \right)} \right]}$ term with an $a = \frac{1}{2 b} \frac{T}{\theta_{0}}\left(n+\frac{1}{2} \right)$ identification to
\begin{align}
e^{-\pi b\left[ \frac{1}{2\pi \frac{T}{\theta_{0}}\left(n+\frac{1}{2} \right)} \right]} 
= \sqrt{\frac{a}{\pi}} \int_{-\infty}^{+\infty} \cos(x) e^{- \frac{1}{2b} \frac{T}{\theta_{0}} \left(n+\frac{1}{2} \right) x^2 }dx.
\end{align}
Next, we use
\begin{align}
\int_{-\infty}^{+\infty} e^{-\left(n+\frac{1}{2}\right)y^2}dy = \sqrt{\frac{\pi}{n+\frac{1}{2}} }
\end{align} 
to lift the $\sqrt{\frac{1}{n+\frac{1}{2}} }$ in to the exponent. Then,
\begin{align}
W_{-} &=  \sum_{n\geq 0}\frac{1}{n+\frac{1}{2}} e^{-\pi b\left[  2\pi \frac{T}{\theta_{0}}\left(n+\frac{1}{2} \right)\right]}
\sqrt{\frac{a}{\pi}} \int_{-\infty}^{+\infty} \cos(x) e^{- \frac{1}{2b} \frac{T}{\theta_{0}} \left(n+\frac{1}{2} \right) x^2 }dx
\\
&=
\sqrt{\frac{1}{2\pi b} \frac{T}{\theta_{0}}}
\sum_{n\geq 0}\frac{1}{\sqrt{n+\frac{1}{2} }} e^{-\pi b\left[  2\pi \frac{T}{\theta_{0}}\left( n+\frac{1}{2} \right)\right]}
 \int_{-\infty}^{+\infty} \cos(x) e^{- \frac{1}{2b} \frac{T}{\theta_{0}} \left(n+\frac{1}{2} \right) x^2 }dx
\\
&=
\frac{T}{\pi}
\sqrt{\frac{1}{2 b \theta_{0}} }
\sum_{n\geq 0}
e^{-\pi b\left[  2\pi \frac{T}{\theta_{0}}\left( n+\frac{1}{2} \right)\right]}
\int_{-\infty}^{+\infty} e^{-T\left(n+\frac{1}{2}\right)y^2}dy 
 \int_{-\infty}^{+\infty} \cos(x) e^{- \frac{1}{2b} \frac{T}{\theta_{0}} \left(n+\frac{1}{2} \right) x^2 }dx
\end{align}
We then use
\begin{align}
\sum_{n\geq 0} e^{-g\left(n+\frac{1}{2}\right)} = e^{-\frac{g}{2}}\frac{e^{g}}{e^{g} -1} = \frac{1}{2\sinh\left( \frac{g}{2}\right)}
\end{align}
for $g>0$. In our case 
\begin{align}
g &= 2\pi^2 b \frac{T}{\theta_{0}} +  \frac{T}{2b\theta_{0}}x^2 + T y^2
\\
&
=   \frac{4\pi^2T}{\omega_{\mathrm{B}}} +  \frac{T\omega_{\mathrm{B}}}{4\theta_{0}^2}x^2 + T y^2.
\end{align}
Then, 
\begin{align}
W_{-} &= 
\frac{T}{2\pi \theta_{0}} 
\sqrt{\omega_{\mathrm{B}}} 
\int_{-\infty}^{+\infty} dy 
\int_{-\infty}^{+\infty} dx 
\frac{\cos(x) }{\sinh\left( \frac{4\pi^2T}{\omega_{\mathrm{B}}} +  \frac{T\omega_{\mathrm{B}}}{4\theta_{0}^2}x^2 + T y^2\right)}
\\
&=
\frac{T}{2\pi } 
\int_{-\infty}^{+\infty} dy 
\int_{-\infty}^{+\infty} dx 
\frac{\cos(x) }{\sinh\left[ \frac{4\pi^2T}{\omega_{\mathrm{B}}} +  \frac{T\omega_{\mathrm{B}}}{4\theta_{0}^2}(x^2 + y^2) \right]}.
\end{align}
This path is not an easy one, but it is consistent with the obtained $T=0$ limit.

%-----------------------------------------------------------------------------------------------------------
\subsection{Conventional Lifshits-Kosevich expression for the temperature dependence}\label{sectionLK}
%-----------------------------------------------------------------------------------------------------------
As a remark, let us contrast our calculation with the standard Lifshits-Kosevich expression \cite{LK,LP} for the temperature dependence of dHvA oscillation of a metal.
We derive it from scratch. We consider two-dimensional non-interacting fermion system described by $\epsilon_{\bf k}=\frac{{\bf k}^2}{2m} - \mu$ in perpendicular magnetic field.
Free energy of the system in the magnetic field is now 
\begin{align}
F = - T \int_{{\bf k}} \ln\left( 1+e^{-\frac{\epsilon_{\bf k}}{T}}\right) \rightarrow - T\omega_{\mathrm{B}}\nu \sum_{n} \ln\left( 1+e^{-\frac{\epsilon_{n}}{T}}\right),
\end{align}
where $\epsilon_{n} = \omega_{\mathrm{B}}\left(n+\frac{1}{2}\right) - \mu$. We again utilize the Poisson summation formula
\begin{align}
F = - T\omega_{\mathrm{B}}\nu \int_{0}^{\infty} \ln\left( 1+e^{-\frac{\epsilon_{x}}{T}}\right) dx  - T\omega_{\mathrm{B}}\nu\sum_{p}  \int_{0}^{\infty}  e^{i2\pi p x}\ln\left( 1+e^{-\frac{\epsilon_{x}}{T}}\right) dx,
\end{align}
from where we study only the second term,
\begin{align}
F_{\mathrm{osc}} \equiv
-T\omega_{\mathrm{B}}\nu\sum_{p}  \int_{0}^{\infty}  e^{i2\pi p x}\ln\left( 1+e^{-\frac{\epsilon_{x}}{T}}\right) dx 
&
\rightarrow 
-T\omega_{\mathrm{B}}\nu \sum_{p}\frac{\omega_{\mathrm{B}}}{2\pi p i T}  \int_{0}^{\infty}  \frac{e^{i2\pi p x}}{ e^{\frac{\epsilon_{x}}{T}} +1 } dx 
\\
&
=
-T\omega_{\mathrm{B}}^2\nu  \sum_{p}\frac{1}{2\pi p i }e^{i2\pi p \left( \frac{\mu}{\omega_{\mathrm{B}}} - \frac{1}{2} \right)}  \int_{-\frac{\mu}{T} + \frac{\omega_{\mathrm{B}}}{2T}}^{\infty}  \frac{e^{i2\pi p \frac{T}{\omega_{\mathrm{B}}} z}}{ e^{z} +1 } dz 
\end{align}

We observe that an integral defining the temperature dependence of the dHvA oscillation amplitude instead of Eqs. (\ref{W}), (\ref{W-}), and (\ref{Wresult}) equals (for example see Ref. \cite{LP}) now to 
\begin{align}
I_{\mathrm{LK}}(\alpha) = \int_{-\infty}^{+\infty}\frac{e^{i\alpha z}}{e^{z}+1} dz,
\end{align}
where $\alpha = 2\pi p \frac{T}{\omega_{\mathrm{B}}}$. The contour of integration is the same as in Fig. (\ref{fig:figSM4}), and the residues are at $z_{n} = 2\pi  i \left( n+\frac{1}{2}\right)$ with $n>0$. 
Then,
\begin{align}\label{LKtemperature}
I_{\mathrm{LK}}(\alpha) 
&
= - 2\pi i \sum_{n=0} e^{-2\pi \left(n+\frac{1}{2} \right)\alpha} 
\\
&
=  - 2\pi i e^{-\pi \alpha} \sum_{n=0} e^{-2\pi n \alpha} 
\\
&
= - \frac{\pi i }{\sinh(\pi \alpha)},
\end{align}
where we used 
\begin{align}
\frac{1}{\sinh(\pi \alpha)} = 2 e^{-\pi \alpha}\sum_{n = 0} e^{-2\pi n \alpha} .
\end{align}
Then for $p=\pm 1$
\begin{align}
F^{(p=\pm 1)}_{\mathrm{osc}} & = 
-T\omega_{\mathrm{B}}^2 \nu \frac{1}{2\pi i }e^{i2\pi \left( \frac{\mu}{\omega_{\mathrm{B}}} - \frac{1}{2} \right)}  \int_{-\frac{\mu}{T} + \frac{\omega_{\mathrm{B}}}{2T}}^{\infty}  \frac{e^{i2\pi \frac{T}{\omega_{\mathrm{B}}} z}}{ e^{z} +1 } dz  + \mathrm{c.c.}
\\
&
=
-T\omega_{\mathrm{B}}^2\nu\frac{1}{2\pi i }e^{i2\pi \left( \frac{\mu}{\omega_{\mathrm{B}}} - \frac{1}{2} \right)} I_{\mathrm{LK}}\left( 2\pi \frac{T}{\omega_{\mathrm{B}}} \right) + \mathrm{c.c.}
\\
&
=
\frac{\omega_{\mathrm{B}}^3}{2\pi^2}\nu \cos\left( 2\pi \frac{\mu}{\omega_{\mathrm{B}}} - \pi \right) \frac{2 \pi^2 \frac{T}{\omega_{\mathrm{B}}}}{\sinh\left(2 \pi^2 \frac{T}{\omega_{\mathrm{B}}} \right)}.
\end{align}
We see that there is no quantum oscillations as a function of temperature in metallic case. At zero temperature $\frac{2 \pi^2 \frac{T}{\omega_{\mathrm{B}}}}{\sinh\left(2 \pi^2 \frac{T}{\omega_{\mathrm{B}}} \right)} \rightarrow 1$, and the oscillating part of the free energy is proportional to the $\omega_{\mathrm{B}}^3$.
Therefore, the magnetization is propotional to $\propto \omega_{\mathrm{B}}\mu \sin\left( 2\pi \frac{\mu}{\omega_{\mathrm{B}}} - \pi \right)$ for the case when $\mu \gg \omega_{\mathrm{B}}$.

%--------------------------------------------------------------------------------------------------------------------------------
\subsection{Three-dimensional case}
%--------------------------------------------------------------------------------------------------------------------------------
Let us study an integral which appears due to the integraion over the $k_{z}$ momentum,
\begin{align}
I=
\int_{0}^{+\infty} e^{i\alpha z^2} dz 
\end{align}
which is calculated using the contour integration, for $\alpha>0$,
\begin{align}
\int_{0}^{+\infty} e^{i\alpha z^2} dz + \int_{(1+i)\infty}^{0}e^{i\alpha z^2}dz = 0,
\end{align}
where integral over the arc vanishes.
Therefore,
\begin{align}
I = -\int_{(1+i)\infty}^{0}e^{i\alpha z^2}dz ,
\end{align}
we make a change of variable to $z= u e^{i\frac{\pi}{4}}$, then
\begin{align}
I = e^{i\frac{\pi}{2}}\int_{0}^{\infty}e^{-\alpha u^2}du = e^{i\frac{\pi}{2}} \sqrt{\frac{\pi}{2\alpha}}.
\end{align}
Therefore, it is straightforward to generalize calculation of the dHvA oscillations to the three-dimensional case.

\end{widetext}

%--------------------------------------------------------------------------------------------------------------------------------------------------------------
%--------------------------------------------------------------------------------------------------------------------------------------------------------------
%--------------------------------------------------------------------------------------------------------------------------------------------------------------
%--------------------------------------------------------------------------------------------------------------------------------------------------------------
%--------------------------------------------------------------------------------------------------------------------------------------------------------------


\begin{references}


\bibitem{LiScience2014} G. Li, Z. Xiang, F. Yu, T. Asaba, B. Lawson, P. Cai, C. Tinsman, A. Berkley,
S. Wolgast, Y.S. Eo, D.-J. Kim, C. Kurdak, J.W. Allen, K. Sun, X.H. Chen,
Y.Y. Wang, Z. Fisk, and L. Li, Science {\bf 346}, 1208 (2014).
%\textit{Two-dimensional Fermi surfaces in Kondo insulator SmB6}

\bibitem{SuchitraScience2015} B.S. Tan, Y.-T. Hsu, B. Zeng, M.C. Hatnean, N. Harrison, Z. Zhu,
M. Hartstein, M. Kiourlappou, A. Srivastava, M.D. Johannes, T.P. Murphy,
J.-H. Park, L. Balicas, G.G. Lonzarich, G. Balakrishnan, and S.E. Sebastian, Science {\bf 349}, 287 (2015).
%\textit{Unconventional Fermi surface in an insulating state}

\bibitem{SuchitraNatPhys2017} M. Hartstein, W.H. Toews, Y.-T. Hsu, B. Zeng, X. Chen, M.C. Hatnean, Q.R. Zhang, S. Nakamura, A.S. Padgett, G. Rodway-Gant, J. Berk, M.K. Kingston, G.H. Zhang, M.K. Chan, S. Yamashita, T. Sakakibara, Y. Takano, J.-H. Park, L. Balicas, N. Harrison, N. Shitsevalova, G. Balakrishnan, G. G. Lonzarich, R. W. Hill, M. Sutherland, and S.E. Sebastian, Nature Physics {\bf 14}, 166 (2018). 
%\textit{Fermi surface in the absence of a Fermi liquid in the Kondo insulator SmB6}

\bibitem{SuchitraScience2020} M. Hartstein, H. Liu, Y.-T. Hsu, B.S. Tan, M.C. Hatnean, G. Balakrishnan,
and S.E. Sebastian, iScience {\bf 23}, 101632 (2020).
%\textit{Intrinsic Bulk Quantum Oscillations in a Bulk Unconventional Insulator SmB6}

\bibitem{BaskaranArxiv} G. Baskaran, arXiv: 1507.03477 (2015).
%\textit{Majorana Fermi Sea in Insulating SmB6: A proposal and a Theory of Quantum Oscillations in Kondo Insulators}

\bibitem{SodemannChowdhurySenthil} I. Sodemann, D. Chowdhury, and T. Senthil, Phys. Rev. B {\bf 97}, 045152 (2017).
%\textit{Quantum oscillations in insulators with neutral Fermi surfaces}

\bibitem{VarmaPRB} C.M. Varma, Phys. Rev. B {\bf 102}, 155145 (2020).
%\textit{Majoranas in mixed-valence insulators}

\bibitem{KnolleCooperPRL15} J. Knolle and N.R. Cooper, Phys. Rev. Lett. {\bf 115}, 146401 (2015).
%\textit{ Quantum oscillations without a fermi surface and the anomalous de Haas-Van Alphen effect.} 

\bibitem{TopodHvA2016} L. Zhang, X.-Y. Song, and F. Wang, Phys. Rev. Lett. {\bf 116}, 046404 (2016)
%\textit{Quantum Oscillation in Narrow-Gap Topological Insulators}

\bibitem{AlisultanovJETP2016}  Z.Z. Alisultanov JETP Letters {\bf 104}, 187 (2016).
%\textit{ On quantum oscillations in a tunable graphene bilayer }

\bibitem{PalPRB2016} H.K. Pal, F. Piechon, J.-N. Fuchs, M. Goerbig, and G. Montambaux, Phys. Rev. B {\bf 94}, 125140 (2016).
%\textit{Chemical potential asymmetry and quantum oscillations in insulators}

\bibitem{PalArxiv2022} G. Singh and H. Pal, Phys. Rev. B {\bf 108}, L201103 (2023).
%\textit{Effect of many-body interaction on de Haas-van Alphen oscillations in insulators}

\bibitem{Hewson} A.C. Hewson \textit{"The Kondo Problem to Heavy Fermions"}, Cambridge University Press, 1993.

\bibitem{KeldyshKopaev}  L.V. Keldysh and Yu.V. Kopaev, Soviet Physics - Solid State. {\bf 6}, 2219 (1965).
%\textit{Possible instability of the semimetallic state toward Coulomb interaction"}




\bibitem{Miyake} K. Miyake, Physica B {\bf 186-188}, 115 (1993).
%\textit{de Haas-van Alphen oscillations in superconducting states as a probe of gap anisotropy}

\bibitem{AlloccaCooper} A. Allocca and N. Cooper, SciPost Phys. {\bf 12}, 123 (2022).
%\textit{Quantum oscillations in interaction-driven insulators}


\bibitem{LK} I.M. Lifshitz and A.M. Kosevich. Sov. Phys. JETP {\bf 2}, 636 (1956).

\bibitem{LP} E.M. Lifshits and L.P. Pitaevskii, \textit{Statistical Physics, Part 2: Course of Theoretical Physics - Vol. 9}, Elsevier 2014.




\bibitem{Tinkham} M. Tinkham, \textit{Introduction to Superconductivity}, Dover Publications (Mineola, New York, 1996).


%\bibitem{AokiJPSJ2014} H. Aoki, N. Kimura, and T. Terashima
%Journal of the Physical Society of Japan {\bf 83}, 072001 (2014).
%\textit{Fermi Surface Properties, Metamagnetic Transition and Quantum Phase Transition of CeRu2Si2 and Its Alloys Probed by the dHvA Effect}

\bibitem{Li_PRX2022}
Z. Xiang, K.-W. Chen, L. Chen, T. Asaba, Y. Sato, N. Zhang, D. Zhang, Y. Kasahara, F. Iga, W.A. Coniglio, Y. Matsuda, J. Singleton, and L. Li,
Phys. Rev. X {\bf 12}, 021050 (2022).
%\textit{Hall Anomaly, Quantum Oscillations and Possible Lifshitz Transitions in Kondo Insulator YbB12: Evidence for Unconventional Charge Transport}





\bibitem{ExpInAs/GaSb1} D. Xiao, C.-X. Liu, N. Samarth, and L.-H. Hu, Phys. Rev. Lett. {\bf 122}, 186802 (2019).
%\textit{Anomalous Quantum Oscillations of Interacting Electron-Hole Gases in Inverted Type-II InAs=GaSb Quantum Wells}


\bibitem{ExpInAs/GaSb2} Z. Han, T. Li, L. Zhang, G. Sullivan, and R.-R. Du, Phys. Rev. Lett. {\bf 123}, 126803 (2019).
%\textit{Anomalous Conductance Oscillations in the Hybridization Gap of InAs/GaSb Quantum Wells}


\bibitem{SpivakZyuzinCobden} B.Z. Spivak, A.Yu. Zyuzin, and D.H. Cobden, Phys. Rev. Lett. {\bf 95}, 226804 (2005).
%\textit{Mesoscopic Oscillations of the Conductance of Disordered Metallic Samples as a Function of Temperature}

\bibitem{DzeroSunGalitskiColeman} M. Dzero, K. Sun, V. Galitski, and P. Coleman, Phys. Rev. Lett. {\bf 104}, 106408 (2010).
%\textit{Topological Kondo insulators}

\bibitem{SovietTI1}B. A. Volkov and O. A. Pankratov, Pis’ma Zh. Eksp. Teor. Fiz. {\bf 42}, 145 (1985) [JETP Lett. {\bf 42}, 178 (1985)].

\bibitem{SovietTI2} O.A. Pankratov, S.V. Pakhomov, and B.A. Volkov, Solid State Communications, {\bf 61}, 93 (1987).
%\textit{Supersymmetry in heterojunctions: Band-inverting contact on the basis of Pb1-xSnxTe and Hg1-xCdxTe}


\bibitem{Kamenev} A. Kamenev, \textit{Field theory of non-equilibrium systems}, Cambridge, University Press, 2012.

\bibitem{SchwieteFinkelstein} G. Schwiete and A.M. Finkel'stein Phys. Rev. B 90, 155441 (2014). 
%\textit{Renormalization group analysis of thermal transport in the disordered Fermi liquid}.








\end{references}
\end{document}